\documentclass[12pt]{article}
\usepackage{epsfig,amssymb,amsmath,psfrag,subfigure,rotate}

\allowdisplaybreaks
\newcommand{\insertfig}[2]{\includegraphics[width=#1cm]{#2}}

\def \be  {\begin{equation}}
\def \ee  {\end{equation}}
\def \ba  {\begin{eqnarray}}
\def \ea  {\end{eqnarray}}
\def \baa {\begin{eqnarray*}}
\def \eaa {\end{eqnarray*}}
\def \lab #1 {\label{#1}}

\newcommand\re[1]{(\ref{#1})}
\def\d{\hbox{{d}\kern-.20em\hbox{l}}}

\def \matrix #1 {\left(\begin{array}{cc} #1 \end{array}\right)}

\def \tr {\mathop{\rm tr}\nolimits}
\def \res{\mathop{\rm res}\nolimits}

\newcommand \vev [1] {\langle{#1}\rangle}

\newcommand \ket [1] {|{#1}\rangle}
\newcommand \bra [1] {\langle {#1}|}
\newcommand{\bit}[1]{\mbox{\boldmath$#1$}}

\newcommand{\ft}[2]{{\textstyle\frac{#1}{#2}}}
\begin{document}

\begin{titlepage}

\thispagestyle{empty}

\vspace*{3cm}

\centerline{\large \bf Nonsinglet pentagons and NMHV amplitudes}
\vspace*{1cm}

\centerline{\sc A.V.~Belitsky}

\vspace{10mm}

\centerline{\it Department of Physics, Arizona State University}
\centerline{\it Tempe, AZ 85287-1504, USA}

\vspace{2cm}

\centerline{\bf Abstract}

\vspace{5mm}

Scattering amplitudes in maximally supersymmetric gauge theory receive a dual description in terms of the expectation value of the super Wilson loop 
stretched on a null polygonal contour. This makes the analysis amenable to nonperturbative techniques. Presently, we elaborate on a refined form of the 
operator product expansion in terms of pentagon transitions to compute twist-two contributions to NMHV amplitudes. To start with, we provide a novel 
derivation of scattering matrices starting from Baxter equations for flux-tube excitations propagating on magnon background. We propose bootstrap 
equations obeyed by pentagon form factors with nonsinglet quantum numbers with respect to the R-symmetry group and provide solutions to them to 
all orders in 't Hooft coupling. These are then successfully confronted against available perturbative calculations for NMHV amplitudes to four-loop
order.

\end{titlepage}

\setcounter{footnote} 0

\newpage

\pagestyle{plain}
\setcounter{page} 1

{\footnotesize \tableofcontents}

\newpage

\section{Introduction}

Scattering amplitudes encode interaction of asymptotic states in a theory and are thus of pa\-ra\-mount importance for unravelling
underlying particle dynamics as a function of kinematical variables $s_n$ and their coupling constant $g^2$, $A = A (s_n; g^2)$. Typically, 
one restricts oneself to a fixed (rather low) order in $g^2$ and computes them exactly in $s_n$ using conventional perturbative
or unitarity based techniques \cite{Dixon:2013uaa}. In this case however, the goal of uncovering  nonperturbative dependence on the coupling 
is completely out of reach. One can attempt to consider instead a particular kinematical setup and attempt an all-order resummation that corresponds to
selection of particular types of Feynman diagrams that dominate in a given region. The problem with these is that the calculation of 
subleading kinematical corrections is highly problematic due to the fact that one has to enlarge the class of graphs involved in the
analysis.

These complications were recently overcome in maximally sypersymmetric gauge theory where an equivalent description of superamplitudes $\mathcal{A}$ 
was uncovered in the form of the vacuum expectaction value of the super Wilson loop $\vev{\mathcal{W}}$ stretched on a closed polygonal contour formed by 
particles' light-like momenta \cite{Mason:2010yk,CaronHuot:2010ek,Belitsky:2011zm} generalizing a deep insight for their bosonic counterparts 
\cite{Alday:2007hr,Drummond:2007cf,Brandhuber:2007yx}. A near-collinear limit for the latter can be reformulated as an operator product expansion 
\cite{Alday:2010ku,Gaiotto:2010fk,Gaiotto:2011dt} for the superloop in terms of nonlocal operators representing excitations of the so-called color flux tube 
\cite{Alday:2007mf,Basso:2013vsa,Basso:2013aha,Basso:2014koa}. These enter the series exponentially suppressed with the power proportional to their twist,
\begin{align}
\label{GenericOPE}
\vev{\mathcal{W}} = {\rm e}^{- \tau} f_1 + {\rm e}^{- 2 \tau} f_2 + \dots
\, .
\end{align}
Each contribution is determined by a correlation function of the aforementioned operators that are built up by a $\Pi$-shaped Wilson link operator with  
elementary field insertions in the middle segment \cite{Belitsky:2011nn,Sever:2012qp}. The renormalization group evolution of $f_i = f_i (\tau)$ in the time 
variable $\tau$ is governed by anomalous dimensions of the flux-tube excitations \cite{Korchemsky:1995be,Belitsky:2006en,Beisert:2006ez,Basso:2010in}, 
also known as Gubser-Klebanov-Polyakov excitations in the dual string picture \cite{Gubser:2002tv}. At one-loop order, the problem of calculating these 
energies can be mapped into an integrable noncompact open spin chain \cite{Belitsky:2011nn,Belitsky:2014rba}. The corresponding eigenfunctions play 
an indispensable role as well, since their overlap in different conformal frames defines multiparticle transitions and coupling of flux-tube excitations to the 
Wilson loop contour \cite{Basso:2013aha,Belitsky:2014rba}. These were dubbed pentagon transitions \cite{Basso:2013vsa}, since they receive a clear 
geometrical interpretation in a tessellation scheme of the Wilson loop in question. It turns out that these multiparticle pentagons factorize into single-particle 
transitions as a natural consequence of underlying integrable dynamics. The advantage of this refined form of the operator product expansion is 
that it can be promoted to all orders in 't Hooft coupling $g^2 = g_{\scriptscriptstyle\rm YM}^2 N/(4 \pi)^2$ becoming a truly nonperturbative framework for computing 
the near-collinear limit of scattering amplitudes \cite{Basso:2013vsa,Basso:2013aha,Basso:2014koa}. One of the ingredients in the construction, the spectrum of 
flux-tube excitations was solved to all orders in coupling in Ref.\ \cite{Basso:2010in} making use of the integrability of the dilatation operator in $\mathcal{N} = 4$ 
super Yang-Mills theory \cite{Beisert:2005fw}. The second building block, the pentagons, were conjectured to obey a system of axiomatic equations \cite{Basso:2013vsa} 
which have their roots in the integrable scattering dynamics of flux-tube excitations \cite{Belitsky:2006en,Dorey:2011gr,Basso:2011rc,Basso:2013pxa,Fioravanti:2013eia}. 
The solution to these lead to remarkable predictions for single $f_1$ \cite{Basso:2013aha} and two-particle $f_2$ \cite{Basso:2014koa} contributions to the bosonic 
Wilson loop \re{GenericOPE}, which is known to provide an equivalent description for maximal helicity-violating (MHV) amplitudes. One-particle contributions were 
also conjectured for some components of the superloop \cite{Basso:2013aha}. When expanded at weak coupling, the operator product expansion results were shown 
to be in striking agreement with available multiloop calculations \cite{Dixon:2011pw,Dixon:2014voa}.

\begin{figure}[t]
\begin{center}
\mbox{
\begin{picture}(0,170)(160,0)
\put(0,-150){\insertfig{15}{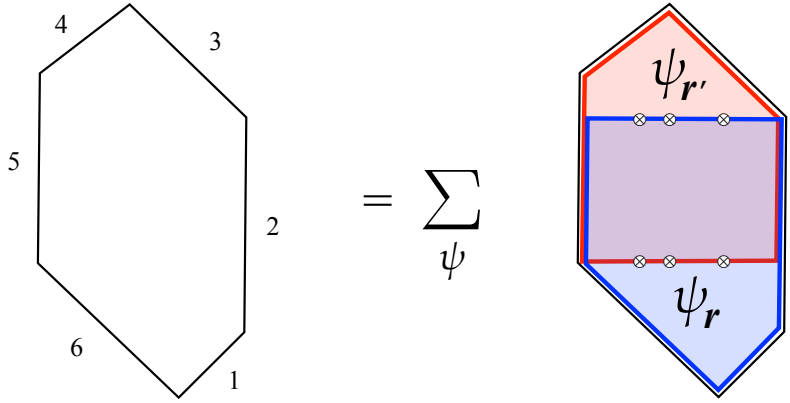}}
\end{picture}
}
\end{center}
\caption{ \label{HexagonTesselation} Hexagon Wilson loop decomposition into pentagon transitions.}
\end{figure}

In the present paper, we will focus on the operator product expansion for NMHV amplitudes, specifically on twist-two effects. We will choose the hexagon as a case
of study. According to the approach of Ref.\ \cite{Basso:2013aha}, an appropriately subtracted hexagon super Wilson loop admits the decomposition
\begin{align}
\mathcal{W}_6
= 
\sum_{\psi_1, \psi_2}
\bra{0} \widehat{\mathcal{P}} \ket{\psi_{\bit{\scriptstyle r}}} \bra{\psi_{\bit{\scriptstyle r}}} 
{\rm e}^{- \tau \widehat{H} + i \sigma \widehat{P} + i \phi \widehat{J}} 
\ket{\psi_{\bit{\scriptstyle r}'}} \bra{\psi_{\bit{\scriptstyle r}'}}
\widehat{\mathcal{P}} \ket{0}
\, ,
\end{align}
where $\widehat{\mathcal{P}}$ is a pentagon operator that creates a (multi)particle state $\psi_{\bit{\scriptstyle r}}$ in the representation $\bit{r}$ with respect
to the $R$-symmetry group SU(4) from the vacuum.  The state then propagates with the `evolution' operator $ {\rm e}^{- \tau \widehat{H} + i \sigma \widehat{P} 
+ i \phi \widehat{J}}$ and then gets absorbed back into the vacuum. The above representation is driven by a particular tessellation of the hexagon into
three null squares as demonstrated in Fig.\ \ref{HexagonTesselation}. The operators $\widehat{H}$, $\widehat{P}$ and $\widehat{J}$ are the
remaining conformal symmetries of the middle square and thus the variables $\tau$, $\sigma$ and $\phi$ parametrize all conformally inequivalent hexagons.
The eigenvalues of the conformal symmetry generators on the intermediate states 
\begin{align*}
\widehat{H} \ket{\psi_{\bit{\scriptstyle r}}} = E (g) \ket{\psi_{\bit{\scriptstyle r}}}
\, , \qquad
\widehat{P} \ket{\psi_{\bit{\scriptstyle r}}} = p (g) \ket{\psi_{\bit{\scriptstyle r}}}
\, , \qquad
\widehat{J} \ket{\psi_{\bit{\scriptstyle r}}} = m \ket{\psi_{\bit{\scriptstyle r}}}
\, ,
\end{align*}
correspond to energy, momentum and angular momentum of contributing flux-tube excitation and they are known to all order in 't Hooft coupling $g$ as we 
already emphasized earlier. It is the form factors 
\begin{align*}
F (0| \psi_{\bit{\scriptstyle r}}) \equiv \bra{0} \widehat{\mathcal{P}} \ket{\psi_{\bit{\scriptstyle r}}} 
\end{align*}
which require a dedicated study. The single-particle contributions were addressed in full detail in Ref.\  \cite{Basso:2013aha}, while two-particle states in
the singlet channel of SU(4) were considered in \cite{Basso:2014koa}. In this paper, we will be interested in two-particle form factors with nonsinglet quantum 
numbers with respect to the R-symmetry group, with the emphasis on two-fermion and scalar-gluon intermediate state $\ket{\psi_\alpha} = 
\ket{{\rm fermion}, {\rm fermion}}, \ket{{\rm hole}, {\rm gauge}}$. Along this way, we will be able to unravel the structure of certain Grassmann components of the
super Wilson loops and, as a consequence of duality, corresponding NMHV amplitudes.

To prepare the playground for this analysis, we will construct from scratch scattering matrices for the main flux-tube excitations. This problem was addressed
in the literature independently in Ref.\ \cite{Fioravanti:2013eia} and \cite{Basso:2013pxa,Basso:2013aha,Basso:2014koa} to all orders in 't Hooft coupling. 
However, the results of the former approach are not  well-suited for our needs, while the second one lacks some of the S-matrices necessary for our
considerations. Our approach while different in the starting point and intermediate manipulations will yield results which admit the same final form as in the 
second framework allude to above. Namely, we will use Baxter equations for spin-chain excitations propagating on magnon background as a main tool to
find flux-tube equations. This will be accomplished in Sect.\ \ref{FluxTubeEquationsSection}. Next, in Sect.\ \ref{SmatricesSection}, we will use the same 
formalism to deduce the S-matrices for any two-particle scattering. Since the solution to the bootstrap equations for pentagon transitions requires scattering matrices in 
the mirror kinematics, we address this issue in detail in Sect.~\ref{MirrorKinematicsSection}. The appendices review details of the mirror transformation for flux-tube 
excitations that was rather concise in previous papers on the subject. Finally, we turn to the construction of pentagon form factors for two-particle states
in Sect.\ \ref{PentagonsSection} by proposing a set of equations that they have to obey. These functional equations are then solved exactly and compared 
with available perturbative data for hexagon NMHV amplitude yielding a complete agreement.

\section{From magnon Baxter to flux tube equations}
\label{FluxTubeEquationsSection}

A complimentary, and actually the original, view \cite{Belitsky:2006en,Alday:2007mf,Beisert:2006ez,Basso:2010in} on flux-tube excitations emerges from the 
study of single-trace local operators with large conformal spin. In this framework, a flux-tube excitation arises as an insertion of an elementary field $X$ into the 
string of covariant derivatives $D_+$ projected on the light cone,
\begin{align}
\label{TwistOp}
O_X = \tr \left[ Z (0) D_+ \dots D_+ X(0) D_+ \dots D_+ Z (0) \right]
\, . 
\end{align}
Here the scalar $Z$-fields merely serve as a source/sink for the gauge flux. We will rely below on this picture since we can then immediately adopt the all-order
Bethe Ansatz equations \cite{Beisert:2005fw} available for this problem. We will derive the spectrum generating flux-tube equations in the next few sections. However,
compared to the analysis of Ref.\ \cite{Basso:2010in} that heavily relied on the introduction of the so-called counting function, we will use the formalism of the Baxter 
equation as a central point of the consideration. The latter appears to be more readily suited for the analysis of the large-spin asymptotics naturally incorporating 
particle-hole transformations  as was demonstrated earlier at one-loop in Ref.\ \cite{Belitsky:2006en} and to all-orders for the hole in Ref.\ \cite{Belitsky:2006wg}. Below, 
we generalize this to the entire spectrum of fundamental flux-tube excitations and their bound states.

\subsection{Hole/scalar excitation}

To start with, let us recall that the Baxter equation encoding the spectrum of anomalous dimensions of the local operator of the type \re{TwistOp} with
$S$ covariant derivatives and $X = Z^{L-2}$ scalar field insertions distributed along the string of derivatives  takes the form \cite{Belitsky:2006wg}
\begin{align}
\label{Deltas}
\Delta_+ (u^+) Q (u+i) + \Delta_- (u^-) Q (u-i) = t (u) Q (u)
\, .
\end{align}
Here $Q$ is a Baxter polynomial in the rapidity variable $u$ with its zeroes determined by the Bethe roots $u_j = u_j (g)$
\begin{align}
Q (u) = \prod_{j = 1}^S (u - u_j)
\, .
\end{align}
The dressing factors
\begin{align}
\label{ExplicitDeltaPM}
\Delta_\pm (u) = x^L 
\prod_{j=1}^S
\left(
1 - \frac{g^2}{x x_j^\mp}
\right)^{-2}
{\rm e}^{- i \Theta (x, x_j)}
\, , 
\end{align}
depend on the Zhukowski variable
\begin{align}
x = x[u] = \ft12 \left( u + \sqrt{u^2 - (2g)^2} \right)
\end{align}
with conventional notation introduced for $x^\pm \equiv x [u^\pm]$ and $u^\pm = u \pm \ft{i}{2}$. The magnon phase $\Theta (x, x_j)$ reads  
\cite{Beisert:2006ib,Beisert:2006ez}
\begin{align}
\label{ThetaPhase}
\Theta (x, y)
=
4
\sum_{m = 1}^\infty \sum_{n = 0}^\infty (- 1)^{n + m} \mathcal{Z}_{2m, 2n+1} (g) 
\left[ 
\frac{q_{2n+1} (y)}{x^{2m}} - \frac{q_{2m} (y)}{x^{2 n + 1}}
\right]
\, ,
\end{align}
with expansion coefficients admitting a representation in terms of integrals of Bessel functions $J_n$,
\begin{align}
\mathcal{Z}_{m, n} (g)
=
g^{m + n} \int_0^\infty \frac{dt}{t} \frac{J_{m} (2 g t) J_{n} (2 g t)}{{\rm e}^t - 1}
\, , 
\end{align}
and the $q_n$ being the kernels of the conserved spin chain charges
\begin{align}
q_n (x) =(x^+)^{- n} - (x^-)^{- n}
\, . 
\end{align}
The last ingredient in the right-hand side of Eq.\ \re{Deltas} is the transfer matrix $t(u)$. In the large-spin limit, it was demonstrated by a thorough 
analysis in Ref.\ \cite{Belitsky:2006en} that it scales as $t \sim O(S^2)$ for $S \to \infty$ and thus only half of the Baxter equation
survive at leading order in the inverse spin expansion.

\subsubsection{Flux-tube equations}

To proceed further, let us take the logarithm of both sides of Eq.\ \re{Deltas} and perform a Fourier transformation of the dressing factors, e.g., 
\begin{align}
\label{GammaEq}
\ln \Delta_+ (u) = L \ln x + \int_0^\infty \frac{dt}{t} {\rm e}^{i u t} \Gamma (2 g t)
\, .
\end{align}
assuming that $\Im{\rm m} [u]>0$ for convergence. Introducing even and odd combinations of the $\Gamma$-function 
\begin{align}
\label{EvenOggComponents}
\Gamma_{\pm} (t) = \ft{1}{2} \left[ \Gamma (t) \pm \Gamma (-t) \right]
\, ,
\end{align}
we can expand each of them in a Neumann series over the Bessel functions \cite{Kotikov:2006ts,Benna:2006nd,Basso:2007wd},
\begin{align}
\label{NeumannSeries}
\Gamma_+ (t) = 2 \sum_{n \geq 1} (2 n) J_{2 n} (t) \Gamma_{2 n}
\, , \qquad
\Gamma_- (t) = 2 \sum_{n \geq 1} (2 n-1) J_{2 n-1} (t) \Gamma_{2 n-1}
\, .
\end{align}
These form orthogonal bases of functions for even/odd indices
\begin{align}
\label{OrthogonalityBessel}
\int_0^\infty \frac{dt}{t} J_{2n} (t) J_{2m} (t) = \frac{\delta_{nm}}{2 (2n)}
\, , \qquad
\int_0^\infty \frac{dt}{t} J_{2n+1} (t) J_{2m+1} (t) = \frac{\delta_{nm}}{2 (2n+1)}
\, .
\end{align}
This way, Eq.\ \re{GammaEq} can be rewritten as
\begin{align}
\label{Eq15}
2 \sum_{n \geq 1} \Gamma_n \left( \frac{i g}{x} \right)^n
=
- 2 \sum_{j=1}^S \ln \left( 1 - \frac{g^2}{x x_j^\mp} \right) - i \sum_{j=1}^S \Theta (x, x_j)
\, ,
\end{align}
where we used the value of the integral
\begin{align}
\int_0^\infty \frac{dt}{t} {\rm e}^{i u t} J_n (2 g t) = \left( \frac{i g}{x} \right)^n
\, ,
\end{align}
valid for $\Im{\rm m} [u] >0$. We can easily extract equations for individual expansions coefficients $\Gamma_n$ by integrating Eq.\
\re{Eq15} over a closed contour in $x$, which translates into a contour in $u = x + g^2/x$ which encircles the cut $[-2g, 2g]$ in 
the complex plane, namely
\begin{align}
\label{Gamman}
\Gamma_n
=
- \frac{1}{2 \pi i} \oint \frac{dx}{x} \left( \frac{x}{i g} \right)^n \ln Q (u^+)
+ d_n
\, ,
\end{align}
where we employed the equality between the integrals
\begin{align}
\label{FromLogXtoLogU}
\oint \frac{dx}{x} \left( \frac{x}{i g} \right)^n \ln \left( 1 - \frac{g^2}{x x_j^-} \right)  
=
\oint \frac{dx}{x} \left( \frac{x}{i g} \right)^n \ln \left( u^+ - u_j \right)  
\, ,
\end{align}
in the first term and $d_n$'s are the coefficient induced by the magnon dressing phase
\begin{align}
d_n \equiv
- \frac{1}{4 \pi} \sum_{j=1}^S \oint \frac{dx}{x} \left( \frac{x}{i g} \right)^n \Theta (x, x_j)
\, .
\end{align}
Making use of the explicit form of the latter given by the right-hand side of Eq.\ \re{ThetaPhase}, we can find $d_n$'s for even and odd value of 
$n$ as follows
\begin{align}
\label{Evends}
d_{2n}
&
=
\int_0^\infty \frac{dt}{t} \frac{J_{2n} (2gt)}{{\rm e}^t - 1}
\left[
\Gamma_- (2gt) + \bar\Gamma_- (2gt)
\right]
\, , \\
\label{Oddds}
d_{2n + 1}
&
=
\int_0^\infty \frac{dt}{t} \frac{J_{2n + 1} (2gt)}{{\rm e}^t - 1}
\left[
\Gamma_+ (2gt) - \bar\Gamma_+ (2gt)
\right]
\, ,
\end{align}
where we relied on the relation between the conserved charges and Fourier coefficients $\Gamma_n$ as can be easily established 
from the ratio of the dressing factors $\Delta_+(u)/\Delta_-(u)$,
\begin{align}
&
\sum_{j=1}^S q_{2n} (x_j) = (- 1)^{n + 1} (2 n) g^{- 2n} \left( \Gamma_{2n} - \bar{\Gamma}_{2n} \right)
\, , \\
&
\sum_{j=1}^S q_{2n+1} (x_j) = i (- 1)^{n} (2 n + 1) g^{- 1 - 2n} \left( \Gamma_{2n + 1} + \bar{\Gamma}_{2n + 1} \right)
\, .
\end{align}
It might appear that Eqs.\ \re{Evends} and \re{Oddds} are vacuous since they seem to define $d$'s in terms of $d$'s. 
However, a naked eye inspection of the last term in Eq.\ \re{Gamman} immediately demonstrates that for even $n$, $d_n$'s are 
real, $\bar{d}_{2n} = d_{2n}$, while for odd $n$, they are imaginary, i.e., $\bar{d}_{2n+1} = - d_{2n+1}$. This implies that in the 
right-hand side of \re{Evends} and \re{Oddds}, only the first term in Eq.\ \re{Gamman} induces a nontrivial contribution. This
observation was first made in Ref.\ \cite{Basso:2011rc}. Thus, the equations for even and odd $\Gamma_n$ take the form
\begin{align}
\label{InterEvenGamma}
&
\Gamma_{2n} - \int_0^\infty \frac{dt}{t} 
\frac{J_{2n} (2gt)}{{\rm e}^t - 1} \left[ \Gamma_- (2gt) + \bar{\Gamma}_- (2gt) \right]
=
-
\frac{1}{2 \pi i} \oint \frac{dx}{x} \left( \frac{x}{i g} \right)^{2n} \ln Q (u^+)
\, , \\
\label{InterOddGamma}
&
\Gamma_{2n - 1} - \int_0^\infty \frac{dt}{t} 
\frac{J_{2n - 1} (2gt)}{{\rm e}^t - 1} 
\left[ \Gamma_+ (2gt) - \bar{\Gamma}_+ (2gt) \right]
=
-
\frac{1}{2 \pi i} \oint \frac{dx}{x} \left( \frac{x}{i g} \right)^{2n - 1} \ln Q (u^+)
\, ,
\end{align}
respectively. Analogous equations for $\bar{\Gamma}$ are simply obtained from the above by complex conjugation. 

As a next step, we have to solve the Baxter equation \re{Deltas} for arbitrary $\Gamma$'s and substitute its solution into the right-hand sides 
of above equations, which once solved will determine the unknown functions $\Gamma$. As we will demonstrate below, see Eqs.\ \re{HalfBEQplus}-\re{HalfBaxterHole}, 
the solution to the half-Baxter equation admits a factorized form
\begin{align}
\label{Qdecomposition}
Q (u) = Q^\Gamma (u) \delta Q (u)
\, ,
\end{align}
where
\begin{align}
\label{Qgamma}
\ln Q^\Gamma (u) = \int_0^\infty \frac{dt}{t} \frac{ {\rm e}^{i u t} \Gamma (2gt) - 2gt \Gamma_1}{{\rm e}^t - 1}
\, ,
\end{align}
is valid in the upper half-plane, and we used the definition
\begin{align}
J_n (2 g t) = \frac{1}{2 \pi i} \oint \frac{dx}{x} {\rm e}^{i u t} \left( \frac{x}{i g} \right)^n = \frac{1}{2 \pi i} \oint \frac{dx}{x} {\rm e}^{i u t} \left( \frac{g}{i x} \right)^n
\, .
\end{align}
Then decomposing $\Gamma$, as before, in even and odd components according to Eq.\ \re{EvenOggComponents},
we can rewrite Eqs.\ \re{InterEvenGamma} and \re{InterOddGamma} as follows
\begin{align}
\Gamma_{2 n} 
&+  \int_0^\infty \frac{dt}{t} \frac{J_{2n} (2gt)}{{\rm e}^t - 1} \left[ \Gamma_+ (2gt) - \bar\Gamma_- (2gt) \right]
=
\chi_{2n}
\, , \\
\Gamma_{2 n - 1} 
&+  \int_0^\infty \frac{dt}{t} \frac{J_{2n - 1} (2gt)}{{\rm e}^t - 1} \left[ \Gamma_- (2gt) - \bar\Gamma_+ (2gt) \right]
=
\chi_{2n-1}
\, , 
\end{align}
where the inhomogeneities are
\begin{align}
\label{ChiSourceEven}
\chi_{2n}
&
\equiv
-\frac{1}{2 \pi i} \oint \frac{dx}{x} \left( \frac{x}{i g} \right)^{2n} \ln \delta Q (u^+)
\, , \\
\label{ChiSourceOdd}
\chi_{2n-1}
&
\equiv
-\frac{1}{2 \pi i} \oint \frac{dx}{x} \left( \frac{x}{i g} \right)^{2n - 1} \ln \delta Q (u^+)
\, .
\end{align}

To make contact with the analysis in Ref.\ \cite{Basso:2010in}, we introduce the following linear combinations of the function $\Gamma$
and its complex conjugate
\begin{align}
\label{GammaTogamma}
\gamma (t) = \ft{1}{2} 
\left[ 
\Gamma (t) + \bar\Gamma (t)
\right]
\, , \qquad
\widetilde\gamma (t) = \ft{i}{2} 
\left[ 
\Gamma (t) - \bar\Gamma (t)
\right]
\, .
\end{align}
Forming appropriate linear combination of equations derived above, we find the final form of the flux-tube equations
\begin{align}
\label{GKPequation1}
\gamma_n
&
+
\int_0^\infty \frac{dt}{t} \frac{J_n (2gt)}{{\rm e}^t - 1} \left[ \gamma_+ (2gt) - (-1)^n \gamma_- (2gt) \right] = \kappa_n
\, , \\
\label{GKPequation2}
\widetilde\gamma_n
&
+
\int_0^\infty  \frac{dt}{t} \frac{J_n (2gt)}{{\rm e}^t - 1} \left[ \widetilde\gamma_- (2gt) + (-1)^n \widetilde\gamma_+ (2gt) \right] = \widetilde\kappa_n
\, , 
\end{align}
where we also formed appropriate linear combinations of the sources as well
\begin{align}
\label{KappaChiSources}
\kappa_n \equiv \ft{1}{2} [\chi_n + \bar\chi_n]
\, , \qquad
\widetilde\kappa_n \equiv \ft{i}{2} [\chi_n - \bar\chi_n]
\, .
\end{align}
For future reference, we can immediately cast the flux-tube equations into the form
\begin{align}
\label{DefEq1}
\int_0^\infty \frac{dt}{t} J_{2n} (2gt) \left[ \frac{\gamma^{\star}_{+,v} (2gt)}{1 - {\rm e}^{-t}} -  \frac{\gamma^{\star}_{-,v} (2gt)}{{\rm e}^t - 1} \right]
&
=
\kappa^{\star}_{2n} (v)
\, , \\
\label{DefEq2}
\int_0^\infty \frac{dt}{t} J_{2n - 1} (2gt) \left[  \frac{\gamma^{\star}_{-,v} (2gt)}{1 - {\rm e}^{- t}} + \frac{\gamma^{\star}_{+,v} (2gt)}{{\rm e}^t - 1} \right]
&
=
\kappa^{\star}_{2n - 1} (v)
\, ,
\end{align}
and
\begin{align}
\label{DefEqTilde1}
\int_0^\infty \frac{dt}{t} J_{2n} (2gt) \left[ \frac{\widetilde\gamma^{\star}_{+,v} (2gt)}{1 - {\rm e}^{-t}} +  \frac{\widetilde\gamma^{\star}_{-,v} (2gt)}{{\rm e}^t - 1} \right]
&
=
\widetilde\kappa^{\star}_{2n} (v)
\, , \\
\label{DefEqTilde2}
\int_0^\infty \frac{dt}{t} J_{2n - 1} (2gt) \left[ \frac{\widetilde\gamma^{\star}_{-,v} (2gt)}{1 - {\rm e}^{- t }} -  \frac{\widetilde\gamma^{\star}_{+,v} (2gt)}{{\rm e}^{t} -1} \right]
&
=
\widetilde\kappa^{\star}_{2n - 1} (v)
\, ,
\end{align}
for even and odd indices, respectively, by using the inverse transformation in the Neumann series \re{NeumannSeries} for the first term in the left-hand side of 
\re{GKPequation1} and \re{GKPequation2}.

\subsubsection{Formal solution of flux-tube equations}

Let us construct an infinite series representation for the solution to Eqs.\ \re{GKPequation1} and \re{GKPequation2}. Let us start from the first one.
Since $\gamma (t)$ and $\widetilde\gamma (t)$ have the same parity as $\Gamma_+ (t)$ and $\Gamma_- (t)$ in Eq. \re{NeumannSeries}, 
respectively, they can be expanded into the same Neumann series. This allows us to write \re{GKPequation1} as a matrix equation 
\cite{Kotikov:2006ts,Benna:2006nd,Basso:2007wd}
\begin{align}
\left(
\delta_{nm}
+ K_{nm}
\right)
\gamma_m
=
\kappa_n
\, ,
\end{align}
where the summation over the repeated index is implied over the range of all positive integers $m \in \mathbb{Z}^+$. We introduced
here a notation for the infinite matrix \cite{Kotikov:2006ts,Benna:2006nd,Basso:2007wd}
\begin{align}
K_{nm} 
\equiv 
2 m (-1)^{m (n+1)} \int_0^\infty \frac{dt}{t} \frac{J_n (2gt) J_m (2gt)}{{\rm e}^t - 1}
\, ,
\end{align}
and we used used an identity to combine contributions of different parity $1 + (-1)^m- (-1)^n + (-1)^{n+m} = 2 (-1)^{m (n+1)}$.
We can formally invert the equations for $\gamma$ and analogous one for $\widetilde\gamma$ to find
\begin{align}
\label{gammaFormalSolution}
\gamma_n = [1 + K]_{nm}^{-1} \kappa_m
\, , \qquad
\widetilde\gamma_n = (-1)^n [1 + K]_{nm}^{-1} (-1)^m \widetilde\kappa_m
\, .
\end{align}
This representation is particularly useful for perturbative calculations of the functions in question.

\subsubsection{Fixing hole inhomogeneity}

Let us now solve the Baxter equation in the large-spin limit. As was demonstrated in Refs.\ \cite{Korchemsky:1995be,Belitsky:2006en},
in this case one can ignore half of the Baxter equation. At weak coupling, the accuracy of the resulting approximation is $O (1/S^2)$,
while at strong coupling, it is only $O (1/\ln S)$,
\begin{align}
Q (u) = Q_+ (u) + Q_- (u)
=
Q_+ (u) \left( 1 + O (1/S^2) \right)
\end{align}
for $\Im{\rm m} [u] > 0$. This will suffice for our purposes since we will be after leading and first subleading effects
in the large-spin expansion. We are interested in the dynamics of the small hole in the distribution of the Bethe roots. Thus, to this end
it is enough to consider a length $L=3$ spin chain since the small hole is accompanied by two large holes whose rapidities scale as
$\pm S/\sqrt{2}$. The transfer matrix admits the form for $u \sim O (S^0)$,
\begin{align}
t (u) = - \eta^2 (u - v)
\, ,
\end{align}
where $v$ is the hole rapidity and $\eta^2 = S^2 + O(S)$. The resulting equation that one has to solve reads
\begin{align} 
\label{HalfBEQplus}
\ln Q_+ (u + i) - \ln Q_+ (u) = - \int_0^\infty \frac{dt}{t} {\rm e}^{i u^+ t} \Gamma (2gt)  + \ln (- \eta^2) - 3 \ln x^+ + \ln (u-v)
\, ,
\end{align}
with $x^+ = x[u^+]$.
Decomposing $Q_+$ as in Eq.\ \re{Qdecomposition} with $\delta Q_+$ factorized further in the vacuum $Q_+^{\O}$ and hole $Q_{+,v}^{\rm h}$ functions,
$\delta Q_+ = Q_+^{\O} Q_{+,v}^{\rm h}$, each of them obeys respective equations
\begin{align}
\ln \frac{Q^\Gamma_+ (u+i)}{Q^\Gamma_+ (u)} 
&
= - \int_0^\infty \frac{dt}{t} {\rm e}^{i u^+ t} \Gamma (2 g t)
\, , \\
\ln \frac{Q^{\O}_+ (u+i)}{Q^{\O}_+ (u)} 
&
=
\ln (- \eta^2) - 2 \ln x^+
\, , \\
\label{HalfBaxterHole}
\ln \frac{Q^{\rm h}_{+,v} (u+i)}{Q^{\rm h}_{+,v} (u)} 
&
=
\ln\frac{u -v}{x^+}
\, .
\end{align}
These can be easily solved one-by-one with the solution to the first one that has already been used in Eq.\ \re{Qgamma}.
We are now in a position to decompose the  functions $\Gamma$ in terms of the vacuum solution $\Gamma^{\O}$, 
when there are no small holes and the remainder describing a single scalar excitation propagating along the chain,
\begin{align}
\Gamma (t) = \Gamma^{\O} (t) + \Gamma^{\rm h}_v (t)
\, .
\end{align}
Let us now turn to finding these.

\subsubsection{Vacuum solution}

Notice that $\Gamma^{\O}$ can be found by solving length-two spin chain. Since it corresponds to twist-two operators, it is real and therefore
\begin{align}
\gamma^{\O} (t) = \Gamma^{\O} (t)
\, , \qquad
\widetilde\gamma^{\O} (t) = 0
\, .
\end{align}
The vacuum solution was found in Ref.\ \cite{Basso:2011rc} and reads
\begin{align}
\label{VacuumSolution}
\ln Q^{\O}_+ (u^+)
= - 2 i u \ln \bar\eta - 2 \int_0^\infty \frac{dt}{t} \frac{{\rm e}^{i u t} J_0 (2 g t) - i u t - 1}{{\rm e}^t - 1}
\, ,
\end{align}
where here and below $\bar\eta = \eta {\rm e}^{\gamma_{\rm E}}$ with $\gamma_{\rm E}$ being the Euler constant.
Then, we immediately find that $\gamma^{\O}$ obeys Eq.\ \re{GKPequation1}, i.e., 
\begin{align}
\label{VacuumFluxTube}
\gamma_n^{\O}
&
+
\int_0^\infty \frac{dt}{t} \frac{J_n (2gt)}{{\rm e}^t - 1} \left[ \gamma^{\O}_+ (2gt) - (-1)^n \gamma^{\O}_- (2gt) \right] = \kappa^{\o}_n
\, ,
\end{align}
with the source
\begin{align}
\label{kappaO}
\kappa^{\O}_n 
= 
2 g \delta_{n,1} \ln \bar\eta + 2 \int_0^\infty \frac{dt}{t} \frac{J_n (2gt) J_0 (2gt) - gt \delta_{n,1}}{{\rm e}^t - 1}
\, .
\end{align}
Notice that compared to Ref.\ \cite{Basso:2011rc} we did not decompose the vacuum functions/sources in terms of $O (\ln S)$ and $O(S^0)$ 
contributions but rather kept them as one function. Since the vacuum solution is independent of the type of excitation that propagates on it,
it will be unique for all sectors studied below as well.

\subsubsection{Hole solution}

The solution to the half-Baxter equation for the hole excitation \re{HalfBaxterHole} reads
\begin{align}
\ln Q^{\rm h}_{+,v} (u^+)
=
\ln c^{\rm h}_+
+
\int_0^\infty \frac{dt}{t} \frac{{\rm e}^{i  u t}}{{\rm e}^t - 1} 
\left(
{\rm e}^{- i v^+ t} - J_0 (2gt)
\right)
\, .
\end{align}
Here we introduced an arbitrary constant $c_+$ that accounts for the ambiguity in solving the finite-difference equation. It will be fixed later on
from the condition of zero quasimomentum carried by excitations around the chain since these are the only physical degrees of freedom in the
problem.

Finally, decomposing the $\gamma$ and $\widetilde\gamma$ functions in terms of the vacuum and hole parts, the latter obey Eqs.\ \re{GKPequation1} 
and \re{GKPequation2} with the sources that can be easily derived from \re{ChiSourceEven} and \re{ChiSourceOdd} by forming the linear 
combinations \re{KappaChiSources}. They read explicitly 
\begin{align}
\label{HoleSourceKappa}
\kappa_n^{\rm h} (v) 
&= - \int_0^\infty \frac{dt}{t} \frac{J_n (2gt)}{{\rm e}^t - 1} \left( {\rm e}^{t/2} \cos(v t) - J_0 (2gt) \right)
\, , \\ 
\label{HoleSourcetildeKappa}
\widetilde\kappa_n^{\rm h} (v)  
&= - \int_0^\infty \frac{dt}{t} \frac{J_n (2gt)}{{\rm e}^t - 1} {\rm e}^{t/2} \sin(v t) 
\, .
\end{align}
They are in agreement with Ref.\ \cite{Basso:2010in}.

\subsubsection{Integral form of flux-tube equations: holes}
\label{IntegralFTeqHoles}

We can rewrite the flux tube equations in a form that will be indispensable for analytic continuation in hole rapidity that will be used later in the paper. 
Namely, from the Jacobi-Anger relation we can find the expansion of trigonometric functions in terms of the Neumann series
\begin{align}
\label{JAcos}
\cos (t \bar{u}) - J_0 (t) = 2 \sum_{n \geq 1} J_{2n} (t) \cos\left( 2n\varphi\right)
\, , \\
\label{JAsin}
\sin (t \bar{u}) = 2 \sum_{n \geq 1} J_{2n-1} (t) \sin\left( (2n-1)\varphi\right)
\, ,
\end{align}
where $\bar{u} = \sin \varphi$ with $|\bar{u}| \leq 1$. Then multiplying both sides of the above Eqs.\ \re{DefEq1} and \re{DefEq2} 
by $\cos ( 2n \varphi )$ and $\sin ( (2n-1)\varphi)$, respectively, and summing over $n$, we get for $\star = {\rm h}$
\begin{align}
\label{AnotherHoleFTgammaPlus}
&
\int_0^\infty \frac{dt}{t} \left( \cos(ut) 
- J_0 (2gt) \right) \gamma^{\rm h}_{+, v} (2gt)
\\
&\qquad\qquad\qquad\quad
= -
\int_0^\infty \frac{dt}{t} \frac{ \cos (ut) - J_0 (2gt)  }{ {\rm e}^t - 1} 
\left[
\gamma^{\rm h}_v (- 2gt)
+
{\rm e}^{t/2} \cos (vt) - J_0 (2gt)
\right]
\, , \nonumber\\
\label{AnotherHoleFTgammaMinus}
&
\int_0^\infty \frac{dt}{t} \sin(ut) \gamma^{\rm h}_{-, v} (2gt)
\\
&\qquad\qquad\qquad\quad
= -
\int_0^\infty \frac{dt}{t } \frac{\sin (ut)}{{\rm e}^t - 1}
\left[
\gamma^{\rm h}_v (2gt)
+
{\rm e}^{t/2} \cos (vt) - J_0 (2gt)
\right]
\, . \nonumber
\end{align}
Similarly, from the flux tube equations in the form \re{DefEqTilde1} and \re{DefEqTilde2}, we find making use of Eqs.\ \re{JAcos} and \re{JAsin},
\begin{align}
\label{AnotherHoleFTgammaTildePlus}
&
\int_0^\infty \frac{dt}{t} \left( \cos(ut) 
- J_0 (2gt) \right) \widetilde\gamma^{\rm h}_{+, v} (2gt)
\\
&\qquad\qquad\qquad\quad
= -
\int_0^\infty \frac{dt}{t} \frac{ \cos (ut) - J_0 (2gt) }{{\rm e}^t - 1}
\left[
\widetilde\gamma^{\rm h}_v (2gt)
+
{\rm e}^{t/2} \sin (vt) 
\right]
\, , \nonumber\\
\label{AnotherHoleFTgammaTildeMinus}
&
\int_0^\infty \frac{dt}{t} \sin(ut) \widetilde\gamma^{\rm h}_{-, v} (2gt)
\\
&\qquad\qquad\qquad\quad
= -
\int_0^\infty \frac{dt}{t} \frac{\sin (ut)}{{\rm e}^t - 1}
\left[
- \widetilde\gamma^{\rm h}_v (- 2gt)
+
{\rm e}^{t/2} \sin (vt)
\right]
\, . \nonumber
\end{align}
Both results appeared previously in Ref.\ \cite{Basso:2013pxa}. Similar equations can be derived for all other excitations and are 
summarized in Appendix \ref{OtherIntegralFTeqs}.

\subsection{Large fermion}

The Baxter equation for the single fermion excitation reads
\begin{align}
\label{FermionBaxterEq}
\Delta_- (u^-) Q^{\rm F} (u - i) (x^+ -x_{\rm F})
+
\Delta_+ (u^+) Q^{\rm F} (u + i) (x^- -x_{\rm F})
=
t (u) Q (u)
\, ,
\end{align}
as can be derived by solving the nesting in the sl$(2|1)$ sector of the complete su$(2,2|4)$ super spin chain Bethe Ansatz equations \cite{Beisert:2005fw}, 
see Ref.\ \cite{Basso:2011rc}. Here $\Delta_\pm$'s are given in Eq.\ \re{ExplicitDeltaPM}. As was demonstrated in the leading order analysis of Ref.\ 
\cite{Gaiotto:2010fk}, see Appendix B.2 there, compared to the one for the bosonic sector alluded to above, the transfer matrix possesses (with exponential 
precision) two extra zeroes located at $u^\pm_F$ of the fermionic root $u_F$. Generalizing this analysis to all orders in 't Hooft coupling as suggested by the left-hand 
side of the above Baxter equation, we can bring $t(u)$ to the form
\begin{align}
t (u) = (x^- - x_{\rm F}) (x^+ - x_{\rm F}) \tau (u)
\, ,
\end{align}
where $x \equiv x[u]$ and the dependence on the fermion rapidity $v$ enters through $x_{\rm F} \equiv x [v]$.
In the large-spin limit $\tau (u)$ is dominated by the two large holes, so it simply scales as $\tau (u) \simeq - \eta^2$ up to subleading terms. Neglecting
one half of the Baxter equation, we find
\begin{align}
\Delta^{\rm F}_+ (u^+) Q^{\rm F}_+ (u + i) = - \eta^2 (u - v^-) Q^{\rm F}_+  (u)
\, , 
\end{align}
where
\begin{align}
\Delta^{\rm F}_+ (u) = \Delta_+ (u) \left( 1 - \frac{g^2}{x_{\rm F} x} \right)
\, .
\end{align}
We introduce the function $\Gamma^{\rm F} (t)$ as follows
\begin{align}
\ln \Delta^{\rm F}_+ (u) = L \ln x + \int_0^\infty \frac{dt}{t} {\rm e}^{i u t} \Gamma^{\rm F} (2gt)
\, ,
\end{align}
with $L = 3$ and valid in the upper half-plane of $u$. Substituting the Neumann expansion for $\Gamma^{\rm F} (t)$, we immediately obtain
\begin{align}
\Gamma_n^{\rm F}
=
\frac{1}{4 \pi i} \oint \frac{dx}{x} \left( \frac{x}{i g} \right)^n \ln \left( 1 - \frac{g^2}{x x_{\rm F}} \right)
-
\frac{1}{2 \pi i} \oint \frac{dx}{x} \left( \frac{x}{i g} \right)^n \ln Q^{\rm F}_+ (u^+)
+ d_n
\, .
\end{align}
The first term can be easily calculated to be
\begin{align}
\frac{1}{4 \pi i} \oint \frac{dx}{x} \left( \frac{x}{i g} \right)^n \ln \left( 1 - \frac{g^2}{x x_{\rm F}} \right)
=
-
\frac{1}{2 n} \left( \frac{g}{i x_{\rm F}} \right)^n
\, .
\end{align}
Now using the following integral representations for even and odd $n$,
\begin{align*}
\left( \frac{g}{x_{\rm F}}\right)^{2 n} 
&= 2 n (-1)^n \int_0^\infty \frac{dt}{t} J_{2n} (2gt) \cos(v t)
\, , \\
\left( \frac{g}{x_{\rm F}}\right)^{2 n - 1} 
&= (2 n - 1) (-1)^{n - 1} \int_0^\infty \frac{dt}{t} J_{2n - 1} (2gt) \sin(v t)
\, ,
\end{align*}
we find, respectively,
\begin{align}
\label{GammanevenFermion}
\Gamma^{\rm F}_{2n} 
&
- \int_0^\infty \frac{dt}{t} J_{2n} (2gt) \frac{\Gamma^{\rm F}_- (2gt) + \bar\Gamma^{\rm F}_- (2gt)}{{\rm e}^t - 1}
\\
&=
- \frac{1}{2} \int_0^\infty \frac{dt}{t} J_{2n} (2gt) \cos(v t)
-
\frac{1}{2 \pi i} \oint \frac{dx}{x} \left( \frac{x}{i g} \right)^{2n} \ln Q^{\rm F}_+ (u^+)
\, , \nonumber\\
\label{GammanoddFermion}
\Gamma^{\rm f}_{2n - 1} 
&
- \int_0^\infty \frac{dt}{t} J_{2n - 1} (2gt) \frac{\Gamma^{\rm F}_+ (2gt) - \bar\Gamma^{\rm F}_+ (2gt)}{{\rm e}^t - 1}
\\
&=
\frac{i}{2} \int_0^\infty \frac{dt}{t} J_{2n - 1} (2gt) \sin(v t)
-
\frac{1}{2 \pi i} \oint \frac{dx}{x} \left( \frac{x}{i g} \right)^{2n - 1} \ln Q^{\rm F}_+ (u^+)
\, . \nonumber
\end{align}
Breaking up the Baxter function into $Q^\Gamma$ given by \re{Qgamma} in terms of the vacuum solution and the fermion excitation,
we deduce for the latter
\begin{align}
\ln Q_{+,v}^{\rm F} (u^+)
=
\int_0^\infty \frac{dt}{t} \frac{{\rm e}^{i u t}}{{\rm e}^t - 1} \left( {\rm e}^{- i v t} - J_0 (2gt) \right)
\, .
\end{align}
Combining the latter solution with the first term in the right-hand side in above equations \re{GammanevenFermion} and \re{GammanoddFermion}, we get 
the inhomogeneities for fermions for $\Gamma^{\rm F}_v$ in the decomposition $\Gamma^{\rm f} = \Gamma^{\O} + \Gamma^{\rm F}_v$,
\begin{align}
\label{LFsource1}
\kappa^{\rm F}_{2n} (v) 
&= - \int_0^\infty \frac{dt}{t} \frac{J_{2n} (2gt)}{{\rm e}^t - 1}
\left(
\cos(v t) - J_0 (2gt)
\right)
-
\frac{1}{2}  \int_0^\infty \frac{dt}{t}  J_{2n} (2gt) \cos(v t)
\, , \\
\kappa^{\rm F}_{2n - 1} (v) 
&= - \int_0^\infty \frac{dt}{t} \frac{J_{2n - 1} (2gt)}{{\rm e}^t - 1}
\left(
\cos(v t) - J_0 (2gt)
\right)
\, , \\
\widetilde\kappa^{\rm F}_{2n} (v) 
&= - \int_0^\infty \frac{dt}{t} \frac{J_{2n} (2gt)}{{\rm e}^t - 1}
\sin(v t)
\, , \\
\widetilde\kappa^{\rm F}_{2n - 1} (v) 
&= - \int_0^\infty \frac{dt}{t} \frac{J_{2n - 1} (2gt)}{{\rm e}^t - 1}
\sin(v t) 
-
\frac{1}{2}  \int_0^\infty \frac{dt}{t}  J_{2n - 1} (2gt) \sin(v t)
\, .
\label{LFsource4}
\end{align}
These sources coincide with the ones for the so-called large fermion, in terminology of Ref.\ \cite{Basso:2010in}.

\subsection{Small fermion}

We can write analogously the Baxter equation for the small fermion. However, we are going to choose a different route. Since the large and small fermions 
are related to each other via an analytic continuation through the cut on the real axis \cite{Basso:2010in}, we will merely use it in order to find the sources in the 
corresponding flux-tube equation. Namely, following the steps outlined in the Appendix \ref{LargeSmallAppendix}, we can easily conclude,
\begin{align}
\label{Skappa1}
\kappa^{\rm f}_{2n} (v) 
&= \frac{1}{2}  \int_0^\infty \frac{dt}{t}  J_{2n} (2gt) \cos(v t)
\, , \\
\label{Skappa2}
\widetilde\kappa^{\rm f}_{2n - 1} (v) 
&=
\frac{1}{2}  \int_0^\infty \frac{dt}{t}  J_{2n - 1} (2gt) \sin(v t)
\, ,
\end{align}
while $\kappa^{\rm f}_{2n - 1} (v) = 0$ and $\widetilde\kappa^{\rm f}_{2n} (v) = 0$, in agreement with Ref.\ \cite{Basso:2010in}.

\subsection{Gauge field and bound states}
\label{GaugeBaxterSection}

Finally, we turn to the gauge field and its bound states. In complete analogy with consideration for other excitations, the resolved nested Bethe Ansatz equations, see 
Eq.\ (B.16) of Ref.\ \cite{Basso:2011rc}, can be cast in the form of a Baxter equation, namely
\begin{align}
\label{GaugeBaxter}
\Delta_- (u^-) Q^{\rm g} (u - i) \left( x_{\rm g}^{[- \ell]} - x^+ \right) \left( 1 - \frac{g^2}{x_{\rm g}^{[- \ell]} x^-} \right)
&+
\Delta_+ (u^+) Q^{\rm g} (u + i) \left( x_{\rm g}^{[+ \ell]} - x^- \right) \left( 1 - \frac{g^2}{x_{\rm g}^{[+ \ell]} x^+} \right)
\nonumber\\
&=
t (u) Q^{\rm g} (u)
\, ,
\end{align}
with $x = x[u]$ and $x_{\rm g}^{[\pm \ell]} = x[v \pm \ft{i}{2} \ell]$. The dressing factor is the same as in the Baxter equation for the hole \re{ExplicitDeltaPM}.
As in the fermionic case addressed above, the transfer matrix $t (u)$ has two zeroes which should be factorized
\begin{align}
t(u) = \left( x_{\rm g}^{[- \ell]} - x^+ \right) \left( x_{\rm g}^{[+ \ell]} - x^- \right) \tau (u)
\, .
\end{align}
Now, in the large-spin limit we can safely ignore one of the terms in the Baxter equation. For $\Im{\rm m}[u]>0$, we find
\begin{align}
\label{HalfGaugeBaxter}
\Delta^{\rm g}_+ (u^+) Q^{\rm g}_+ (u + i) = - \eta^2 \left( u - v^{[- \ell - 1]} \right) Q^{\rm g}_+ (u)
\, ,
\end{align}
where we used the fact that we have two large holes in the game so that $\tau (u) \simeq - \eta^2$ and we defined
\begin{align}
\label{GdeltaToDelta}
\Delta^{\rm g}_+ (u) \equiv \left( 1 - \frac{g^2}{x_{\rm g}^{[- \ell]} x} \right)  \left( 1 - \frac{g^2}{x_{\rm g}^{[- \ell]} x} \right)  \Delta_+ (u) 
\, ,
\end{align}
with $\Delta_+$ given by Eq.\ \re{ExplicitDeltaPM}. Introducing a new function $\Gamma^{\rm g}$, we write $\Delta^{\rm g}_+$ in the form
\begin{align}
\label{DeltaGauge}
\Delta^{\rm g}_+ (u)
=
L \ln x 
+ 
\int_0^\infty \frac{dt}{t} {\rm e}^{i u t} \Gamma^{\rm g} (2gt)
\, , 
\end{align}
where $\Im{\rm m} [u] > 0$.  Expanding the $\Gamma^{\rm g} (t)$ in the Neumann series \re{NeumannSeries}, we find 
\begin{align}
\Gamma^{\rm g}_n = - \frac{1}{2 n} \left( \frac{g}{i x^{[- \ell]}_{\rm g}} \right)^n  - \frac{1}{2 n} \left( \frac{g}{i x^{[+ \ell]}_{\rm g}} \right)^n
-
\frac{1}{2 \pi i} \oint \frac{dx}{x} \left( \frac{x}{i g} \right)^n \ln Q^{\rm g}_+ (u^+) + d_n
\, ,
\end{align}
where we relied on Eq.\ \re{FromLogXtoLogU} to compute the integrals of the logarithms of prefactors in the right-hand side of Eq.\ \re{GdeltaToDelta}.
Since the functional form of Eq.\ \re{HalfGaugeBaxter} does not differ from the one of the hole, the dressing terms $d_n$ take the form \re{Evends} and
\re{Oddds} where one obviously has to replace $\Gamma$ by $\Gamma^{\rm g}$. Finally, using 
\begin{align}
\left( \frac{g}{i x^{[\pm \ell]}_{\rm g}} \right)^n = n \int_0^\infty \frac{dt}{t} J_n (\mp 2gt) {\rm e}^{\pm i v t - \ell t/2}
\, ,
\end{align}
we can bring $\Gamma^{\rm g}_n$ to the form
\begin{align}
\label{GammaGluon}
\Gamma^{\rm g}_n
=
- \frac12 \int_0^\infty \frac{dt}{t} J_n (2gt) {\rm e}^{- \ell t/2} \left( {\rm e}^{- i v t} + (- 1)^n {\rm e}^{i v t} \right)
- \frac{1}{2 \pi i} \oint \frac{dx}{x} \left( \frac{x}{i g} \right)^n \ln Q^{\rm g}_+ (u^+) +  d_n
\, , 
\end{align}
where the first term will be an additional contribution to the sources of the flux-tube equations. Decomposing $Q^{\rm g}_+$ as before $Q^{\rm g}_+ = Q_+^\Gamma 
Q_+^{\O} Q_{+,v}^{\rm g}$, we find that $Q_+^\Gamma$ and $Q_+^{\o}$ have the same form as \re{Qgamma} and \re{VacuumSolution}, respectively. While for the 
gluon bound state, we have
\begin{align}
\ln Q_{+,v}^{\rm g} (u^+) = \int_0^\infty \frac{dt}{t} \frac{{\rm e}^{i u t}}{{\rm e}^t - 1} \left( {\rm e}^{- i v t - \ell t/2} - J_0 (2gt) \right)
\, .
\end{align}
So the right-hand side of this equation together with the first term in Eq.\ \re{GammaGluon} cumulatively generate the inhomogeneous terms in the flux-tube
equations
\begin{align}
\kappa_{2n}^{\rm g} (v) 
&
=
- \int_0^\infty \frac{dt}{t} \frac{J_{2n} (2gt)}{1 - {\rm e}^{- t}} \left( {\rm e}^{- \ell t/2} \cos(v t) - J_0 (2gt) \right)
\, , \\
\kappa_{2n-1}^{\rm g} (v) 
&
=
- \int_0^\infty \frac{dt}{t} \frac{J_{2n - 1} (2gt)}{{\rm e}^t - 1} \left( {\rm e}^{- \ell t/2} \cos(v t) - J_0 (2gt) \right)
\, , \\
\widetilde\kappa_{2n}^{\rm g} (v) 
&
=
- \int_0^\infty \frac{dt}{t} \frac{J_{2n} (2gt)}{{\rm e}^t - 1} {\rm e}^{- \ell t/2} \sin(v t) 
\, , \\
\widetilde\kappa_{2n-1}^{\rm g} (v) 
&
=
- \int_0^\infty \frac{dt}{t} \frac{J_{2n - 1} (2gt)}{1 - {\rm e}^{- t}} {\rm e}^{- \ell t/2} \sin(v t) 
\, .
\end{align}
For a single gauge field, i.e., $\ell = 1$, the sources can be found from the above general relations and cast into the concise form
\begin{align}
\label{GluonKappa}
\kappa_{n}^{\rm g} (v) 
&
=
- \int_0^\infty \frac{dt}{t} \frac{J_n (2gt)}{{\rm e}^{t} - 1} \left( {\rm e}^{(-1)^n t/2} \cos(v t) - J_0 (2gt) \right)
\, , \\
\label{GluonKappaTilde}
\widetilde\kappa_{n}^{\rm g} (v) 
&
=
- \int_0^\infty \frac{dt}{t} \frac{J_n (2gt)}{{\rm e}^t - 1} {\rm e}^{- (-1)^n t/2} \sin(v t) 
\, , 
\end{align}
where we added and subtracted ${\rm e}^{\mp t} J_0 (2gt)$ making use of  the orthogonality conditions \re{OrthogonalityBessel}. Again
these agree with Ref.\ \cite{Basso:2010in}.

\section{S-matrices}
\label{SmatricesSection}

Let us now turn to the derivation of scattering matrices between different flux-tube excitations. We will focus on their dynamical part which enters as
overall scalar prefactors accompanied by a matrix encoding SU(4) tensor structure for particles charged with the respect to the R-symmetry group,
like holes and (anti)fermions. To accomplish this goal we will rely on the Bethe-Yang equations derived in Ref.\ \cite{Basso:2011rc}. Since the solutions 
constructed above have an unfixed integration constants, we have to provide an additional condition to fix the ambiguity. The latter is provided by the 
vanishing of the quasimomentum of excitations propagating along the closed spin chain since this is ensured by the cyclicity of the single trace operators. 
The resulting condition 
\begin{align}
\frac{1}{2 \pi i} \oint \frac{dx}{x} \ln \frac{Q^{\star}_+ (u^+)}{Q^{\star}_- (u^-)} = 0
\, , 
\end{align}
is written in terms of the solutions $Q^{\star}_{\pm} (u)$ to the half-Baxter equations valid in the upper/lower half-plane of $u$ for the $\star$-excitation.

We start our analysis by reproducing expressions already available in the literature \cite{Basso:2013pxa,Basso:2013aha,Basso:2014koa}, namely, for hole-hole, 
fermion-fermion and gauge-gauge fields in Sections \ref{HHsmatrix}, \ref{FFsmatrix} and \ref{GGsmatrix}, respectively. At the same time, we report new 
results\footnote{In a different framework these were derived in Ref.\  \cite{Fioravanti:2013eia}.} for scattering phases involving excitations of different kinds, 
i.e., hole-large (small) fermion, hole-gauge and large/small fermion-gauge fields in Sections \ref{HFsmatrix} (\ref{holeSmallSmatrix}), \ref{HGsmatrix} and 
\ref{GFsmatrix}, respectively.

\subsection{Hole-hole S-matrices}
\label{HHsmatrix}

Combining all ingredients together, the solutions to half-Baxter equation for holes (with $\Im{\rm m} [u] \gtrless 0$) are
\begin{align}
\ln Q_+ (u^+)
&
=
\ln c_+ - 2i u \ln \eta
-
2 \int_0^\infty \frac{dt}{t} \frac{{\rm e}^{iut} J_0 (2gt) - iut - 1}{{\rm e}^t - 1}
\\
&
+
\int_0^\infty \frac{dt}{t} \frac{{\rm e}^{iut} \Gamma (2gt) - 2gt \Gamma_1}{{\rm e}^t - 1}
+
\int_0^\infty \frac{dt}{t ({\rm e}^t - 1)} {\rm e}^{iut}
\left(
{\rm e}^{- i v^+ t} - J_0 (2gt)
\right)
\, , \nonumber\\
\ln Q_- (u^-)
&
=
\ln c_- + 2i u \ln \eta
-
2 \int_0^\infty \frac{dt}{t} \frac{{\rm e}^{-iut} J_0 (2gt) + iut - 1}{{\rm e}^t - 1}
\\
&
+
\int_0^\infty \frac{dt}{t} \frac{{\rm e}^{-iut} \bar\Gamma (2gt) - 2gt \bar\Gamma_1}{{\rm e}^t - 1}
+
\int_0^\infty \frac{dt}{t ({\rm e}^t - 1)} {\rm e}^{-iut}
\left(
{\rm e}^{i v^- t} - J_0 (2gt)
\right)
\, .
\end{align}
The zero-momentum  condition fixes the ratio of the constants to be
\begin{align}
0 
&=
\frac{1}{2 \pi i} \oint \frac{dx}{x} \ln \frac{Q_+ (u^+)}{Q_- (u^-)}
=
\ln\frac{c_+}{c_-}
\\
&
+
2i \int_0^\infty \frac{dt}{t ({\rm e}^t - 1)}
\left(
J_0 (2gt) \widetilde\gamma^{\rm h}_v (2gt) - 2gt \widetilde\gamma^{\rm f}_{v, 1}
\right)
+
2i \int_0^\infty \frac{dt}{t ({\rm e}^t - 1)}
{\rm e}^{t/2} \sin(vt) J_0 (2gt)
\, . \nonumber 
\end{align}

Next, we introduce the Bethe-Yang function that is written in the form \cite{Basso:2011rc}
\begin{align}
Y^{\rm h} (u, v) = \frac{Q_- (u)}{Q_+ (u)}
=
{\rm e}^{i P_{\rm h} (u)} S_{\rm hh} (u, v)
\, ,
\end{align}
with
\begin{align}
\label{HoleMomentum}
P_{\rm h} (u)
=
4 u \ln \eta
+
4 \int_0^\infty \frac{dt}{t ({\rm e}^t - 1)} 
\left(
{\rm e}^{t/2} J_0 (2gt) \sin (vt) - u t
\right)
-
2i \int_0^\infty \frac{dt}{t ({\rm e}^t - 1)} {\rm e}^{t/2} \gamma^{\O} (2gt) \sin (u t)
\, ,
\end{align}
being the hole propagation phase. More precisely, it gets decomposed into $P_{\rm h} (u) = p_{\rm h} (u) \ln \bar\eta + \delta p_{\rm h} (u)$ where $p_{\rm h}$ 
is the momentum of the scalar flux-tube excitation \cite{Basso:2010in}. The S-matrix can be easily read off from the rest and can be brought to the form
\begin{align}
\label{hhSmatrix}
S_{\rm hh} (u, v) = \exp \left( 2 i \sigma_{\rm hh} (u, v) - 2 i f^{(1)}_{\rm hh} (u, v) + 2 i f^{(2)}_{\rm hh} (u, v) \right)
\, ,
\end{align}
where
\begin{align}
\label{Sigmahh}
\sigma_{\rm hh} (u,v)
=
\int_0^\infty \frac{dt}{t ({\rm e}^t - 1)}
\left[
{\rm e}^{t/2} J_0 (2gt) \sin(ut) - {\rm e}^{t/2} J_0 (2gt) \sin(vt) - {\rm e}^t \sin((u-v)t)  
\right]
\end{align}
and
\begin{align}
f^{(1)}_{\rm hh} (u, v) 
&
=  \int_0^\infty \frac{dt}{t ({\rm e}^t - 1)} {\rm e}^{t/2} \sin (u t) \gamma^{\rm h}_v (2gt)
\, , \\
f^{(2)}_{\rm hh} (u, v) 
&
=  \int_0^\infty \frac{dt}{t ({\rm e}^t - 1)} \left( {\rm e}^{t/2} \cos (u t) - J_0 (2gt) \right) \widetilde\gamma^{\rm h}_v (2gt)
\, .
\end{align}
This coincides with the result of Ref.\ \cite{Basso:2013pxa}.

\subsection{Fermion S-matrices}
\label{FFsmatrix}

Next we turn to fermions and start with the ones that carry large momentum. The solutions to the large fermonic half-Baxter equation \re{FermionBaxterEq} 
valid in the upper/lower half-plane read
\begin{align}
\label{QplusF}
\ln Q^{\rm F}_+ (u^+)
&=
\ln c_+^{\rm F}
- 2i u \ln \eta
-
2\int_0^\infty \frac{dt}{t} \frac{{\rm e}^{i u t} J_0 (2gt) - i u t - 1}{{\rm e}^t - 1}
\\
&
+
\int_0^\infty \frac{dt}{t} \frac{{\rm e}^{i u t} \Gamma^{\rm F} (2gt) - 2 g t \Gamma^{\rm F}_1}{{\rm e}^t - 1}
+
\int_0^\infty \frac{{\rm e}^{i u t}}{{\rm e}^t - 1}
\left(
{\rm e}^{- i v t} - J_0 (2gt)
\right)
\, , \nonumber\\
\label{QminusF}
\ln Q^{\rm F}_- (u^-)
&=
\ln c_-^{\rm F}
+ 2i u \ln \eta
-
2\int_0^\infty \frac{dt}{t} \frac{{\rm e}^{- i u t} J_0 (2gt) + i u t - 1}{{\rm e}^t - 1}
\\
&
+
\int_0^\infty \frac{dt}{t} \frac{{\rm e}^{- i u t} \bar\Gamma^{\rm F} (2gt) - 2 g t \bar\Gamma^{\rm F}_1}{{\rm e}^t - 1}
+
\int_0^\infty \frac{{\rm e}^{- i u t}}{{\rm e}^t - 1}
\left(
{\rm e}^{i v t} - J_0 (2gt)
\right)
\, . \nonumber
\end{align}
Again, the vanishing of the quasimomentum  fixes the ratio of the integration constants
\begin{align}
0 
&= 
\frac{1}{2 \pi i}
\oint \frac{dx}{x} \ln \frac{Q^{\rm F}_+ (u^+)}{Q^{\rm F}_- (u^-)}
=
\ln \frac{c_+^{\rm F}}{c_-^{\rm F}}
\\
&+
2 i \int_0^\infty \frac{dt}{t ({\rm e}^t - 1)}
\left(
\widetilde\gamma^{\rm F}_v (2gt) J_0 (2gt) - 2gt \widetilde\gamma^{\rm F}_{v, 1}
\right)
+ 2i \int_0^\infty \frac{dt}{t ({\rm e}^t - 1)} J_0(2gt) \sin(vt)
\, . \nonumber
\end{align}
Here we used the linear combinations introduced in Eq.\ \re{GammaTogamma} and the fact that
\begin{align}
\gamma^{\rm F} (t) = \gamma^{\O} (t) + \gamma^{\rm F}_v (t)
\, , \qquad
\widetilde\gamma^{\rm F} (t) = \widetilde\gamma^{\rm F}_v (t)
\, ,
\end{align}
since the vacuum solution is insensitive to the sign of the rapidity $v$. So that while $\gamma^{\rm F} (t)$ receives contributions from both the vacuum and large 
fermion excitation,  $\widetilde\gamma^{\rm F} (t)$ only describes the latter.

\subsubsection{Hole--large-fermion S-matrix}
\label{HFsmatrix}

The Bethe-Yang function defined in terms of the solutions \re{QplusF} and \re{QminusF} to the fermionc Baxter equation \re{FermionBaxterEq}
\begin{align}
Y^{\rm h} (u) = \frac{Q^{\rm F}_- (u)}{Q^{\rm F}_+ (u)}
\, ,
\end{align}
can be brought to the form similar to the one for the hole studied above,
\begin{align}
Y^{\rm h} (u) = {\rm e}^{i P_{\rm h} (u)} S_{\rm hF} (u, v)
\, .
\end{align}
Here the propagating phase of the hole $P_{\rm h}$ was determined earlier in Eq.\ \re{HoleMomentum}, while the scattering matrix between the hole
and large fermion excitations with rapidities $u$ and $v$, respectively, reads
\begin{align} 
\label{HoleLargeFermSmatrix}
S_{\rm hF} (u, v) = \exp \left( 2 i \sigma_{\rm hF} (u, v) - 2 i f^{(1)}_{\rm hF} (u, v) + 2 i f^{(2)}_{\rm hF} (u, v) \right)
\, , 
\end{align}
with
\begin{align}
\sigma_{\rm hF} (u, v) = \int_0^\infty \frac{dt}{t ({\rm e}^t - 1)}
\left[
{\rm e}^{t/2} J_0 (2gt) \sin (ut) - J_0 (2gt) \sin (vt) - {\rm e}^{t/2} \sin ((u-v)t)
\right]
\, ,
\end{align}
and
\begin{align}
\label{f1hF}
f^{(1)}_{\rm hF} (u, v)
&=
\int_0^\infty \frac{dt}{t ({\rm e}^t - 1)} {\rm e}^{t/2} \sin (ut) \gamma^{\rm F}_v (2gt)
\, , \\
\label{f2hF}
f^{(2)}_{\rm hF} (u, v)
&=
\int_0^\infty \frac{dt}{t ({\rm e}^t - 1)} \left( {\rm e}^{t/2} \cos(ut) - J_0 (2gt) \right) \widetilde\gamma^{\rm F}_v (2gt)
\, .
\end{align}
Using the formal solutions to the flux-tube equations \re{gammaFormalSolution}, the scattering phases can be cast in the form that was 
quoted in Ref.\ \cite{Basso:2014koa}.

\subsubsection{Large-fermion--large-fermion S-matrix}

The Bethe-Yang function for propagation of a large fermion with rapidity $u$ reads \cite{Basso:2011rc},
\begin{align}
Y^{\rm F} (u) = \frac{Q^{\rm F}_- (u^-)}{Q^{\rm F}_+ (u^+)} \left[ \frac{\Delta^{\rm F}_- (u)}{\Delta^{\rm F}_+ (u)} \right]^{1/2}
\, .
\end{align}
From the representation 
\begin{align}
Y^{\rm F} (u) = {\rm e}^{i P_{\rm F} (u)} S_{\rm FF} (u, v)
\, ,
\end{align}
we can immediately identify the dynamical phase
\begin{align}
P_{\rm F} (u)
=
4 u \ln \eta 
&
+ 4 \int_0^\infty \frac{dt}{t ({\rm e}^t - 1)}
\left(
J_0 (2gt) \sin(ut) - ut
\right)
\\
&
-
2 \int_0^\infty \frac{dt}{t ({\rm e}^t - 1)} \sin(ut) \gamma^{\O} (2gt)
-
\int_0^\infty \frac{dt}{t} \sin(ut) \gamma_-^{\O} (2gt)
\, , \nonumber
\end{align}
and the large-fermion--large-fermion S-matrix as well \cite{Basso:2014koa}
\begin{align}
S_{\rm FF} (u, v)
=
\exp\left(
2i \sigma_{\rm FF} (u, v)
-
2i f^{(1)}_{\rm FF} (u, v)
+
2i f^{(2)}_{\rm FF} (u, v)
\right)
\, ,
\end{align}
where
\begin{align}
\sigma_{\rm FF} (u, v)
=
\int_0^\infty
\frac{dt}{t ({\rm e}^t - 1)}
\left[
J_0 (2gt) \sin (ut) - J_0 (2gt) \sin (vt) - \sin((u-v)t)
\right]
\, ,
\end{align}
and
\begin{align}
f^{(1)}_{\rm FF} (u, v)
&
=
\int_0^\infty \frac{dt}{t}
\sin(ut)
\left[
\frac{\gamma^{\rm F}_v (2gt)}{{\rm e}^t - 1} + \frac{1}{2} \gamma^{\rm F}_{-,v} (2gt)
\right]
\, , \nonumber\\ 
f^{(2)}_{\rm FF} (u, v)
&
=
\int_0^\infty \frac{dt}{t}
\left(
\cos(ut) - J_0 (2gt)
\right)
\left[
\frac{\widetilde\gamma^{\rm F}_v (2gt)}{{\rm e}^t - 1} + \frac{1}{2} \widetilde\gamma^{\rm F}_{+,v} (2gt)
\right]
\, . \nonumber
\end{align}
Making use of the exchange relations that we come to discuss next, one can demonstrate that $f^{(1)}_{\rm FF} (u, v) = f^{(2)}_{\rm FF} (v, u)$.

\subsubsection{Exchange relations}
\label{ExchangeRelationsSection}

To find scattering matrices for small fermion, instead of using corresponding Baxter equation and Bethe-Yang functions, we will rely on the
analytic continuation from the large to small fermion sheet in the complex rapidity plane, as reviewed at length in Appendix \ref{LargeSmallAppendix}. 

The starting point is Eq.\ \re{HoleLargeFermSmatrix}. However, to perform the analytic continuation in fermion rapidity $v$, it is instructive to cast 
$f$'s into a form where dependence on $v$ becomes explicit. This is achieved via the so-called exchange relations \cite{Basso:2009gh}. First, 
expanding $\gamma$ in the Neumann series, we find for \re{f1hF}
\begin{align}
\label{HFExchangef1Eq1}
f^{(1)}_{\rm hF} (u, v) 
= 
- 2 \sum_{n \geq 1} (2n) \gamma^{\rm F}_{2n} (v) \widetilde\kappa^{\rm h}_{2n} (u)
- 2 \sum_{n \geq 1} (2n - 1) \gamma^{\rm F}_{2n - 1} (v) \widetilde\kappa^{\rm h}_{2n - 1} (u)
\, ,
\end{align}
where we used the definition of the hole sources \re{HoleSourcetildeKappa}.
Replacing the sources via the flux-tube equation \re{DefEqTilde1} and \re{DefEqTilde2}, we find the representation
\begin{align}
f^{(1)}_{\rm hF} (u, v) 
=
-
\int_0^\infty \frac{dt}{t} \widetilde\gamma^{\rm h}_{-, u} (2gt)
\left[
\frac{\gamma^{\rm F}_{-,v} (2gt)}{1 - {\rm e}^{-t}} 
+
\frac{\gamma^{\rm F}_{+,v} (2gt)}{{\rm e}^t - 1} 
\right]
-
\int_0^\infty \frac{dt}{t} \widetilde\gamma^{\rm h}_{+, u} (2gt)
\left[
\frac{\gamma^{\rm F}_{+,v} (2gt)}{1 - {\rm e}^{-t}} 
-
\frac{\gamma^{\rm F}_{-,v} (2gt)}{{\rm e}^t - 1} 
\right]
\, .
\end{align}
Finally expanding $\gamma^{\rm h}$ in Neumann series and replacing the coefficients accompanying $\gamma^{\rm h}_{2n}$ and 
$\gamma^{\rm h}_{2n - 1}$ by the large fermion sources making use of the flux-tube equation, we ultimately deduce
\begin{align}
\label{HFExchangef1Eq2}
f^{(1)}_{\rm hF} (u, v) 
=
- 2 \sum_{n \geq 1} (2n) \widetilde\gamma^{\rm h}_{2n} (u) \kappa^{\rm F}_{2n} (v)
- 2 \sum_{n \geq 1} (2n-1) \widetilde\gamma^{\rm h}_{2n-1} (u) \kappa^{\rm F}_{2n-1} (v)
\, ,
\end{align}
so that
\begin{align}
\label{f1hFfermion}
f^{(1)}_{\rm hF} (u, v) 
=
\int_0^\infty \frac{dt}{t} 
\frac{\widetilde\gamma^{\rm h}_u (2gt)}{{\rm e}^t - 1}
\left(
\cos (vt) - J_0 (2gt)
\right)
+
\frac{1}{2}
\int_0^\infty \frac{dt}{t} 
\widetilde\gamma^{\rm h}_{+,u} (2gt) \cos (vt) 
\, .
\end{align}
Analogous considerations yield two equivalent series representation for $f^{(2)}_{\rm hg}$
\begin{align}
\label{HFExchangef2Eq2}
f^{(2)}_{\rm hF} (u, v) 
&=
- 2 \sum_{n \geq 1} (2n) \widetilde\gamma^{\rm F}_{2n} (u) \kappa^{\rm h}_{2n} (v)
- 2 \sum_{n \geq 1} (2n-1) \widetilde\gamma^{\rm F}_{2n-1} (u) \kappa^{\rm h}_{2n-1} (v)
\\
&=
- 2 \sum_{n \geq 1} (2n) \gamma^{\rm h}_{2n} (u) \widetilde\kappa^{\rm F}_{2n} (v)
- 2 \sum_{n \geq 1} (2n-1) \gamma^{\rm h}_{2n-1} (u) \widetilde\kappa^{\rm F}_{2n-1} (v)
\, ,
\end{align}
with the last relation easily brought to the desired form
\begin{align}
\label{f2hFfermion}
f^{(2)}_{\rm hF} (u, v) 
=
\int_0^\infty \frac{dt}{t} 
\frac{\gamma^{\rm h}_u (2gt)}{{\rm e}^t - 1}
\sin (vt) 
+
\frac{1}{2}
\int_0^\infty \frac{dt}{t} 
\gamma^{\rm h}_{-,u} (2gt) \sin (vt) 
\, .
\end{align}

\subsubsection{Hole--small-fermion S-matrix}
\label{holeSmallSmatrix}

As we advertised above, we will find the hole--small-fermion S-matrix by analytically continuing in femion rapidity $v$ from the large to the small
sheet, see Fig.\ \ref{FermionAnalytic}. Obviously, only the last terms in Eqs.\ \re{f1hFfermion} and \re{f2hFfermion} require special attention for their 
proper analytic continuation. Let us demonstrate it for $f^{(1)}_{\rm hF} (u, v)$. For $|v|>2g$ on the real axis, we have
\begin{align}
\label{gammahExp}
\int_0^\infty \frac{dt}{t} 
\widetilde\gamma^{\rm h}_{+,u} (2gt) \cos (vt) 
=  
\int_0^\infty \frac{dt}{t} 
\widetilde\gamma^{\rm h}_{+,u} (2gt) {\rm e}^{- ivt}
\, ,
\end{align}
where the right-hand side is real, so that the corresponding integral can be continued into the lower half-plane. We move inside the strip $|\Im{\rm m} [v]| < 2g$,
as shown in Fig.\ \ref{FermionAnalytic}, and then have to cross the cut and pass to the lower sheet of the Riemann surface. In order to achieve this, we use the
following, easy-to-prove generic relation for a test function $\gamma(t)$
\begin{align}
\label{AnContInt}
\int_0^\infty \frac{dt}{t} {\rm e}^{- ivt} \gamma(t)
=
-
\int_0^\infty \frac{dt}{t} {\rm e}^{ivt} \gamma(- t)
+
2
\int_0^\infty \frac{dt}{t} \cos (vt) \gamma_+ (t)
-
2i \int_0^\infty \frac{dt}{t} \sin(vt) \gamma_- (t)
\, .
\end{align}
Applying this to \re{gammahExp} and adding
\begin{align}
\int_0^\infty \frac{dt}{t} J_0 (2gt) \widetilde\gamma^{\rm h}_{+,v} (2gt)
=
0
\end{align}
to the right-hand side for convergence, we can use the flux-tube equation \re{AnotherHoleFTgammaTildePlus} to write
\begin{align}
\int_0^\infty \frac{dt}{t} 
\widetilde\gamma^{\rm h}_{+,u} (2gt) {\rm e}^{- ivt}
=
-
\int_0^\infty \frac{dt}{t} 
\widetilde\gamma^{\rm h}_{+,u} (2gt) {\rm e}^{ivt}
- 2 \int_0^\infty \frac{dt}{t} \frac{\cos(vt) - J_0 (2gt)}{{\rm e}^t - 1} \left( \widetilde\gamma^{\rm h}_u (2gt) - J_0 (2gt) \right)
\, .
\end{align}
Returning to the real axis, we replace the exponent ${\rm e}^{ivt}$ in the first term in the right-hand side by $\cos(vt)$. Thus, we find
\begin{align}
\label{f1Ltof}
f^{(1)}_{\rm hF} (u, v_{\circlearrowright})
= f^{(1)}_{\rm hf} (u, v) - \int_0^\infty \frac{dt}{t} \frac{{\rm e}^{t/2} \sin(ut)}{{\rm e}^t - 1} \left( \cos(vt) - J_0 (2gt) \right)
\, ,  
\end{align}
where $v_{\circlearrowright}$ stands for $v$ on the small fermion sheet of the Riemann surface. The small-fermion phase $f^{(1)}_{\rm hf}$ is
\begin{align}
f^{(1)}_{\rm hf} (u, v) = - \frac{1}{2} \int_0^\infty \frac{dt}{t} \cos(vt) \widetilde\gamma^{\rm h}_{+, u} (2gt)
\, .
\end{align}
As we can see from its explicit representation, this function possesses only one cut on the real axis, as anticipated, see Fig.\ \ref{FermionAnalytic}. It 
becomes implicit however, if we use exchange relations and trade the $u$ dependence for the $v$ dependence
\begin{align}
f^{(1)}_{\rm hf} (u, v) = \int_0^\infty \frac{dt}{t} \frac{\sin(ut) {\rm e}^{t/2}}{{\rm e}^t - 1} \gamma^{\rm f}_{v} (2gt)
\, .
\end{align}
But can be seen to be identical to the previous form by means of the flux-tube equations from Appendix \ref{SmallFTappendix}.
Finally, the analytic continuation of $f^{(2)}_{\rm hF} (u, v)$ is accomplished analogously and reads
\begin{align}
\label{f2Ltof}
f^{(2)}_{\rm hF} (u, v_{\circlearrowright})
= 
f^{(2)}_{\rm hf} (u, v) - \int_0^\infty \frac{dt}{t} \frac{\sin(vt)}{{\rm e}^t - 1} \left( {\rm e}^{t/2} \cos(ut) - J_0 (2gt) \right)
\, ,  
\end{align}
where the phase admits the two equivalent representations
\begin{align}
f^{(2)}_{\rm hf} (u, v) 
= 
- \frac{1}{2} \int_0^\infty \frac{dt}{t} \sin (vt) \gamma^{\rm h}_{-, u} (2gt)
=
\int_0^\infty \frac{dt}{t} \frac{{\rm e}^{t/2} \cos (ut) - J_0 (2gt)}{{\rm e}^t - 1} \widetilde\gamma^{\rm f}_{v} (2gt)
\, .
\end{align}
Adding these results together, we conclude that the extra terms in Eqs.\ \re{f1Ltof} and \re{f2Ltof} cancel $\sigma_{\rm hF} (u,v)$, and thus the S-matrix for 
the small fermion scattering on the hole is
\begin{align}
S_{\rm hf} (u, v) = S_{\rm hF} (u, v_{\circlearrowright})
=
\exp \left( 
- 2 i f^{(1)}_{\rm hf} (u, v) + 2 i f^{(2)}_{\rm hf} (u, v)
\right)
\, ,
\end{align}
where  $\sigma_{\rm hf} (u,v) =  0$.

\subsubsection{Small-fermion--large-fermion S-matrix}
\label{SLSmatrix}

To obtain the small-fermion--large-fermion S-matrix, we now analytically continue in the rapidity $u$. All we need to this end is the analogue of the
flux-tube equations \re{AnotherHoleFTgammaPlus} -- \re{AnotherHoleFTgammaTildeMinus} but for the large fermions. They are collected in
Appendix \ref{LargeFTappendix}, such that
\begin{align}
f^{(1)}_{\rm FF} (u_{\circlearrowright}, v)
&= 
f^{(1)}_{\rm fF} (u, v) - \int_0^\infty \frac{dt}{t} \frac{\sin(ut)}{{\rm e}^t - 1} \left( \cos(vt) - J_0 (2gt) \right)
\, ,  \\
f^{(2)}_{\rm FF} (u_{\circlearrowright}, v)
&= 
f^{(2)}_{\rm fF} (u, v) - \int_0^\infty \frac{dt}{t} \frac{\cos (ut) - J_0 (2gt)}{{\rm e}^t - 1} \sin(vt) 
\end{align}
where
\begin{align}
f^{(1)}_{\rm fF} (u, v) 
&
= - \frac{1}{2} \int_0^\infty \frac{dt}{t} \sin (ut) \gamma^{\rm F}_{-, v} (2gt)
\nonumber\\
\label{f1fF}
&
= \int_0^\infty \frac{dt}{t} \frac{\widetilde\gamma^{\rm f}_u (2gt)}{{\rm e}^t - 1} \left( \cos(vt) - J_0 (2gt) \right)
+
\frac{1}{2} \int_0^\infty \widetilde\gamma^{\rm f}_{+,u} (2gt) \cos(vt)
\
\, , \\
f^{(2)}_{\rm fF} (u, v) 
&
= - \frac{1}{2} \int_0^\infty \frac{dt}{t} ( \cos (ut) - J_0 (2gt) ) \widetilde\gamma^{\rm F}_{+, v} (2gt)
\nonumber\\
\label{f2fF}
&=
\int_0^\infty \frac{dt}{t} 
\frac{\gamma^{\rm f}_u (2gt)}{{\rm e}^t - 1}
\sin (vt) 
+
\frac{1}{2}
\int_0^\infty \frac{dt}{t} 
\gamma^{\rm f}_{-,u} (2gt) \sin (vt) 
\, .
\end{align}
Here we used exchange relations after the second equality signs. These are derived making use of the results in Sect.\ \ref{ExchangeRelationsSection}
by replacing the superscript ${\rm h} \to {\rm f}$. The latter representation in above equations will be indispensable for analysis in the
following section. From here we conclude that the $S_{\rm fF}$-matrix is
\begin{align}
\label{Ssmalllarge}
S_{\rm fF} (u, v)
=
S_{\rm FF} (u_{\circlearrowright}, v) 
=
\exp \left( - 2i f^{(1)}_{\rm fF} (u, v) + 2i f^{(2)}_{\rm fF} (u, v)  \right)
\, ,
\end{align}
again in agreement with \cite{Basso:2014koa}.

\subsubsection{Small-fermion--small-fermion S-matrix}

To deduce the small-fermion--small-fermion S-matrix, we perform the analytic continuation of \re{Ssmalllarge} in the large fermion
rapidity $v$ making use of the representations \re{f1fF} and \re{f2fF}. We find immediately,
\begin{align}
S_{\rm ff} (u, v)
=
S_{\rm FF} (u_{\circlearrowright}, v_{\circlearrowright})
=
\exp \left( - 2i f^{(1)}_{\rm ff} (u, v) + 2i f^{(2)}_{\rm ff} (u, v)  \right)
\, ,
\end{align}
with the phases
\begin{align}
f^{(1)}_{\rm ff} (u, v) 
&
= - \frac{1}{2} \int_0^\infty \frac{dt}{t} \cos (vt) \widetilde\gamma^{\rm f}_{+, u} (2gt)
\, , \\
f^{(2)}_{\rm ff} (u, v) 
&
= - \frac{1}{2} \int_0^\infty \frac{dt}{t} \sin (vt) \gamma^{\rm f}_{-, u} (2gt)
\, .
\end{align}
To derive these results we have used the flux-tube equations for the small fermions given in Appendix \ref{SmallFTappendix}.
Again, the exchange relations imply the symmetry property $f^{(1)}_{\rm ff} (u, v) = f^{(2)}_{\rm ff} (v, u)$.

\subsection{Gauge fields}
\label{GGsmatrix}

Finally we turn to the gauge field and its bound states and how they scatter on each other and remaining excitations.
As we found in Sect.\ \ref{GaugeBaxterSection}, the solution to the gauge Baxter equation \re{HalfGaugeBaxter} in the upper half plane is
\begin{align}
\label{QgPlus}
\ln Q^{\rm g}_+ (u^+)
&
=
\ln c^{\rm g}_+
-
2 i u \ln \eta
-
2 \int_0^\infty \frac{dt}{t} \frac{{\rm e}^{i u t} J_0 (2gt) - iut - 1}{{\rm e}^t - 1}
\\
&
+
\int_0^\infty \frac{dt}{t} \frac{{\rm e}^{iut} \Gamma^{\rm g} (2gt) - 2gt \Gamma^{\rm g}_1}{{\rm e}^t - 1}
+
\int_0^\infty \frac{dt}{t ({\rm e}^t - 1)} {\rm e}^{iut}
\left(
{\rm e}^{- i v t} {\rm e}^{- \ell t/2} - J_0 (2gt)
\right)
\, , \nonumber
\end{align}
and analogously for the $\Im{\rm m} [u] < 0$,
\begin{align}
\label{QgMinus}
\ln Q^{\rm g}_- (u^-)
&
=
\ln c^{\rm g}_-
+
2 i u \ln \eta
-
2 \int_0^\infty \frac{dt}{t} \frac{{\rm e}^{- i u t} J_0 (2gt) + iut - 1}{{\rm e}^t - 1}
\\
&
+
\int_0^\infty \frac{dt}{t} \frac{{\rm e}^{- iut} \bar\Gamma^{\rm g} (2gt) - 2gt \bar\Gamma^{\rm g}_1}{{\rm e}^t - 1}
+
\int_0^\infty \frac{dt}{t ({\rm e}^t - 1)} {\rm e}^{-iut}
\left(
{\rm e}^{i v t} {\rm e}^{- \ell t/2} - J_0 (2gt)
\right)
\, . \nonumber
\end{align}
The condition of zero quasimomentum fixes the ratio of the unknown constants to be
\begin{align}
0 
&
= \frac{1}{2 \pi i} \oint \frac{dx}{x} \ln \frac{Q^{\rm g}_- (u^-)}{Q^{\rm g}_+ (u^+)}
=
\ln\frac{c^{\rm g}_-}{c^{\rm g}_+}
\\
&+
2i \int_0^\infty \frac{dt}{t ({\rm e}^t - 1)} \left( J_0 (2gt) \widetilde\gamma^{\rm g}_v (2gt) - 2gt \widetilde\gamma^{\rm g}_{v, 1} \right)
+
2i \int_0^\infty \frac{dt}{t ({\rm e}^t - 1)} J_0 (2gt) {\rm e}^{- \ell t/2} \sin(v t)
\, . \nonumber
\end{align}
Here we used the linear combinations introduced in Eq.\ \re{GammaTogamma} and the fact that
\begin{align}
\gamma^{\rm g} (t) = \gamma^{\O} (t) + \gamma^{\rm g}_v (t)
\, , \qquad
\widetilde\gamma^{\rm g} (t) = \widetilde\gamma^{\rm g}_v (t)
\, ,
\end{align}
as for all other propagating states.

\subsubsection{Bound-state--bound-state S-matrix}

Let us find the scattering matrix of a gauge stack on a conjugate, i.e., opposite helicity, gauge stack. The starting point is the gauge Bethe-Yang function, see 
Eq.\ (3.38) in Ref.\ \cite{Basso:2011rc},
\begin{align}
Y^{\rm g} (u)
=
\frac{Q^{\rm g}_- (u - \ft{i}{2} (\ell + 1))}{Q^{\rm g}_+(u + \ft{i}{2} (\ell + 1))}
\left[
\frac{\Delta^{\rm g}_- (u + \ft{i}{2}\ell) \Delta^{\rm g}_- (u - \ft{i}{2} \ell)}{\Delta^{\rm g}_+ (u + \ft{i}{2}\ell) \Delta^{\rm g}_+ (u - \ft{i}{2} \ell)}
\right]^{1/2}
\, .
\end{align}
Substituting \re{DeltaGauge} and the above solutions \re{QgPlus} and \re{QgMinus}, we find after some manipulations
\begin{align}
Y^{\rm g} (u) = {\rm e}^{i P_{\rm g} (u)} S_{{\rm g} {\rm g}'} (u, v)
\, ,
\end{align}
where the $\ell$-stack is associated with rapidity $u$, while the $\ell'$-stack is centered around the rapidity $v$. The propagating dynamical phase is 
expressed in terms of the vacuum solution as follows
\begin{align}
P_{\rm g} (u) = 4 u \ln \eta 
&+ 
4
\int_0^\infty \frac{dt}{t ({\rm e}^t - 1)} 
\left(
{\rm e}^{- \ell t/2} \sin (u t) J_0 (2gt) - ut
\right)
\\
&
-
2
\int_0^\infty \frac{dt}{t} {\rm e}^{- \ell /2} \sin (ut) 
\left[ 
\frac{\gamma_+^{\O} (2gt)}{{\rm e}^t - 1}
+
\frac{\gamma_-^{\O} (2gt)}{1 - {\rm e}^{-t }}
\right]
\, , \nonumber
\end{align}
while the S-matrix is
\begin{align}
S_{{\rm g}{\rm g}'} (u, v)
=
\exp \left( 2 i \sigma_{{\rm g}{\rm g}'} (u, v) - 2i f^{(1)}_{{\rm g}{\rm g}'} (u, v) + 2i f^{(2)}_{{\rm g}{\rm g}'} (u, v) \right)
\, ,
\end{align}
with
\begin{align}
\sigma_{{\rm g}{\rm g}'} (u, v)
=
\int_0^\infty \frac{dt}{t ({\rm e}^t - 1)}
\left[
{\rm e}^{- \ell t/2} \sin (ut) J_0 (2gt) - {\rm e}^{- \ell^\prime t/2} \sin (vt) J_0 (2gt) - {\rm e}^{- (\ell + \ell') t/2} \sin ((u-v)t)
\right]
\, ,
\end{align}
and
\begin{align}
f^{(1)}_{{\rm g}{\rm g}'} (u,v)
&
=
\int_0^\infty \frac{dt}{t} {\rm e}^{- \ell t/2} \sin (u t)
\left[
\frac{\gamma_{+,v}^{\rm g} (2gt)}{{\rm e}^t - 1}
+
\frac{\gamma_{-,v}^{\rm g} (2gt)}{1 - {\rm e}^{- t}}
\right]
\, , \\
f^{(2)}_{{\rm g}{\rm g}'} (u,v)
&
=
\int_0^\infty \frac{dt}{t} 
\left(
{\rm e}^{- \ell t/2} \cos (u t)
-
J_0 (2gt)
\right)
\left[
\frac{\widetilde\gamma_{-,v}^{\rm g} (2gt)}{{\rm e}^t - 1}
+
\frac{\widetilde\gamma_{+,v}^{\rm g} (2gt)}{1 - {\rm e}^{- t}}
\right]
\, .
\end{align}

\subsubsection{Hole-bound-state S-matrix}
\label{HGsmatrix}

Let us introduce the Bethe-Yang function
\begin{align}
Y^{\rm h} (u) = \frac{Q^{\rm g}_- (u)}{Q^{\rm g}_+ (u)}
\, ,
\end{align}
determined by the functions $Q_\pm^{\rm g}$ that are the solution to the gauge Baxter equation \re{GaugeBaxter} in the upper/lower half plane.
The $Y^{\rm h}$ admits the representation
\begin{align}
Y^{\rm h} (u) = {\rm e}^{i P_{\rm h} (u)} S_{\rm hg} (u, v)
\, ,
\end{align}
where $P_{\rm h}$ is defined in Eq.\ \re{HoleMomentum}, while the scattering matrix does not depend on the gauge-field helicity and reads
\begin{align}
S_{\rm hg} (u, v) = \exp \left( 2i \sigma_{\rm hg} (u, v) - 2i f^{(1)}_{\rm hg} (u,v) + 2i f^{(2)}_{\rm hg} (u,v) \right)
\end{align}
with
\begin{align}
\label{Sigmahg}
\sigma_{\rm hg} (u, v)
=
\int_0^\infty \frac{dt}{t ({\rm e}^t - 1)} 
\left[
{\rm e}^{t/2} J_0 (2gt) \sin(ut) 
-
{\rm e}^{- \ell t/2} J_0 (2gt) \sin (vt) - {\rm e}^{- (\ell - 1) t/2} \sin ((u - v)t) 
\right]
\, ,
\end{align}
and
\begin{align}
\label{f1hg}
f^{(1)}_{\rm hg} (u,v)
&
=
\int_0^\infty \frac{dt}{t ({\rm e}^t - 1)} 
{\rm e}^{t/2} \sin(u t) \gamma^{\rm g}_v (2gt) 
\, , \\
\label{f2hg}
f^{(2)}_{\rm hg} (u,v)
&
=
\int_0^\infty \frac{dt}{t ({\rm e}^t - 1)} 
\left( {\rm e}^{t/2} \cos(u t) - J_0 (2gt) \right)
\widetilde\gamma^{\rm g}_v (2gt) 
\, .
\end{align}

For $\ell = 1$ we obtain the hole-gauge field scattering matrix, i.e, the one that will appear in the analysis of nonsinglet pentagon
transitions, so we deal with them below separately. Namely, the phases of the S-matrix are determined by Eqs.\ \re{f1hg} and \re{f2hg}.
Since this will be the main object of our consideration that follows, we can cast these functions in a series form that will be useful in
studies of exchange relations and perturbative expansions. Namely, expanding the $\gamma$ in the Neumann series
\begin{align}
\gamma^{\rm g}_v (t) = 2 \sum_{n \geq 1} n J_n (t) \gamma^{\rm g}_{n, v}
\, ,
\end{align}
and solving for $\gamma^{\rm g}_{n, v}$ from Eq.\ \re{gammaFormalSolution}, we immediately find a generic representation
\begin{align}
\label{f1series}
f^{(1)}_{{\rm pp}^\prime} (u, v)
&
=
- 2 \widetilde\kappa_n^{\rm p} (u) Q_{nl} [1 + K]_{lm}^{-1} \kappa_m^{{\rm p}^\prime} (v) 
\, , \\
\label{f2series}
f^{(2)}_{{\rm pp}^\prime} (u, v)
&
=
- 2 \kappa_n^{\rm p} (u) (-1)^n Q_{nl} [1 + K]_{lm}^{-1} (-1)^m \widetilde\kappa_m^{{\rm p}^\prime} (v)
\, ,
\end{align}
with ${\rm p} ={\rm  h}$ and ${\rm p}^\prime = {\rm g}$ for the case at hand, and where 
\begin{align}
Q_{nl} = n \delta_{nl}
\, ,
\end{align}
and summation over the repeated indices is implied to run from one to infinity.

\subsubsection{Gauge bound-state--fermion S-matrix}
\label{GFsmatrix}

Without further ado, we quote the S-matrices for scattering of a helicity-one gauge bound state on a helicity-one-half fermion.
These take the same generic form as any other scattering phases 
\begin{align}
S_{\star{\rm g}} (u, v) = \exp \left( 2i \sigma_{\star{\rm g}} (u, v) - 2 i f^{(1)}_{\star{\rm g}} (u, v) + 2 i f^{(2)}_{\star{\rm g}} (u, v) \right)
\, ,
\end{align}
with $\star = {\rm F, f}$ corresponding to the large and small fermion, respectively. Here the $f$-functions admit an infinite-series representation 
\re{f1series} and \re{f2series}. For the large fermion ($\star = {\rm F}$), these can be cast in the integral form
\begin{align}
f^{(1)}_{\rm Fg} (u, v)
&
=
\int_0^\infty \frac{dt}{t} \frac{\gamma^{\rm g}_v (2gt)}{{\rm e}^t - 1} \sin (ut) 
+
\frac{1}{2}
\int_0^\infty \frac{dt}{t} \gamma^{\rm g}_{-, v} (2gt) \sin (ut) 
\, , \\
f^{(2)}_{\rm Fg} (u, v)
&
=
\int_0^\infty \frac{dt}{t} \frac{\widetilde\gamma^{\rm g}_v (2gt)}{{\rm e}^t - 1} \left( \cos (ut) - J_0(2gt) \right)
+
\frac{1}{2}
\int_0^\infty \frac{dt}{t} \widetilde\gamma^{\rm g}_{+, v} (2gt) \cos (ut) 
\, ,
\end{align}
while the explicit $\sigma$-phase reads
\begin{align}
\sigma_{{\rm Fg}} (u, v) 
=
\int_0^\infty \frac{dt}{t ({\rm e}^t - 1)}
\left[
J_0 (2gt) \sin (ut) - {\rm e}^{- \ell t/2} J_0 (2gt) \sin (vt) -  {\rm e}^{- \ell t/2} \sin ((u - v)t) 
\right] 
\, .
\end{align}
The analytic continuation to the small fermion sheet produces the answer
\begin{align}
f^{(1)}_{\rm fg} (u, v)
&
= -
\frac{1}{2}
\int_0^\infty \frac{dt}{t} \gamma^{\rm g}_{-, v} (2gt) \sin (ut) 
\, , \\
f^{(2)}_{\rm fg} (u, v)
&
= -
\frac{1}{2}
\int_0^\infty \frac{dt}{t} \widetilde\gamma^{\rm g}_{+, v} (2gt) \cos (ut) 
\, ,
\end{align}
with $\sigma_{{\rm fg}} (u, v) = 0$. This concludes our discussion of scattering matrices for main flux-tube excitations.

\section{Mirror kinematics}
\label{MirrorKinematicsSection}

As will be clear from the following section, the flux-tube form factors  arise as solutions to a set of bootstrap equations that involve
transformation to the mirror kinematics. The latter was studied in detail in Ref.\ \cite{Basso:2011rc} for holes and gauge bound states, 
while the more intricate fermionic case was addressed just recently in Ref.\ \cite{Basso:2014koa}. In this section we will study the mirror 
transformation and crossing for scattering matrices. For pedagogical purposes, we will make the analysis rather detailed. However, before 
going to the next section the reader is invited to take a lightning-quick course on mirror transformations 
reviewed in Appendices \ref{ScalarMirrorApp} and \ref{GaugeMirrorApp}.

Following the outline of the previous section, we again start by reproducing known phases in the mirror kinematics \cite{Basso:2013pxa,Basso:2013aha,Basso:2014koa}
for hole-hole, gauge-gauge and fermion-fermion scattering in Sections \ref{HHmirrorsmatrix}, \ref{GGmirrorsmatrix} and \ref{FFmirrorsmatrix}, respectively.
At the same time, we provide a new study of the mirror kinematics for hole-gauge and hole-fermion S-matrices in, accordingly, Sections \ref{GHmirrorsmatrix1} 
(\ref{GHmirrorsmatrix2}) and \ref{HFmirrorsmatrix}. These form the basic building blocks for the calculation of particular NMHV components of the superamplitude addressed 
later in Section \ref{NMHVhexagon}.

\subsection{Hole-hole S-matrix}
\label{HHmirrorsmatrix}

Here we will revisit the analytical continuation of the hole-hole scattering matrix to the mirror kinematics, though it was discussed
at length in Ref.\ \cite{Basso:2013pxa}, since we will use intermediate results for the hole-gauge field analysis, that are of primary 
interest to this work later in this section. Let us take the hole-hole S-matrix \re{hhSmatrix} and analytically continue it to the mirror 
point $u^\gamma = u + i$. 

Let us start with $f^{(1)}_{\rm hh} (u,v)$
\begin{align}
f^{(1)}_{\rm hh} (u, v) 
= 
\int_0^\infty \frac{dt}{t ({\rm e}^t - 1)} {\rm e}^{t/2} \sin (ut) \gamma^{\rm h}_v (2gt)
\, .
\end{align}
Its analytic continuation will be done stepwise as explained in Appendix \ref{ScalarMirrorApp}. First, we continue $f^{(1)}_{\rm hh}$ 
just below $u^+ = u + \ft{i}{2}$, such that it reads
\begin{align}
f^{(1)}_{\rm hh} (u^+ - i 0_+, v)
=
\int_0^\infty \frac{dt}{t ({\rm e}^t - 1)} \sin (ut) \gamma^{\rm h}_v (2gt)
+
\frac{i}{2}
\int_0^\infty \frac{dt}{t} {\rm e}^{-iut} \gamma^{\rm h}_v (2gt)
\, .
\end{align}
We cannot simply use the right-hand side to analytically continue above $\Im{\rm m}[u] > 0$ since the last term does not converge there. 
However using \re{AnContInt}, we can rewrite the last term in $f^{(1)}_{\rm hh} (u^+ - i 0_+, v)$ first as
\begin{align}
\label{ContEq1}
\int_0^\infty \frac{dt}{t} {\rm e}^{-iut} \gamma^{\rm h}_v (2gt)
&
=
-
\int_0^\infty \frac{dt}{t} {\rm e}^{iut} \gamma^{\rm h}_v (- 2gt)
\\
&
+
2 \int_0^\infty \frac{dt}{t} \left( \cos(ut) - J_0 (2gt) \right) \gamma^{\rm h}_{+,v} (2gt)
-
2i \int_0^\infty \frac{dt}{t} \sin(ut) \gamma^{\rm h}_{-,v} (2gt)
\, , \nonumber
\end{align}
such that the first term in Eq.\ \re{ContEq1} can be continued above $\Im{\rm m}[u] = 0$. However, for the last two terms this cannot be done in the current 
form. The continuation to $u^\gamma$ can, however, be easily accomplished using the flux-tube equations in the form \re{AnotherHoleFTgammaPlus} and 
\re{AnotherHoleFTgammaMinus}. Substituting the latter into Eq.\ \re{ContEq1}, we find
\begin{align}
f^{(1)}_{\rm hh} (u^+ + i 0_+, v)
&
=
- \frac{i}{2} \int_0^\infty \frac{dt}{t} {\rm e}^{iut} \gamma^{\rm h}_v (-2gt)
-
i \int_0^\infty \frac{dt}{t ({\rm e}^t - 1)} \left( \cos(ut) - J_0 (2gt) \right) \gamma^{\rm h}_v (-2gt)
\\
&
-
i \int_0^\infty \frac{dt}{t ({\rm e}^t - 1)} \left( {\rm e}^{-iut} - J_0 (2gt) \right) \left( {\rm e}^{t/2} \cos(vt) - J_0 (2gt) \right)
\, . \nonumber
\end{align}
In this form, we have sufficient suppression in the integrand to continue the integral to $u^\gamma = u + i$. After a little algebra, we find
\begin{align}
f^{(1)}_{\rm hh} (u^\gamma, v)
=
- i f^{(4)}_{\rm hh} (u, v)
-
i \int_0^\infty \frac{dt}{t ({\rm e}^t - 1)} 
\left( {\rm e}^{-iut + t/2} - J_0 (2gt) \right)  \left( {\rm e}^{t/2} \cos(vt) - J_0 (2gt) \right)
\, , \nonumber
\end{align}
where we introduced
\begin{align}
f^{(4)}_{\rm hh} (u, v)
=
\int_0^\infty \frac{dt}{t ({\rm e}^t - 1)}
\left( {\rm e}^{t/2} \cos(ut) - J_0 (2gt) \right) \gamma^{\rm h}_v (- 2gt)
\, .
\end{align}

The consideration of $f^{(2)}_{\rm hh}$ is analogous and is done again in two successive steps. First, we obtain
\begin{align}
f^{(2)}_{\rm hh} (u^+ - i 0_+, v) 
&= 
\int_0^\infty \frac{dt}{t ({\rm e}^t - 1)}
\left(
\cos (ut) - J_0 (2gt)
\right) \widetilde\gamma^{\rm h}_v (2gt)
+
\frac{1}{2}
\int_0^\infty \frac{dt}{t} {\rm e}^{-iut} \widetilde\gamma^{\rm h}_v (2gt)
\\
&
= 
\int_0^\infty \frac{dt}{t ({\rm e}^t - 1)}
\left(
\cos (ut) - J_0 (2gt)
\right) \widetilde\gamma^{\rm h}_v (2gt)
-
\frac{1}{2}
\int_0^\infty \frac{dt}{t} {\rm e}^{iut} \widetilde\gamma^{\rm h}_v (- 2gt)
\nonumber\\
&
+
\int_0^\infty \frac{dt}{t} \left( \cos(ut) - J_0 (2gt) \right) \widetilde\gamma^{\rm h}_{+,v} (2gt)
-
i \int_0^\infty \frac{dt}{t} \sin(ut) \widetilde\gamma^{\rm h}_{-,v} (2gt)
\, .
\end{align}
Here to be able to analytically continue the last two term, we used \re{AnotherHoleFTgammaTildePlus} and \re{AnotherHoleFTgammaTildeMinus}.
The last step is now trivial and we obtain at $u^\gamma$,
\begin{align}
f^{(2)}_{\rm hh} (u^\gamma, v)
=
- i f^{(3)}_{\rm hh} (u, v) 
-
i \int_0^\infty \frac{dt}{t ({\rm e}^t - 1)} 
\left( {\rm e}^{-iut + t/2} - J_0 (2gt) \right)  {\rm e}^{t/2} \sin(vt) 
\, , \nonumber
\end{align}
with 
\begin{align}
f^{(3)}_{\rm hh} (u, v) 
= 
\int_0^\infty \frac{dt}{t ({\rm e}^t - 1)} {\rm e}^{t/2} \sin(ut) \widetilde\gamma^{\rm h}_v (- 2gt)
\, .
\end{align}

Finally, we turn to $\sigma_{\rm hh}$ given in Eq.\ \re{Sigmahh}. As we can see from its integral representation, the first and last term require special care 
before we can safely analytically continue the integrand to $u^\gamma$. To this end, we break up the integral in two pieces,
\begin{align}
\sigma_{\rm hh} (u, v)
=
\int_0^\infty \frac{dt}{t ({\rm e}^t - 1)}
\big[
\sin(u^- t) J_0 (2gt) 
&
- {\rm e}^{t/2} \sin (vt) J_0 (2gt)- {\rm e}^{t/2} \sin ((u^- - v)t)
\big]
\nonumber\\
&
+ \frac{i}{2}
\int_0^\infty \frac{dt}{t} \left( {\rm e}^{- iut - t/2} J_0 (2gt) - {\rm e}^{-i(u-v)t} \right)
\, .
\end{align}
Making use of \re{AnContInt}, we find the analytic continuation for the first integral in the second line for $u \to u^+$
\begin{align}
\label{I1hh}
I_1 (u^+)
\equiv
\int_0^\infty \frac{dt}{t} {\rm e}^{- iut} J_0(2gt)
&=
-
\int_0^\infty \frac{dt}{t} {\rm e}^{iut} J_0(2gt)
+
2 \int_0^\infty \frac{dt}{t} J_0 (2gt)
\nonumber\\
&=
-
\int_0^\infty \frac{dt}{t} {\rm e}^{iut} J_0(2gt)
+
2 \int_0^\infty \frac{dt}{t} J_0^2 (2gt)
\, ,
\end{align}
where we used the fact that
\begin{align}
\int_0^\infty \frac{dt}{t} J_0 (2gt) (\cos(ut) - 1) 
=
\int_0^\infty \frac{dt}{t} J_0 (2gt) (\cos(ut) - J_0 (2gt)) 
= 0
\, ,
\end{align}
for $|u| < 2g$, such that
\begin{align}
I_1 (u^\gamma) = 
-
\int_0^\infty \frac{dt}{t} {\rm e}^{iut - t/2} J_0(2gt)
+
2 \int_0^\infty \frac{dt}{t} J_0^2 (2gt)
\, .
\end{align}
While for the second contribution, we easily get
\begin{align}
\label{I2hh}
I_2 (u-v)
=
\int_0^\infty  \frac{dt}{t} {\rm e}^{-i(u - v)t}
=
\int_0^\infty \frac{dt}{t} {\rm e}^{-i(u-v-i)t} - \ln \frac{u - v}{u - v - i}
\, .
\end{align}
Substituting Eq.\ \re{I1hh} into \re{I2hh}, we immediately find for $u \to u^\gamma = u + i$
\begin{align}
\sigma_{\rm hh} (u^\gamma, v) = \sigma_{\rm hh} (u, v) - \frac{i}{2} \ln \frac{u - v}{u - v + i} - i \int_0^\infty \frac{dt}{t} \left[ \cos ((u-v)t) - J_0^2 (2gt) \right]
\, .
\end{align}

\vspace{10pt}

\noindent Putting everything together, we deduce the mirror hole-hole S-matrix
\begin{align}
S_{\rm hh} (u^\gamma, v)
=
\frac{u - v}{u - v + i}
\exp \left( 2 \widehat\sigma_{\rm hh} (u, v) + 2 f^{(3)}_{\rm hh} (u, v) - 2 f^{(4)}_{\rm hh} (u, v) \right)
\, ,
\end{align}
where $f^{(3)}_{\rm hh}$ and $f^{(4)}_{\rm hh}$ were introduced above, while $\widehat\sigma_{\rm hh}$ is determined by
\begin{align}
\label{SigmahhAst}
\widehat\sigma_{\rm hh} (u, v)
=
\int_0^\infty \frac{dt}{t ({\rm e}^t - 1)}
\left[
{\rm e}^{t/2} \left( \cos(ut) + \cos(vt) \right) J_0 (2gt) - \cos((u-v)t) - {\rm e}^t J_0^2 (2gt)
\right]
\, .
\end{align}
This is the result advertised in Ref.\ \cite{Basso:2013pxa}.

\subsection{Gauge-gauge S-matrix}
\label{GGmirrorsmatrix}

Let us construct the mirror S-matrix for gauge field bound states. It was discussed previously in Refs.\ \cite{Basso:2013vsa} and \cite{Basso:2013aha}. 
The path to the mirror kinematics was  worked out in Ref.\ \cite{Basso:2011rc} and recollected in Appendix \ref{GaugeMirrorApp}. The
continuation is split into several steps following the contour in the complex rapidity plane as shown in Fig.\ \ref{GluonMirror}. First, we have to cross through 
the cut $[i/2-2g, i/2+2g]$. We will perform this individually for each contribution building up the S-matrix. Then we will combine them and go below the lower 
cut $[- i/2-2g, - i/2+2g]$ before crossing through it to go to the mirror sheet.

Let us start with
\begin{align}
f^{(1)}_{{\rm g}{\rm g}'} (u,v)
&
=
\int_0^\infty \frac{dt}{t} {\rm e}^{- \ell t/2} \sin (u t)
\left[
\frac{\gamma_{+,v}^{\rm g} (2gt)}{{\rm e}^t - 1}
+
\frac{\gamma_{-,v}^{\rm g} (2gt)}{1 - {\rm e}^{- t}}
\right]
\, .
\end{align}
First, we shift $u$ upwards just below the cut $[i/2-2g, i/2+2g]$
\begin{align}
f^{(1)}_{{\rm g}{\rm g}'} (u^+ - i 0_+,v)
&
=
\frac{i}{2} \int_0^\infty \frac{dt}{t} \left( {\rm e}^{- iut} - {\rm e}^{iut - t} \right)
\left[
\gamma_{-,v}^{\rm g} (2gt)
+
\frac{\gamma_v^{\rm g} (2gt)}{{\rm e}^t - 1}
\right]
\, .
\end{align}
To cross the cut $u \to u + i 0_+$ for $|u| < 2g$, the problem arises for the product of the first terms in the two brackets. To overcome the
complication, we use Eq.\ \re{AnContInt} and the flux-tube equation \re{GFMinus}, so that just above the cut, we find
\begin{align}
f^{(1)}_{{\rm g}{\rm g}'} (u^+ + i 0_+,v)
&
=
\frac{i}{2} \int_0^\infty \frac{dt}{t} {\rm e}^{iut} (1 - {\rm e}^{- t})
\left[
\frac{\gamma^{\rm g}_{-, v} (2gt)}{1 - {\rm e}^{-t}} 
+
\frac{\gamma^{\rm g}_{+, v} (2gt)}{{\rm e}^{t} - 1} 
\right]
\\
&
-
\int_0^\infty \frac{dt}{t} \frac{\sin(ut)}{{\rm e}^t - 1}
\left(
{\rm e}^{- t/2} \cos (vt) - J_0 (2gt)
\right)
\, .
\end{align}

Similarly for $f^{(2)}_{{\rm g}{\rm g}'}$ that reads
\begin{align}
f^{(2)}_{{\rm g}{\rm g}'} (u,v)
&
=
\int_0^\infty \frac{dt}{t} 
\left(
{\rm e}^{- \ell t/2} \cos (u t)
-
J_0 (2gt)
\right)
\left[
\frac{\widetilde\gamma_{-,v}^{\rm g} (2gt)}{{\rm e}^t - 1}
+
\frac{\widetilde\gamma_{+,v}^{\rm g} (2gt)}{1 - {\rm e}^{- t}}
\right]
\, ,
\end{align}
we first shift $u \to u + \ft{i}{2} - i 0_+$ just below the cut and then cross it using the equation
\begin{align}
\int_0^\infty \frac{dt}{t} {\rm e}^{-iut} \widetilde\gamma^{\rm g}_{+, v} (2gt)
=
&
- \int_0^\infty \frac{dt}{t} {\rm e}^{iut} \widetilde\gamma^{\rm g}_{+, v} (2gt)
\\
&
- 
2
\int_0^\infty \frac{dt}{t ({\rm e}^t - 1)} 
\left( \cos(ut) - J_0 (2gt) \right)
\left[
\widetilde\gamma^{\rm g}_v (2gt)
+
{\rm e}^{- t/2} \sin (vt) 
\right]
\, . \nonumber
\end{align}
So that just above the cut, we analogously obtain
\begin{align}
f^{(2)}_{{\rm g}{\rm g}'} (u^+ + i 0_+,v)
=
&
- \frac{1}{2}
\int_0^\infty \frac{dt}{t} 
{\rm e}^{iut} \left( 1 - {\rm e}^{-t} \right)
\left[
\frac{\widetilde\gamma_{-,v}^{\rm g} (2gt)}{{\rm e}^t - 1}
+
\frac{\widetilde\gamma_{+,v}^{\rm g} (2gt)}{1 - {\rm e}^{- t}}
\right]
\\
&
-
\int_0^\infty \frac{dt}{t ({\rm e}^t - 1)} 
\left(
\cos (u t) - J_0 (2gt)
\right)
{\rm e}^{- t/2} \sin(vt)
\, .
\end{align}

For $\sigma_{{\rm g}{\rm g}'} (u,v)$ crossing the upper cut is not problematic and thus we simply shift $u \to u^+$ in the integrand.
The subsequent algebra simplifies if one considers the linear combination of all ingredients that form the scattering matrix. Namely,
adding them up together yields
\begin{align}
\ln S_{{\rm g}{\rm g}'} (u^+ + i 0_+,v)
=
&-
\int_0^\infty \frac{dt}{t} {\rm e}^{iut} (1 - {\rm e}^{-t})
\left[
\frac{\gamma_{+,v}^{\rm g} (2gt)}{{\rm e}^t - 1}
+
\frac{\gamma_{-,v}^{\rm g} (2gt)}{1 - {\rm e}^{- t}}
\right]
\nonumber\\
&-
i \int_0^\infty \frac{dt}{t} {\rm e}^{iut} (1 - {\rm e}^{-t})
\left[
\frac{\widetilde\gamma_{-,v}^{\rm g} (2gt)}{{\rm e}^t - 1}
+
\frac{\widetilde\gamma_{+,v}^{\rm g} (2gt)}{1 - {\rm e}^{- t}}
\right]
\nonumber\\
&-
\int_0^\infty \frac{dt}{t} \frac{1 - {\rm e}^{-t}}{{\rm e}^t - 1}
\left[
{\rm e}^{iut} J_0 (2gt) - {\rm e}^{i(u-v)t - t/2}
\right]
\, ,
\end{align}
which clearly demonstrates 
that the exponent of the S-matrix on the Goldstone sheet possesses only a finite number of cuts (i.e., two for gauge fields), same as the energy and momentum
discussed in Appendix \ref{GaugeMirrorApp}. As a next step, we have to analytically continue this expression for $|u|> 2g$ into the lower 
half-plane and shift $u \to u - i$. For the first two lines in the above equation, this was explained in Appendix  \ref{GaugeMirrorApp}. The last term
is purely rational, so its continuation is trivial. One gets then
\begin{align}
\ln S_{{\rm g}{\rm g}'} (u^+ - i,v)
=
&-
\int_0^\infty \frac{dt}{t} {\rm e}^{- iut} (1 - {\rm e}^{-t})
\left[
\frac{\gamma_{+,v}^{\rm g} (2gt)}{1 - {\rm e}^{-t}}
-
\frac{\gamma_{-,v}^{\rm g} (2gt)}{{\rm e}^t - 1}
\right]
\nonumber\\
&-
i \int_0^\infty \frac{dt}{t} {\rm e}^{- iut} (1 - {\rm e}^{-t})
\left[
- \frac{\widetilde\gamma_{-,v}^{\rm g} (2gt)}{1 - {\rm e}^{-t}}
+
\frac{\widetilde\gamma_{+,v}^{\rm g} (2gt)}{{\rm e}^t - 1}
\right]
\nonumber\\
&-
\int_0^\infty \frac{dt}{t}
\left[
{\rm e}^{- iut} J_0 (2gt) - {\rm e}^{- i(u-v)t - t/2}
\right]
- i \ln \frac{u - v - \ft{i}{2}}{u - v + \ft{i}{2}}
\, .
\end{align}
Next, we have to cross the lower cut $[-i/2-2g,-i/2+2g]$, i.e., $u \to u + i 0_+$. Using the gauge flux-tube equations that yield the relations
\begin{align}
\int_0^\infty \frac{dt}{t} {\rm e}^{-iut} \gamma_{+, v}^{\rm g} (2gt)
=
&-
\int_0^\infty \frac{dt}{t} {\rm e}^{iut} \gamma_{+, v}^{\rm g} (2gt)
\\
&-
2
\int_0^\infty \frac{dt}{t ({\rm e}^t - 1)} \left( \cos (ut) - J_0 (2gt) \right) 
\left[
\gamma^{\rm g}_v (- 2gt)
+
{\rm e}^{t/2} \cos (vt) - J_0 (2gt)
\right]
\, ,
\nonumber\\
\int_0^\infty \frac{dt}{t} {\rm e}^{-iut} \widetilde\gamma^{\rm g}_{-, v} (2gt)
=&
\int_0^\infty \frac{dt}{t} {\rm e}^{iut} \widetilde\gamma^{\rm g}_{-, v} (2gt)
+
2 i
\int_0^\infty \frac{dt}{t ({\rm e}^t - 1)} \sin (ut)
\left[
- \widetilde\gamma^{\rm g}_v (- 2gt)
+
{\rm e}^{t/2} \sin (vt)
\right]
\, , \nonumber
\end{align}
as well as with the help of Eq.\ \re{I1hh}, we can safely get across the cut to our final destination by the replacement $u \to u + \ft{i}{2}$, getting 
\begin{align}
S_{{\rm g}{\rm g}'} (u^\gamma ,v)
=\frac{u - v}{u - v + i}
\exp \left( 2 \widehat\sigma_{{\rm g}{\rm g}'} (u, v) + 2 f^{(3)}_{{\rm g}{\rm g}'} (u, v) - 2 f^{(4)}_{{\rm g}{\rm g}'} (u, v) \right)
\, ,
\end{align}
where
\begin{align}
\widehat\sigma_{{\rm g}{\rm g}'} (u, v) = \widehat\sigma_{\rm hh} (u, v)
\, ,
\end{align}
and
\begin{align}
f^{(3)}_{{\rm g}{\rm g}'} (u, v)
&=
\int_0^\infty \frac{dt}{t} {\rm e}^{-t/2} \sin (ut)
\left[
\frac{\widetilde\gamma_{+,v}^{\rm g} (2gt)}{{\rm e}^{t} - 1}
-
\frac{\widetilde\gamma_{-,v}^{\rm g} (2gt)}{1 - {\rm e}^{- t}}
\right]
\, , \\
f^{(4)}_{{\rm g}{\rm g}'} (u, v)
&=
\int_0^\infty \frac{dt}{t} \left( {\rm e}^{-t/2} \cos (ut) - J_0 (2gt) \right)
\left[
\frac{\gamma_{+,v}^{\rm g} (2gt)}{1 - {\rm e}^{- t}}
-
\frac{\gamma_{-,v}^{\rm g} (2gt)}{{\rm e}^{t} - 1}
\right]
\, .
\end{align}

\subsection{Gauge-hole S-matrix in hole rapidity}
\label{GHmirrorsmatrix1}

Now we analyze the hole-gauge field S-matrix that will play a pivotal role in the calculation of NMHV amplitudes. Namely,
we will demonstrate in this section the following identities that it obeys:
\begin{itemize}
\item Crossing: 
\begin{align}
\label{Sprop1}
S_{\rm hg} (u^{2 \gamma}, v) S_{\rm hg} (u, v) = S_{\rm hg} (u, v^{2 \gamma}) S_{\rm hg} (u, v) =  1
\, .
\end{align}
\item Mirror:
\begin{align}
\label{Sprop2}
S_{\rm hg} (u^\gamma, v^\gamma) =  S_{\rm hg} (u, v)
\, .
\end{align}
\item Unitarity:
\begin{align}
\label{Sprop3}
S_{\rm hg} (u, v) S_{\rm gh} (v, u) = 1
\, ,
\end{align}
\end{itemize}
where $S_{\rm gh} (v, u) = S_{\rm hg} (-u, -v)$. We can also establish a chain of useful identities
\begin{align}
S_{\rm hg} (u, v^\gamma) = [S_{\rm hg} (u^\gamma, v)]^{-1} = S_{\rm gh} (v, u^\gamma) = [S_{\rm gh} (v^\gamma, u)]^{-1}
\, ,
\end{align}
where the first equality is a consequence of the mirror transformation and crossing, while the second and the last are consequences of
unitarity.

\subsubsection{Mirror transformation}

Let us analytically continue $f^{(1)}_{\rm hg}$ given in Eq. \re{f1hg}. The first step is the same as for the $f^{(1)}_{\rm hh}$, i.e., for $u \to u + \ft{i}{2}$, we get
\begin{align}
f^{(1)}_{\rm hg} (u^+ - i 0_+, v)
&
=
\int_0^\infty \frac{dt}{t ({\rm e}^t - 1)} \sin (ut) \gamma^{\rm g}_v (2gt)
-
\frac{i}{2}
\int_0^\infty \frac{dt}{t} {\rm e}^{iut} \gamma^{\rm g}_v (- 2gt)
\\
&
+
i \int_0^\infty \frac{dt}{t} \left( \cos(ut) - J_0 (2gt) \right) \gamma^{\rm g}_{+,v} (2gt)
+
 \int_0^\infty \frac{dt}{t} \sin(ut) \gamma^{\rm g}_{-,v} (2gt)
\, .
\end{align}
Substitution of the gauge field flux-tube equations from Appendix \ref{GaugeFTappendix} into $f^{(1)}_{\rm hg} (u + \ft{i}{2}, v)$ gives
\begin{align}
f^{(1)}_{\rm hg} (u^+ + i 0_+, v)
=
- i  f^{(4)}_{\rm hg} (u - \ft{i}{2}, v)
&
-
i \int_0^\infty \frac{dt}{t ({\rm e}^t - 1)} \left( {\rm e}^{-iut} - J_0 (2gt) \right) \left( {\rm e}^{t/2} \cos(vt) - J_0 (2gt) \right)
\nonumber\\
&
+
\int_0^\infty \frac{dt}{t} {\rm e}^{-t/2} \sin (ut) \cos (vt)
\, ,
\end{align}
where
\begin{align}
f^{(4)}_{\rm hg} (u, v)
=
\int_0^\infty \frac{dt}{t ({\rm e}^t - 1)}
\left( {\rm e}^{t/2} \cos(ut) - J_0 (2gt) \right) \gamma^{\rm g}_v (- 2gt)
\, .
\end{align}
Further analytic continuation to $u^\gamma$ is straightforward here.

For the case of $f^{(2)}_{\rm hg}$, the first step of analytic continuation produces
\begin{align}
f^{(2)}_{\rm hg} (u^+ - i 0_+, v) 
&
= 
\int_0^\infty \frac{dt}{t ({\rm e}^t - 1)}
\left(
\cos (ut) - J_0 (2gt)
\right) \widetilde\gamma^{\rm g}_v (2gt)
-
\frac{1}{2}
\int_0^\infty \frac{dt}{t} {\rm e}^{iut} \widetilde\gamma^{\rm g}_v (- 2gt)
\nonumber\\
&
+
\int_0^\infty \frac{dt}{t} \left( \cos(ut) - J_0 (2gt) \right) \widetilde\gamma^{\rm g}_{+,v} (2gt)
-
i \int_0^\infty \frac{dt}{t} \sin(ut) \widetilde\gamma^{\rm g}_{-,v} (2gt)
\, .
\end{align}
Here the last two terms can be worked out using Eqs.\ \re{GFtildePlus} and \re{GFtildeMinus} from Appendix \ref{GaugeFTappendix}
such that
\begin{align}
f^{(2)}_{\rm hg} (u^+ + i 0_+, v)
=
- 
i f^{(3)}_{\rm hg} (u^-, v)
&
-
\int_0^\infty \frac{dt}{t ({\rm e}^t - 1)} \left( {\rm e}^{-iut} - J_0 (2gt) \right) {\rm e}^{t/2} \sin (vt)
\nonumber\\
&+
\int_0^\infty \frac{dt}{t} \left( \cos(ut) - J_0 (2gt) \right) {\rm e}^{-t/2} \sin (vt)
\, ,
\end{align}
with
\begin{align}
f^{(3)}_{\rm hg} (u, v) 
= 
\int_0^\infty \frac{dt}{t ({\rm e}^t - 1)} {\rm e}^{t/2} \sin(ut) \widetilde\gamma^{\rm g}_v (- 2gt)
\, .
\end{align}

Finally, for $\sigma_{\rm hg}$ we split it into two terms
\begin{align}
\sigma_{\rm hg} (u, v)
&
=
\int_0^\infty \frac{dt}{t ({\rm e}^t - 1)} 
\left[
\sin(u^- t) J_0 (2gt) - {\rm e}^{-t/2} \sin (vt) J_0 (2gt) - \sin((u-v)t)
\right]
\\
&
+ \frac{i}{2} \int_0^\infty \frac{dt}{t} {\rm e}^{-iut - t/2} J_0 (2gt)
\, .
\end{align}
While the shift $u \to u^\gamma$ is straightforward for the first line, in the second, one has to use the analytic continuation of the integral given in Eq.\ \re{I1hh} 
first.

Summarizing all of the above, we find the mirror S-matrix
\begin{align}
\label{ShgInuh}
S_{\ast{\rm hg}} (u, v) 
\equiv
S_{\rm hg} (u^\gamma, v) 
= 
\exp \left( 2 \widehat\sigma_{\rm hg} (u, v) + 2 f^{(3)}_{\rm hg} (u, v) - 2 f^{(4)}_{\rm hg} (u, v) \right)
\, ,
\end{align}
where $f^{(3,4)}_{\rm hg}$ were introduced earlier and 
\begin{align}
\widehat\sigma_{\rm hg} (u, v) = \widehat\sigma_{\rm hh} (u, v)
\, ,
\end{align}
with the latter given in \re{SigmahhAst}. In Eq.\ \re{ShgInuh}, we introduced a notation $S_{\ast{\rm hg}} (u, v)$ for the S-matrix with
the hole in the mirror kinematics.

\subsubsection{Crossing}

To establish the crossing relation for the mixed S-matrix, we have to perform a further $\gamma$-shift in $u$. The $f^{(3,4)}$ then read
\begin{align}
f^{(3)}_{\rm hg} (u^\gamma, v)
=
- i f^{(2)}_{\rm hg} (u, v)
&
- i \int_0^\infty \frac{dt}{t ({\rm e}^t - 1)} \left( {\rm e}^{iu^+t} - J_0 (2gt) \right) {\rm e}^{t/2} \sin (vt)
\nonumber\\
&
+ i \int_0^\infty \frac{dt}{t} \left( \cos(u^+t) - J_0 (2gt) \right) {\rm e}^{-t/2} \sin (vt)
\, , \\
f^{(4)}_{\rm hg} (u^\gamma, v)
=
- i f^{(1)}_{\rm hg} (u, v)
&
-
\int_0^\infty \frac{dt}{t ({\rm e}^t - 1)} \left( {\rm e}^{iu^+t} - J_0 (2gt) \right) \left( {\rm e}^{t/2} \cos(vt) - J_0 (2gt) \right)
\nonumber\\
&
+
i \int_0^\infty \frac{dt}{t} {\rm e}^{-t/2} \sin (u^+t) \cos(vt)
\, ,
\end{align}
such that
\begin{align}
\label{SholeCrossing}
S_{\rm hg} (u^{2 \gamma}, v) = \left[ S_{\rm hg} (u, v) \right]^{-1}
\, .
\end{align}

\subsection{Gauge-hole S-matrix in gauge rapidity}
\label{GHmirrorsmatrix2}

In this section, we will do the same consideration as above but now for the gauge rapidity $v$.

\subsubsection{Exchange relations}

To perform the analytic continuation in $v$, we have to make it explicit by means of exchange relations obtained in a manner identical to
the discussion in Sect.\ \ref{holeSmallSmatrix}. Namely, we find
\begin{align}
\label{HoleExchangef1Eq1}
f^{(1)}_{\rm hg} (u, v) 
&= 
- 2 \sum_{n \geq 1} (2n) \gamma^{\rm g}_{2n} (v) \widetilde\kappa^{\rm h}_{2n} (u)
- 2 \sum_{n \geq 1} (2n - 1) \gamma^{\rm g}_{2n - 1} (v) \widetilde\kappa^{\rm h}_{2n - 1} (u)
\, , \\
\label{HoleExchangef1Eq2}
&=
- 2 \sum_{n \geq 1} (2n) \widetilde\gamma^{\rm h}_{2n} (u) \kappa^{\rm g}_{2n} (v)
- 2 \sum_{n \geq 1} (2n-1) \widetilde\gamma^{\rm h}_{2n-1} (u) \kappa^{\rm g}_{2n-1} (v)
\, ,
\end{align}
so that
\begin{align}
f^{(1)}_{\rm hg} (u, v) 
=
\int_0^\infty \frac{dt}{t} 
\left[
\frac{\widetilde\gamma^{\rm h}_{+,u} (2gt)}{1 - {\rm e}^{- t}} 
+
\frac{\widetilde\gamma^{\rm h}_{-,u} (2gt)}{{\rm e}^{t} - 1} 
\right]
\left(
{\rm e}^{- t/2} \cos (vt) - J_0 (2gt)
\right)
\, .
\end{align}
An analogous consideration yields two equivalent series representation for $f^{(2)}_{\rm hg}$ as well
\begin{align}
f^{(2)}_{\rm hg} (u, v) 
&
= 
- 2 \sum_{n \geq 1} (2n) \widetilde\gamma_{2n}^{\rm g} (v) \kappa_{2n}^{\rm h} (u)
- 2 \sum_{n \geq 1} (2n - 1) \widetilde\gamma_{2n - 1}^{\rm g} (v) \kappa_{2n - 1}^{\rm h} (u)
\nonumber\\
\label{HoleExchangef2Eq2}
&
=
- 2 \sum_{n \geq 1} (2n) \gamma_{2n}^{\rm h} (u) \widetilde\kappa_{2n}^{\rm g} (v)
- 2 \sum_{n \geq 1} (2n - 1) \gamma_{2n - 1}^{\rm h} (u) \widetilde\kappa_{2n - 1}^{\rm g} (v)
\, ,
\end{align}
with the latter one that can be cast in the integral form making the dependence on the gluon rapidity manifest,
\begin{align}
f^{(2)}_{\rm hg} (u, v) 
= 
\int_0^\infty \frac{dt}{t} 
\left[
\frac{\gamma^{\rm h}_{+,u} (2gt)}{{\rm e}^{t} - 1} 
+
\frac{\gamma^{\rm h}_{-,u} (2gt)}{1 - {\rm e}^{- t}} 
\right]
{\rm e}^{- t/2} \sin (vt)
\, .
\end{align}
Applying the same analysis to $f^{(3,4)}_{\rm hg}$, we merely list the results in the series representation
\begin{align}
f^{(3)}_{\rm hg} (u, v) 
&=
- 2 \sum_{n \geq 1} (2n) \widetilde\gamma_{2n}^{\rm g} (v) \widetilde\kappa_{2n}^{\rm h} (u)
+ 2 \sum_{n \geq 1} (2n - 1) \widetilde\gamma_{2n - 1}^{\rm g} (v) \widetilde\kappa_{2n - 1}^{\rm h} (u)
\nonumber\\
\label{HoleExchangef3Eq2}
&=
- 2 \sum_{n \geq 1} (2n) \widetilde\gamma_{2n}^{\rm h} (u) \widetilde\kappa_{2n}^{\rm g} (v)
+ 2 \sum_{n \geq 1} (2n - 1) \widetilde\gamma_{2n - 1}^{\rm h} (u) \widetilde\kappa_{2n - 1}^{\rm g} (v)
\, , \\
f^{(4)}_{\rm hg} (u, v) 
&=
- 2 \sum_{n \geq 1} (2n) \gamma_{2n}^{\rm g} (v) \kappa_{2n}^{\rm h} (u)
+ 2 \sum_{n \geq 1} (2n - 1) \gamma_{2n - 1}^{\rm g} (v) \kappa_{2n - 1}^{\rm h} (u)
\nonumber\\
\label{HoleExchangef4Eq2}
&=
- 2 \sum_{n \geq 1} (2n) \gamma_{2n}^{\rm h} (u) \kappa_{2n}^{\rm g} (v)
+ 2 \sum_{n \geq 1} (2n - 1) \gamma_{2n - 1}^{\rm h} (u) \kappa_{2n - 1}^{\rm g} (v)
\, ,
\end{align}
and the integral form
\begin{align}
\label{f3hgGLuon}
f^{(3)}_{{\rm hg}} (u, v)
&=
\int_0^\infty \frac{dt}{t} {\rm e}^{-t/2} \sin (vt)
\left[
\frac{\widetilde\gamma_{+,u}^{\rm h} (2gt)}{{\rm e}^{t} - 1}
-
\frac{\widetilde\gamma_{-,u}^{\rm h} (2gt)}{1 - {\rm e}^{- t}}
\right]
\, , \\
\label{f4hgGLuon}
f^{(4)}_{{\rm hg}} (u, v)
&=
\int_0^\infty \frac{dt}{t} \left( {\rm e}^{-t/2} \cos (vt) - J_0 (2gt) \right)
\left[
\frac{\gamma_{+,u}^{\rm h} (2gt)}{1 - {\rm e}^{- t}}
-
\frac{\gamma_{-,u}^{\rm h} (2gt)}{{\rm e}^{t} - 1}
\right]
\, ,
\end{align}
respectively.

\subsubsection{Mirror kinematics in gluon rapidity}

Following the same footsteps as in the gauge-gauge S-matrix we find by moving to the Goldstone sheet $u \to u^+ + i 0_+$
\begin{align}
f^{(1)}_{\rm hg} (u,v^+)
=
&
- \frac{1}{2}
\int_0^\infty \frac{dt}{t} 
{\rm e}^{ivt} \left( 1 - {\rm e}^{-t} \right)
\left[
\frac{\widetilde\gamma_{-,u}^{\rm h} (2gt)}{{\rm e}^t - 1}
+
\frac{\widetilde\gamma_{+,u}^{\rm h} (2gt)}{1 - {\rm e}^{- t}}
\right]
\nonumber\\
&
-
\int_0^\infty \frac{dt}{t ({\rm e}^t - 1)} 
\left(
\cos (v t) - J_0 (2gt)
\right)
{\rm e}^{t/2} \sin(ut)
\, , \\
f^{(2)}_{\rm hg} (u,v^+)
&
=
\frac{i}{2} \int_0^\infty \frac{dt}{t} {\rm e}^{ivt} (1 - {\rm e}^{- t})
\left[
\frac{\gamma^{\rm h}_{+, u} (2gt)}{{\rm e}^{t} - 1} 
+
\frac{\gamma^{\rm h}_{-, u} (2gt)}{1 - {\rm e}^{-t}} 
\right]
\nonumber\\
&
-
\int_0^\infty \frac{dt}{t ({\rm e}^t - 1)}
\sin(vt)
\left(
{\rm e}^{t/2} \cos (ut) - J_0 (2gt)
\right)
\, .
\end{align}
So that the phases possess only a finite number of cuts by analogy to the gauge field energy and momentum studied in Appendix \ref{ScalarMirrorApp}.
Performing all steps, one finds
\begin{align}
S_{{\rm hg}} (u ,v^{\pm \gamma})
=
\exp \left( \mp 2 \widehat\sigma_{{\rm hg}} (u, v) \mp 2 f^{(3)}_{{\rm hg}} (u, v) \pm 2 f^{(4)}_{{\rm hg}} (u, v) \right)
\, ,
\end{align}
where $f^{(3,4)}_{{\rm hg}} (u, v)$ are given in Eqs.\ \re{f3hgGLuon} and \re{f4hgGLuon}. That is, we immediately conclude that
\begin{align}
\label{Shgmirrior}
S_{\rm hg} (u, v^\gamma) = [ S_{\rm hg} (u^\gamma, v) ]^{- 1}
\, ,
\end{align}
by comparing above to Eq.\ \re{ShgInuh}.

\subsubsection{Crossing in gluon rapidity}

Performing an additional mirror transformation in gluon rapidity, one finds the crossing relation
\begin{align}
S_{\rm hg} (u, v^{2 \gamma}) = [S_{\rm hg} (u, v)]^{-1}
\, .
\end{align}

\subsubsection{Unitarity}

One can easily construct the gauge-hole S-matrix,
\begin{align}
S_{\rm gh} (u, v) = \exp \left( 2i \sigma_{\rm gh} (u,v) -  2i f^{(1)}_{\rm gh} (u, v) + 2i f^{(2)}_{\rm gh} (u, v) \right)
\, ,
\end{align}
where
\begin{align}
\sigma_{\rm gh} (u,v)
=
\int_0^\infty \frac{dt}{t ({\rm e}^t - 1)} 
\left[
{\rm e}^{- t/2} J_0 (2gt) \sin(ut) 
-
{\rm e}^{t/2} J_0 (2gt) \sin (vt) - \sin ((u - v)t) 
\right]
\, ,
\end{align}
while
\begin{align}
f^{(1)}_{\rm gh} (u, v) 
&
= 
- 2 \sum_{n \geq 1} (2n) \widetilde\gamma^{\rm g}_{2n} (u) \kappa^{\rm h}_{2n} (v)
- 2 \sum_{n \geq 1} (2n - 1) \widetilde\gamma^{\rm g}_{2n - 1} (u) \kappa^{\rm h}_{2n - 1} (v)
\, , \\
f^{(2)}_{\rm gh} (u, v) 
&
= 
- 2 \sum_{n \geq 1} (2n) \gamma^{\rm g}_{2n} (u) \widetilde\kappa^{\rm h}_{2n} (v)
- 2 \sum_{n \geq 1} (2n - 1) \gamma^{\rm g}_{2n - 1} (u) \widetilde\kappa^{\rm h}_{2n - 1} (v)
\, .
\end{align}
These obviously obey
\begin{align}
\sigma_{\rm gh} (u,v)
=
-
\sigma_{\rm hg} (v,u)
\, , \qquad
f^{(1)}_{\rm gh} (u, v) 
=
f^{(2)}_{\rm hg} (v, u) 
\, , \qquad
f^{(2)}_{\rm gh} (u, v) 
=
f^{(1)}_{\rm hg} (v, u) 
\, ,
\end{align}
such that
\begin{align}
\label{SUnitarity}
S_{\rm hg} (u, v) S_{\rm gh} (v, u) = 1
\, ,
\end{align}
as expected.

\subsubsection{Mirror transformation}

Finally, one can demonstrate that
\begin{align}
S_{\rm hg} (u^\gamma, v^\gamma) = S_{\rm hg} (u, v) = [ S_{\rm gh} (v, u)]^{-1}
\, ,
\end{align}
by combining Eq.\ \re{Shgmirrior} with \re{SholeCrossing} and the unitarity relation \re{SUnitarity}.

\subsection{Hole-fermion S-matrix in hole rapidity}
\label{HFmirrorsmatrix}

To prepare ourselves to the construction of the mirror S-matrix for fermions, let us perform a mirror transformation in the hole rapidity in the 
hole-fermion $S_{\rm hF}$ matrix \re{HoleLargeFermSmatrix} since it will be an ingredient in the construction of the former. As before, it is
done in two steps. First, we go just below the cut $[-2g + i/2, 2g + i/2]$ and then cross it all the way to $u^\gamma = u + i$ using the integral form 
of flux-tube equations for the large fermion spelled out in Appendix \ref{LargeFTappendix}. This way the combination $f^{(1)}_{\rm hF} - f^{(2)}_{\rm hF}$ 
after the analytic continuation reads
\begin{align}
f^{(1)}_{\rm hF} (u^\gamma) 
&
- f^{(2)}_{\rm hF} (u^\gamma)
\nonumber\\
=
&
- i \int_0^\infty \frac{dt}{t} \frac{\gamma_v^{\rm F} (- 2gt)}{{\rm e}^t - 1} \left[ {\rm e}^{t/2} \cos(u t) - J_0 (2gt) \right]
+ 
i  \int_0^\infty \frac{dt}{t} \frac{\widetilde\gamma_v^{\rm F} (- 2gt)}{{\rm e}^t - 1} {\rm e}^{t/2} \sin(u t) 
\nonumber\\
&
-
i
\int_0^\infty \frac{dt}{t ({\rm e}^t - 1)} 
\left[
{\rm e}^{- i (u - v) t + t/2} - \left( {\rm e}^{- i ut + t/2} + {\rm e}^{i v t} \right) J_0 (2gt) + J_0^2 (2gt)
\right]
-
\frac{i}{2} \ln \left( \frac{x [v]}{u - v + \ft{i}{2}} \right)
, \nonumber
\end{align}
where we substituted the explicit form of the integral
\begin{align*}
\int_0^\infty \frac{dt}{t} \left[ \cos((u - v)t) - J_0 (2gt) \cos (vt) \right] = \ln \left( \frac{x[v]}{u - v} \right)
\, .
\end{align*}
We have to add to this the analytic continuation of $\sigma_{\rm hF}$
\begin{align}
\sigma_{\rm hF} (u^\gamma, v) 
&
= 
\int_0^\infty \frac{dt}{t ({\rm e}^t - 1)}
\left[
{\rm e}^{t/2} J_0 (2gt) \sin (ut) - J_0 (2gt) \sin (vt) - {\rm e}^{t/2} \sin ((u-v)t)
\right] 
\nonumber\\
&
+
i \int_0^\infty \frac{dt}{t} \left[ J_0^2 (2gt) - {\rm e}^{- t/2} \cos ((u-v)t) \right]
+
\frac{i}{2} \ln \left( \frac{u - v + \ft{i}{2}}{u - v - \frac{i}{2}} \right)
\, .
\end{align}
So that combining all of these contributions together, the mirror hole--large-fermion S-matrix finally reads
\begin{align}
\label{MirrorhF}
S_{\ast{\rm hF}} (u, v) \equiv S_{\rm hF} (u^\gamma, v)
= \frac{- g^2}{x [v] (u - v + \ft{i}{2})} 
\exp\left(
2 \widehat\sigma_{\rm hF} (u, v) + 2 f^{(3)}_{\rm hF} (u, v) - 2 f^{(4)}_{\rm hF} (u, v)
\right)
\, ,
\end{align}
with
\begin{align}
\widehat\sigma_{\rm hF} (u, v)
=
\int_0^\infty \frac{dt}{t ({\rm e}^t - 1)}
\left[
\left( {\rm e}^{t/2} \cos(ut) + \cos(vt) \right) J_0 (2gt) - {\rm e}^{t/2} \cos ((u - v)t) - J_0^2 (2gt)
\right] 
\, , 
\end{align}
and
\begin{align}
f^{(3)}_{\rm hF} (u, v)
&
=
\int_0^\infty \frac{dt}{t} \frac{\widetilde\gamma_v^{\rm F} (- 2gt)}{{\rm e}^t - 1} {\rm e}^{t/2} \sin(ut) 
\, , \\
f^{(4)}_{\rm hF} (u, v)
&
=
\int_0^\infty \frac{dt}{t} \frac{\gamma_v^{\rm F} (- 2gt)}{{\rm e}^t - 1} \left[ {\rm e}^{t/2} \cos(ut) - J_0 (2gt) \right]
\, .
\end{align}
We will use this result momentarily.

\subsection{Large-fermion--large-fermion S-matrix}
\label{FFmirrorsmatrix}

A complication with a naive mirror transformation for the fermion was uncovered early on in the analysis of Ref.\ \cite{Basso:2011rc}
which failed to find a path that double Wick rotates the fermionic dispersion relation. This issue was readdressed in Ref.\ \cite{Basso:2014koa}
where it was demonstrated that a transformation of this type is not even possible since it would break U(1) charge assignment for
flux-tube excitations. It was shown instead that there exists an interpolation that brings a twist-one fermion into a composite twist-two
excitation. Then one can use fusion to recover the mirror S-matrix for the original fermion. The outcome of the analysis \cite{Basso:2014koa}
can be summarized in the following mirror S-matrix for two large fermions 
\begin{align}
S_{\ast{\rm FF}} (u, v)
=
S_{{\rm fF}} (u, v) S_{\ast{\rm hF}} (u + \ft{i}{2}, v)
\, ,
\end{align}
with the two ingredients given above in Eqs.\ \re{Ssmalllarge} and \re{MirrorhF}, respectively. A simple calculation immediately gives
\begin{align}
S_{\ast{\rm FF}} (u, v)
=
- \frac{g^2}{x[u] x[v]}
\frac{u - v}{u - v + i} 
 \exp 
\left( 
2 \widehat{\sigma}_{\rm FF} (u, v) + 2 f^{(3)}_{\rm FF} (u, v) - 2 f^{(4)}_{\rm FF} (u, v)
\right)
\, .
\end{align}
Here we used the fact that
\begin{align}
\widehat{\sigma}_{\rm hF} (u + \ft{i}{2}, v) = \widehat{\sigma}_{\rm FF} (u, v) + \frac{1}{2} \ln \left( \frac{u-v}{x[u]} \right)
\, ,
\end{align}
with
\begin{align}
\widehat{\sigma}_{\rm FF} (u, v) 
=
\int_0^\infty \frac{dt}{t ({\rm e}^t - 1)}
\left[
\left( \cos(ut) + \cos(vt) \right) J_0 (2gt) - \cos((u-v) t) - J_0^2 (2gt)
\right]
\, ,
\end{align}
while the phase factors depending on the flux-tube functions are
\begin{align}
f^{(3)}_{\rm FF} (u, v) 
&
= \int_0^\infty \frac{dt}{t} \frac{\widetilde\gamma_v^{\rm F} (- 2gt)}{{\rm e}^t - 1} \sin (ut)
-
\frac{1}{2}  \int_0^\infty \frac{dt}{t} \widetilde\gamma_{-, v}^{\rm F} (2gt) \sin (ut)
\, , \\
f^{(4)}_{\rm FF} (u, v) 
&
= \int_0^\infty \frac{dt}{t} \frac{\gamma_v^{\rm F} (- 2gt)}{{\rm e}^t - 1} \left[ \cos (ut) - J_0 (2gt) \right]
+
\frac{1}{2}  \int_0^\infty \frac{dt}{t} \gamma_{+, v}^{\rm F} (2gt) \cos (ut)
\, .
\end{align}
This object will be indispensable in the construction of the pentagon transitions involving large fermions.

\subsection{Small-fermion--large-fermion S-matrix}

The analytic continuation to the small fermion sheet in rapidity $u$ follows the same footsteps as used in Sect.\ \ref{SLSmatrix}.
Sparing the reader from repetitious details, we just quote the final answer
\begin{align}
S_{\ast{\rm fF}} (u, v)
=
\frac{x[u]}{u - v + i}
 \exp 
\left( 
2 f^{(3)}_{\rm fF} (u, v) - 2 f^{(4)}_{\rm fF} (u, v)
\right)
\, ,
\end{align}
where
\begin{align}
f^{(3)}_{\rm fF} (u, v) 
&
= 
\frac{1}{2}  \int_0^\infty \frac{dt}{t} \widetilde\gamma_{-, v}^{\rm F} (2gt) \sin (ut)
\nonumber\\
&
= \int_0^\infty \frac{dt}{t}  \frac{\widetilde\gamma^{\rm f}_u (- 2gt)}{{\rm e}^t - 1} \sin(vt) - \frac{1}{2} \int_0^\infty  \frac{dt}{t}  \widetilde\gamma^{\rm f}_{-,u} (2gt) \sin (vt)
\, , \\
f^{(4)}_{\rm fF} (u, v) 
&
=
-
\frac{1}{2}  \int_0^\infty \frac{dt}{t} \gamma_{+, v}^{\rm F} (2gt) \cos (ut)
\nonumber\\
&
= \int_0^\infty \frac{dt}{t}  \frac{\gamma^{\rm f}_u (- 2gt)}{{\rm e}^t - 1} ( \cos (vt) - J_0 (2gt) )+ \frac{1}{2} \int_0^\infty  \frac{dt}{t}  \gamma^{\rm f}_{+,u} (2gt) \cos (vt)
\, .
\end{align}
Here we used exchange relations after the second equality sign to make explicit the dependence on $v$ for the subsequent analytic continuation to the small 
fermion sheet. This is done in the next section.

\subsection{Small-fermion--small-fermion S-matrix}

Finally, the analytic continuation to the small fermion sheet in rapidity $v$ yields
\begin{align}
S_{\ast{\rm ff}} (u, v)
=
\frac{u - v}{u - v + i} 
\exp \left( 
2 f^{(3)}_{\rm ff} (u, v) - 2 f^{(4)}_{\rm ff} (u, v)
\right)
\, ,
\end{align}
where
\begin{align}
f^{(3)}_{\rm ff} (u, v) 
&
= 
+
\frac{1}{2}  \int_0^\infty \frac{dt}{t} \widetilde\gamma_{-, u}^{\rm f} (2gt) \sin (vt)
\, , \\
f^{(4)}_{\rm ff} (u, v) 
&
=
-
\frac{1}{2}  \int_0^\infty \frac{dt}{t} \gamma_{+, u}^{\rm f} (2gt) \cos (vt)
\, .
\end{align}

\section{Nonsinglet form factors and pentagons}
\label{PentagonsSection}

As we advertised in the introduction, the goal of this work is to provide a nonperturbative description for twist-two contribution in
the operator product expansion of the null polygonal super Wilson loop with SU(4) nonsinglet quantum numbers in the expansion channel. Our
focus here will be on the particle pairs transforming in the ${\bf 6}$ of SU(4) and carrying one unit of total helicity. These will contribute nontrivially 
to a specific component of the NMHV amplitude that the super Wilson loop is dual to. There are two types of two-particle flux-tube
excitations that  fit the profile. These are either hole-gluon or two-fermion composite states. Let us discuss their coupling to
the Wilson loop contour and provide an educated guess for their form factors. The latter arise as solutions to conjectured axiomatic
equations that are based on interplay with two-dimensional scattering of flux-tube excitations and nontrivial consequences under mirror 
transformations and reflections. These turn out to be so constraining that they allow us to uniquely fix the functional form of form factors on 
particle rapidities as well as the coupling constant. Recently, this approach was successfully applied to form factors that emerge in the 
description of MHV amplitudes \cite{Basso:2013vsa,Basso:2013aha,Basso:2014koa} and leading twist contribution to certain components
of NMHV ones. Presently we generalize these considerations to the NMHV case at twist-two level.

\subsection{Hole-gluon pentagons}
\label{ScalarGluonFF}

\begin{figure}[t]
\begin{center}
\mbox{
\begin{picture}(0,100)(210,0)
\put(0,-150){\insertfig{12}{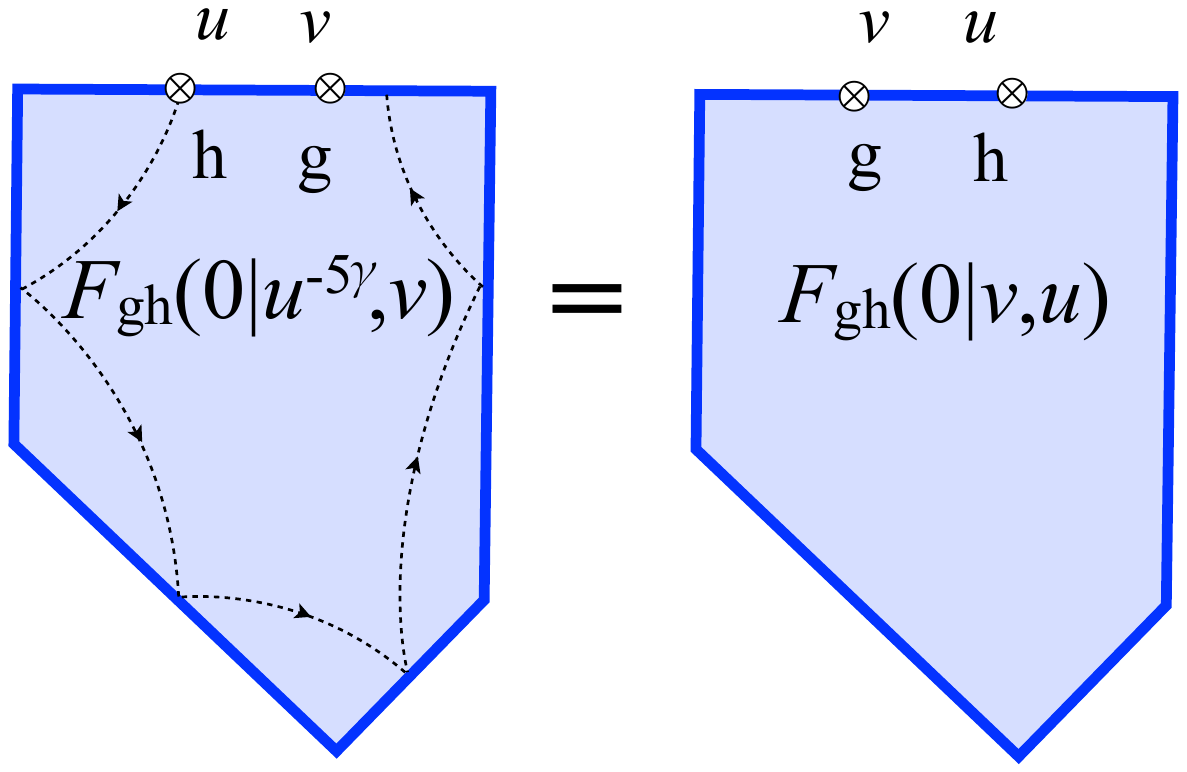}}
\put(260,-150){\insertfig{12}{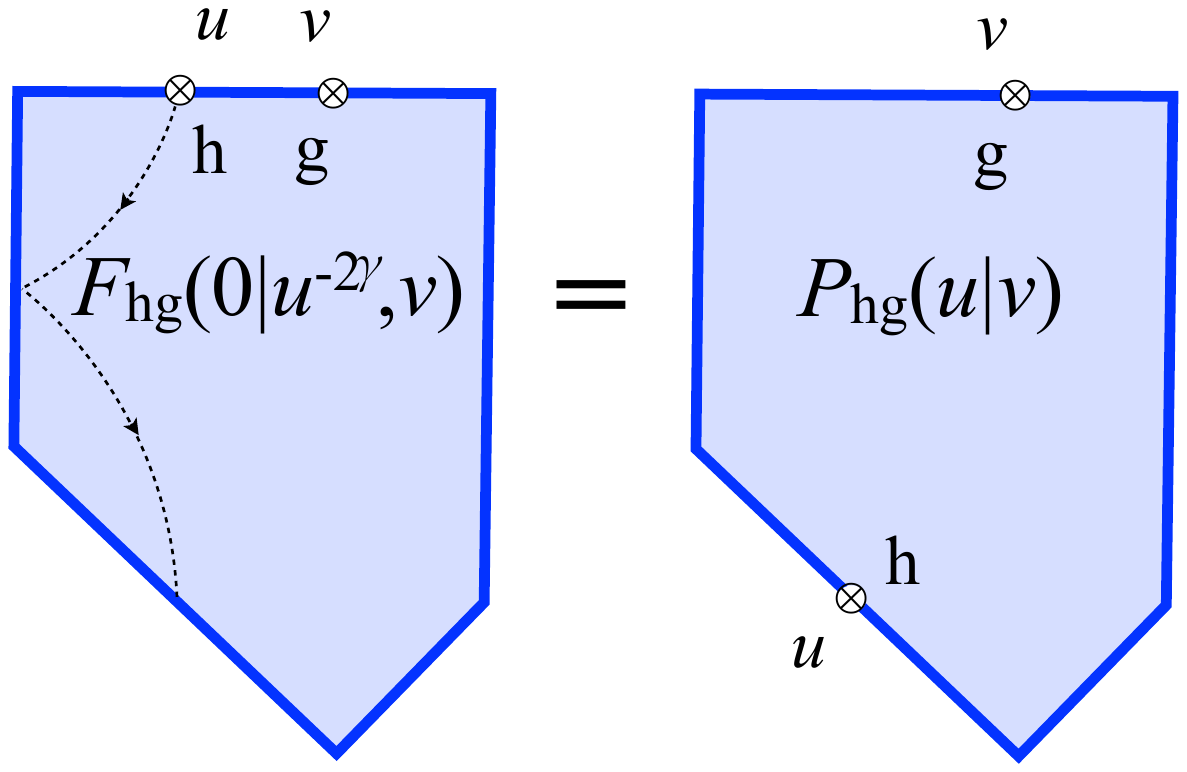}}
\end{picture}
}
\end{center}
\caption{ \label{HoleGluonMirror} Relation between $F_{\rm hg}$ and $F_{\rm gh}$ form factors through a mirror transformation that brings
the excitation in a complete cycle around the contour (left panel). A particular way to obtain the pentagon transition $P_{\rm hg}$ as a double 
mirror transformation in hole rapidity of the hole-gluon form factor (right panel).}
\end{figure}

We start with a simpler hole-gauge field case. Let us introduce the amplitudes for creation of the hole and gluon excitations with positive/negative helicity 
${\rm g/\bar{g}}$ on the top side of the pentagon
\begin{align}
F_{\rm hg} (0|u, v) 
&
= 
\bra{{\rm h}(u) {\rm g}(v)} \widehat{\mathcal{P}} \ket{0}
\, , \qquad
F_{\rm gh} (0|u, v) = \bra{{\rm g}(u) {\rm h}(v)} \widehat{\mathcal{P}} \ket{0}
\, , 
\end{align}
and similarly with ${\rm g}$ being replaced by ${\rm \bar{g}}$.
Permutation of the two excitations sitting on the same side is achieved by acting on it with a scattering matrix, such that the above form factors 
obey the Watson equations \cite{Watson:1954uc}
\begin{align}
\label{WatsonEqs}
F_{\rm hg} (0|u, v) = S_{\rm gh} (v, u) F_{\rm gh} (0|v, u)
\, , \qquad
F_{\rm gh} (0|u, v) = S_{\rm hg} (v, u) F_{\rm hg} (0|v, u)
\, .
\end{align}
Sequential application of these equations yields an identity making use of the unitarity \re{SUnitarity}. These two equations are
not sufficient to solve for the form factors. One has to use extra constraints that arise from the mirror transformations. Namely, moving 
either of the excitations in a complete cycle around the pentagon contour yields the other form factor, see the left panel in Fig.\ \ref{HoleGluonMirror}. 
Namely,
\begin{align}
\label{FFmirror} 
F_{\rm hg}(0|u^{-5 \gamma}, v) = F_{\rm gh} (0|v, u)
\, , \qquad
F_{\rm gh}(0|u, v^{5 \gamma}) = F_{\rm hg} (0|v, u)
\, .
\end{align}
These form factors obey the reflection and cyclicity identities
\begin{align}
F_{\rm hg} (0|u,v) = F_{\rm hg} (-u,-v|0) = F_{\rm gh} (v,u|0) = F_{\rm gh} (0|-v,-u)
\,.
\end{align}
The above equations can be solved to determine the mixed form factors. However, we will choose a different route to proceed. Namely, by shifting one of 
the excitations along the piece-wise contour, we will introduce mixed pentagon transitions
\begin{align}
P_{\rm hg} (u|v) =  \bra{{\rm g}(v)} \widehat{\mathcal{P}} \ket{{\rm h}(u)}
\, , \qquad
P_{\rm gh} (u|v) =  \bra{{\rm h}(v)} \widehat{\mathcal{P}} \ket{{\rm g}(u)}
\, ,
\end{align}
as well as analogous definitions for $P_{\rm h\bar{g}}$ and $P_{\rm \bar{g}h}$. This can be done in a variety of ways
\begin{align}
F_{\rm hg} (0|u^{- 2 \gamma}, v) = P_{\rm hg} (u|v)
\, , \qquad
&
F_{\rm hg} (0|u^{- 3 \gamma}, v) = P_{\rm gh} (v|u)
\, , \\
F_{\rm hg} (0|u, v^{3 \gamma}) = P_{\rm \bar{g}h} (v|u)
\, , \qquad
&
F_{\rm hg} (0|u, v^{2 \gamma}) = P_{\rm hg} (u|v)
\, ,
\end{align} 
See the graphical representation of the first equation in the right panel of Fig.\ \ref{HoleGluonMirror}. Notice that for an odd number of mirror transformations of
the gluon, its helicity gets changed. We conjecture these pentagon transitions to obey the following equations, echoing the axioms of Ref.\ \cite{Basso:2013aha} 
for singlet transitions,
\begin{align}
\label{Axiom1}
P_{\rm hg} (u|v) = S_{\rm hg} (u, v) P_{\rm gh} (v|u)
\, , \qquad
P_{\rm gh} (u|v) = S_{\rm gh} (u, v) P_{\rm hg} (v|u)
\, ,
\end{align}
and mirror identities
\begin{align}
\label{Axiom2}
P_{\rm hg} (u^{-\gamma}|v) &=P_{\rm gh} (v|u)
\, , \qquad
P_{\rm gh} (u^{-\gamma}|v) =P_{\rm h\bar{g}} (v|u)
\, , \\
P_{\rm hg} (u| v^{\gamma}) &=P_{\rm \bar{g}h} (v|u)
\, , \qquad
P_{\rm gh} (u|v^{\gamma}) =P_{\rm hg} (v|u)
\, , \nonumber
\end{align}
as well as the reflection equation
\begin{align}
\label{ReflectionID}
P_{\rm hg} (u|v) = P_{\rm \bar{g}h} (-v|-u)
\, . 
\end{align}
One can easily check the consistency of Eqs.\ \re{Axiom1} with the Watson equations \re{WatsonEqs}, though the former do not follow from the latter. 
Namely, starting from the ratio of form factors, we find
\begin{align}
S_{\rm gh} (v, u) 
&= \frac{F_{\rm hg} (0|u, v)}{F_{\rm gh}(0| v,u)} = \frac{P_{\rm hg} (u^{2 \gamma}| v)}{P_{\rm hg} (u^{- 3 \gamma}|v)}
\nonumber\\
&= \frac{S_{\rm hg} (u^{2\gamma}, v) S_{\rm hg} (u^\gamma, v)}{S_{\rm gh} (v, u^{-2 \gamma}) S_{\rm gh} (v, u^{- \gamma}) S_{\rm gh} (v, u)}
\, , 
\end{align}
which can be verified as a consequence of the identity
\begin{align}
S_{\rm hg} (u^{2\gamma}, v) S_{\rm hg} (u^\gamma, v) S_{\rm hg} (u^{-2 \gamma}, v) S_{\rm hg} (u^{- \gamma}, v) S_{\rm hg} (u, v) = S_{\rm gh} (v, u)
\, .
\end{align}

The following two solutions are consistent with Eqs.\ \re{Axiom1}
\begin{align}
\label{ExplicitPentagons}
P_{\rm hg}^2 (u| v) = w_{\rm hg} (u, v) \frac{S_{\rm hg} (u, v)}{S_{\ast{\rm hg}} (u, v)}
\, , \qquad
P_{\rm gh}^2 (u| v) = w_{\rm hg} (v, u) \frac{S_{\rm gh} (u, v)}{S_{\ast{\rm gh}} (u, v)}
\, ,
\end{align}
with an overall function $w_{\rm hg}(u,v)$. Similar expression holds for the negative helicity gluon which differs from the above by the overall function 
$w_{\rm h\bar{g}}(u,v)$. Here the mirror S-matrix $S_{\ast{\rm hg}}$ was computed in Eq.\ \re{ShgInuh} while $S_{\ast{\rm gh}} (u, v) = S_{{\rm gh}} 
(u^\gamma, v) \equiv S_{\ast{\rm hg}} (v, u)$.  Equations \re{Axiom2} imply that $w_{\rm hg}(u,v)$ and $w_{\rm h\bar{g}}(u,v)$ are mirror invariant in
hole rapidity $w_{\rm hg/\bar{g}} (u,v) = w_{\rm hg/\bar{g}} (u^\gamma, v)$ but related to each other by the mirror transformation in gluon rapidity
$w_{\rm hg} (u,v) = w_{\rm h\bar{g}} (u, v^\gamma)$. Next, using the obvious properties of the S-matrix $S_{\rm hg} (-u, -v) = [S_{\rm hg} (u, v)]^{-1}$ 
and $S_{\rm hg} (-u^\gamma, -v) = S_{\rm hg} (u^\gamma, v)$ (and the same for $S_{\rm gh}$), we find from \re{ReflectionID} that $w$ is an even 
function of its arguments $w (-u, -v) = w (u, v)$. For scalars the mirror is basically a shift $u^\gamma = u + i$, while for gluons, one returns to the 
``same'' point but passing through two cuts such that $x[v^\gamma] = g^2/x[v]$.  We will confirm by an explicit comparison with multiloop perturbative 
data that the simplest solution $w_{\rm hg} (u, v) = 1$ will lead agreement with OPE predictions. This implies that $P_{\rm hg} (u|v) = P_{\rm h\bar{g}} (u|v)$.

Finally making use of the explicit solution \re{ExplicitPentagons} and the properties of the S-matrices, we can relate form factors directly to the pentagon transitions
\begin{align}
F_{\rm hg} (0|u,v) 
&= P_{\rm hg} (u^{2 \gamma}|v) = \frac{1}{P_{\rm hg} (u|v)}
\, , \\
F_{\rm gh} (0|u,v) 
&= P_{\rm \bar{g}h} (u^{2 \gamma}|v) = \frac{1}{P_{\rm gh} (u|v)}
\, .
\end{align}
These obviously solve Eqs.\ \re{WatsonEqs} and \re{FFmirror} making use of the properties of the scattering matrices \re{Sprop1}, \re{Sprop2} and \re{Sprop3}.

\subsection{Fermion-fermion pentagons}

Now we have to find the form factor for same-helicity fermion in the antisymmetric representation, that is ${\bf 6}$. To accomplish this goal, we will start with the
fermion-fermion scattering with open indices transforming with respect to the ${\bf 4}$ of SU(4)
\begin{align}
\left[ S_{\Psi\Psi} (u, v) \right]^{AB}_{CD} = S_{\Psi\Psi} (u, v) \left[ \frac{u - v}{u - v - i} \delta^A_C \delta^B_D - \frac{i}{u - v - i} \delta^A_D \delta^B_C \right]
\, ,
\end{align}
where $\Psi$ cumulatively denotes large ($\Psi = {\rm F}$) and small ($\Psi = {\rm f}$) fermions. The tensor structure in the above equation is determined by the 
fundamental $R$-matrices  \cite{Berg:1977dp} and the overall phase specific to the flux-tube excitations $S_{\Psi\Psi} (u, v)$. The product of two ${\bf 4}$'s gets 
decomposed into symmetric and antisymmetric representations, ${\bf 4} \otimes {\bf 4} = {\bf 6} + {\bf 10}$. These can be projected out with the tensors
\begin{align}
[\Pi^{{\bf 6}}]^{AB}_{CD} = \frac{1}{2} \left(  \delta^A_C \delta^B_D - \delta^A_D \delta^B_C \right)
\, , \qquad
[\Pi^{{\bf 10}}]^{AB}_{CD} = \frac{1}{2} \left(  \delta^A_C \delta^B_D + \delta^A_D \delta^B_C \right)
\, .
\end{align}
The scattering matrix for the two fermions in the ${\bf 6}$ is
\begin{align}
S^{{\bf 6}}_{\Psi\Psi} (u, v)
=
[\Pi^{{\bf 6}}]^{AB}_{CD} \left[ S_{\Psi\Psi} (u, v) \right]_{AB}^{CD}
=
\frac{u - v + i}{u - v - i} S_{\Psi\Psi} (u, v)
\, .
\end{align}
Our ansatz for the two same-helicity fermion form factor in the ${\bf 6}$ of SU(4) reads
\begin{align}
F^{{\bf 6}}_{\Psi\Psi} (0| u, v) = \frac{i}{u - v + i} \frac{1}{P_{\Psi\Psi} (u|v)}
\, ,
\end{align}
which inherits the rational prefactor from the above scattering matrix. The pentagon transition $P_{\Psi\Psi} (u|v)$ defining the above form factor was studied in 
Ref.\ \cite{Basso:2014koa} and conjectured to be
\begin{align}
P^2_{\Psi\Psi} (u|v) = \frac{f_{\Psi\Psi} (u, v)}{(u - v) (u - v + i)} \frac{S_{\Psi\Psi} (u, v)}{S_{\ast\Psi\Psi} (u, v)}
\, ,
\end{align}
where
\begin{align}
f_{\Psi\Psi} (u, v) = \frac{x[u] x[v]}{g^2} \left( 1 - \frac{g^2}{x[u] x[v]} \right)
\, .
\end{align}

Since the index $\Psi$ cumulatively stands for both large and small fermions depending on what sheet of the Riemann surface, shown in Fig.\ \ref{FermionAnalytic},
the corresponding rapidity belongs to, we will display below explicit formulas for all pentagons. Namely, when both rapidities are on the
large sheet, we find
\begin{align}
P_{\rm FF} (u|v) 
&
= \frac{i x[u] x[v]}{g^2 (u - v)} \left(1 - \frac{g^2}{x[u] x[v]} \right)^{1/2}
\\
&\times\exp
\left(
 i \sigma_{\rm FF} - \widehat{\sigma}_{\rm FF}
- i f_{\rm FF}^{(1)} (u, v)
+ i f_{\rm FF}^{(2)} (u, v)
- f_{\rm FF}^{(3)} (u, v)
+ f_{\rm FF}^{(4)} (u, v)
\right)
\, . \nonumber
\end{align}
while when one of them, say $u$, is on the small sheet, we find instead
\begin{align}
\label{fPpentagon}
P_{\rm fF} (u|v) = - \frac{i}{x [u]}\left(1 - \frac{g^2}{x[u] x[v]} \right)^{-1/2}
\exp
\left(
- i f_{\rm fF}^{(1)} (u, v)
+ i f_{\rm fF}^{(2)} (u, v)
- f_{\rm fF}^{(3)} (u, v)
+ f_{\rm fF}^{(4)} (u, v)
\right)
\, .
\end{align}
Finally, when both $u$ and $v$ are on the small fermion sheet, we get
\begin{align}
P_{\rm ff} (u|v) = \frac{i}{u - v}\left(1 - \frac{g^2}{x[u] x[v]} \right)^{1/2}
\exp
\left(
- i f_{\rm ff}^{(1)} (u, v)
+ i f_{\rm ff}^{(2)} (u, v)
- f_{\rm ff}^{(3)} (u, v)
+ f_{\rm ff}^{(4)} (u, v)
\right)
\, .
\end{align}
The first and last one are used to calculate the transition measure,
\begin{align}
\res_{v = u} P_{\rm FF} (u|v) = \frac{i}{\mu_{\rm F} (u)}
\, , \qquad
\res_{v = u} P_{\rm ff} (u|v) = \frac{i}{\mu_{\rm f} (u)}
\, .
\end{align}
They read
\begin{align}
\mu_{\rm F} (u) 
&
=
\frac{g^2}{x[u] \sqrt{x^2[u] - g^2}} \exp \left( \widehat{\sigma}_{\rm FF} (u, u) + f^{(3)}_{\rm FF} (u, u) - f^{(4)}_{\rm FF} (u, u) \right)
\, , \\ 
\mu_{\rm f} (u)
&=
- \frac{x[u]}{\sqrt{x^2[u] - g^2}} \exp \left( f^{(3)}_{\rm ff} (u, u) - f^{(4)}_{\rm ff} (u, u) \right)
\, ,
\end{align}
for large and small fermions, respectively. These results are in agreement with Ref.\ \cite{Basso:2014koa}.

\section{Twist-two contributions to NMHV hexagon}
\label{NMHVhexagon}

In this section, we will use nonsinglet pentagons defined in the previous section to calculate twist-two contributions to NMHV amplitudes.

\subsection{NMHV hexagon observable}

To start with, let us introduce a proper NMHV hexagon observable that will be matched against operator product expansion predictions. To this end, we 
construct the following combination
\begin{align}
\label{HexagonObservable}
\mathcal{W}_{6;1} = g^2 \mathcal{P}_6 W_6^{\rm U(1)} {\rm e}^{R_6}
\, , 
\end{align}
where a power of the 't Hooft coupling in the right-hand side of the equation is introduced in order to match the left-hand side to the definition of the 
superloop in Refs.\ \cite{{Mason:2010yk,CaronHuot:2010ek,Belitsky:2011zm}} such that tree NMHV amplitude corresponds to a one-loop calculation 
on the Wilson loop side etc. In the above expression, $W_6^{\rm U(1)}$ is a ratio of Wilson loops computed in Abelian theory \cite{Gaiotto:2011dt}
\begin{align}
W_6 
= 
\frac{\vev{W_6} \vev{W_{\rm square}}}{\vev{W_{\rm top \ pent}} \vev{W_{\rm bot \ pent}}}
=
\exp\bigg( g^2_{\rm U(1)} X_6 (u, v, w) \bigg)
\, ,
\end{align}
with U(1) coupling being replaced by the all-order cusp anomalous dimension $g^2_{\rm U(1)} = \ft14 \Gamma_{\rm cusp} (g) \simeq g^2 \left( 1 - 
2 \zeta_2 g^2 + \dots \right)$ and the function $X_6$ of conformal cross ratios reads \cite{Gaiotto:2011dt}
\begin{align}
X_6 (u, v, w) = - {\rm Li}_2 (1-u) - {\rm Li}_2 (1-v) - {\rm Li}_2 (1-w) - \ln u \ln w + \ln(1-v) \ln \frac{(1-v)u}{v w}+ 2 \zeta_2
\, .
\end{align}
Here we introduced three conformal cross ratios
\begin{align}
u = \frac{(1234)(4561)}{(1245)(3461)}
\, , \qquad
v = \frac{(2345)(5612)}{(2356)(4512)}
\, , \qquad
w = \frac{(3456)(6123)}{(3461)(5623)}
\, ,
\end{align}
determined by means of four-brackets $(ijkl) = \varepsilon_{ABCD} Z_i^A Z_j^B Z_k^C Z_l^D$ of momentum twistors $Z_i^A$. 

Next, $R_6$ is the hexagon remainder function that starts at two-loop order \cite{DelDuca:2010zg,Goncharov:2010jf} and currently available all the way to four loops 
\cite{Dixon:2013eka,Dixon:2014voa}. 

The main NMHV ingredient in the formula \re{HexagonObservable}
is the NMHV ratio function $\mathcal{P}_6$ given by the quotient of the NMHV component of the superamplitude and its lowest bosonic component
\begin{align}
\mathcal{P}_6 = \mathcal{A}_{6;1}/\mathcal{A}_{6;0}
\, ,
\end{align}
that takes a well-known form \cite{Drummond:2008vq,Dixon:2011nj}
\begin{align}
\mathcal{P}_6^{\rm NMHV} 
&
= \ft12 [(2)+(5)] V (u, v, w) + \ft12 [(3)+(6)] V (v, w, u) + \ft12 [(1) + (4)] V (w, u, v)
\\
&
-
\ft12 [(2)-(5)] \widetilde{V} (u, v, w) + \ft12 [(3)-(6)] \widetilde{V} (v, w, u) + \ft12 [(1) - (4)] \widetilde{V} (w, u, v)
\, , \nonumber
\end{align}
with the $R$-invariants being \cite{Drummond:2008vq,Mason:2009qx}
\begin{align}
(1) \equiv [23456] = \frac{\delta^{0|4} \left(\chi_2 (3456) + \chi_3 (4562) + \chi_4 (5623) + \chi_5 (6234) + \chi_6 (2345) \right)}{(2345)(3456)(4562)(5623)(6234)}
\, ,
\end{align}
and the rest of them obtained by cyclic permutations. Here the functions $V$ and $\widetilde{V}$ admit perturbative expansion, with the even $V$ being nontrivial already at 
tree level, while the odd $\widetilde{V}$ starts only from two loops
\begin{align}
V = 1 + \sum_{\ell \geq 1} g^{2\ell} V^{(\ell)} 
\, , \qquad
\widetilde{V} = \sum_{\ell \geq 2} g^{2\ell} \widetilde{V}^{(\ell)}
\, .
\end{align}
We quote here only the one-loop function
\begin{align}
\label{V1uvw}
V^{(1)} (u,v,w) = {\rm Li}_2 (1-u) + {\rm Li}_2 (1-v) + {\rm Li}_2 (1-w) - \ln u \ln w + \ln (u w) \ln v - 2 \zeta_2
\, ,
\end{align}
to define our normalization, while the  $\ell = 2$ case can be found in Ref.\ \cite{Dixon:2011nj} and the three- and four-loop cases were recently worked out 
in Refs.\ \cite{Dixon:2014iba} and \cite{DixVonHip15}, respectively, and we refer the reader to these papers for explicit expressions.

To make contact with the operator product expansion, we use the hexagon parametrized by the twistors
\begin{align*}
Z_1 
&= ({\rm e}^{\sigma - i \phi/2}, 0, {\rm e}^{\tau + i \phi/2}, {\rm e}^{-\tau+ i \phi/2})
\, , \quad
Z_2 
= (1, 0, 0, 0)
\, , \quad
Z_3 
= (-1, 0, 0, 1)
\, , \\
Z_4 
&= (0, 1, -1, 1)
\, , \qquad\qquad\qquad\qquad
Z_5 
= (0, 1, 0, 0)
\, , \quad
Z_6 
= (0, {\rm e}^{- \sigma - i \phi/2}, {\rm e}^{\tau + i \phi/2}, 0)
\, , 
\end{align*}
with assignments of labels as shown in Fig.\ \ref{HexagonTesselation}. The conformal cross ratios read in this case
\begin{align}
u &= {\rm e}^{2 \tau - 2 \sigma} v w
\, , \\
v &= 1/(1 + {\rm e}^{2 \tau}) 
\, , \\
w &= 1/(1 + {\rm e}^{2 \sigma} + 2 \cos(\phi) {\rm e}^{\sigma - \tau} + {\rm e}^{- 2 \tau})
\, .
\end{align}
Notice that $u$, $w$ do not interchange under $\sigma \to -\sigma$, but rather under $\sigma \to \sigma^\prime \equiv - \sigma + \ln (1 + {\rm e}^{-2 \tau})$, 
i.e., $u (\sigma^\prime) = w (\sigma)$ and $w (\sigma^\prime) = u (\sigma)$. 

Since we are after contributions with the quantum numbers of the scalar-gluon pair, the most natural choice for the NMHV component\footnote{The fastest 
access to any component of one-loop non-MHV amplitudes can be achieved by using the package of Ref.\ \cite{Bourjaily:2013mma} instead.} in question 
would be the $\chi_1 \chi_3 \chi_4 \chi_6$ one that was used in Ref.\ \cite{Basso:2013aha} to constrain the twist-one contribution from scalars. However, 
as we will comment on it shortly, this component will not be good from the point of view of pentagon expansion. Instead, we consider $\chi_1^2 \chi_4^2$. 
In this case, the nontrivial effect arises from the following $R$-invariants, expanded to order ${\rm e}^{-2\tau}$, 
\begin{align}
\label{RinvExpansion}
(2) + (5) 
&
= 
-
\frac{\chi_1^2 \chi_4^2}{\cosh(\sigma)}
\left[
{\rm e}^{- \tau} - \frac{(1 + \sinh(2 \sigma) )\cos(\phi)}{\cosh (\sigma)}  {\rm e}^{- 2 \tau}  + \dots
\right]
+ \dots
\, ,\\
(3)+(6)
&
=
- 2 \cos(\phi) {\rm e}^{-2\tau}  \chi_1^2 \chi_4^2
+ \dots
\, , \qquad
(2) - (5) 
=
(3) - (6) 
=
- 2 i \sin (\phi) {\rm e}^{-2 \tau} \chi_1^2 \chi_4^2 + \dots
\, , \nonumber
\end{align}
with the rest inducing no contribution to $\chi_1^2 \chi_4^2$. The collinear expansion for $V^{(1)}$ in Eq.\ \re{V1uvw} reads for three permutation of its arguments
\begin{align*}
V^{(1)} (u, v, w) 
&= - 4 \tau \left[ \sigma - \ln (1 + {\rm e}^{2 \sigma}) \right] + 4 \sigma  \ln (1 + {\rm e}^{2 \sigma}) - 2  \ln^2 (1 + {\rm e}^{2 \sigma}) 
\nonumber\\
&\qquad\qquad\! + 4 {\rm e}^{- \tau} \frac{\cos(\phi)}{\cosh(\sigma)} 
\left[
\tau 
-
\sigma {\rm e}^{\sigma}  \sinh(\sigma) + \sinh^2 (\sigma)  \ln (1 + {\rm e}^{2 \sigma}) 
\right]
+ \dots
\, , \\
V^{(1)} (v, w, u) 
&= + 4 \sigma \tau - 4 {\rm e}^{- \tau} \cos(\phi) \left[ \sigma {\rm e}^{\sigma} - \cosh(\sigma) \ln(1 + {\rm e}^{2 \sigma}) \right]
+ \dots
\, , \\
V^{(1)} (w, u, v) 
&= - 4 \sigma \tau - 4 {\rm e}^{- \tau} \cos(\varphi) \left[ \sigma {\rm e}^{\sigma} - \cosh(\sigma) \ln(1 + {\rm e}^{2 \sigma}) \right]
+ \dots
\, ,
\end{align*}
while for $X_6$ it is
\begin{align}
X_6 (u, v, w) = 4 {\rm e}^{- \tau} \cos (\phi)  \left[ \sigma {\rm e}^{\sigma} - \cosh(\sigma) \ln(1 + {\rm e}^{2 \sigma}) \right] + \dots
\, .
\end{align}
Assembling everything together, we find for twist-two contribution\footnote{Further terms in the perturbative expansion are a way too cumbersome to
be presented here. They are summarized in the accompanying Mathematica notebook.} to \re{HexagonObservable},
\begin{align}
\label{R6componentExp2tau}
\mathcal{W}_{6;1} 
&
= \chi_1^2 \chi_4^2 {\rm e}^{- 2 \tau}  \frac{\cos(\phi)}{2 \cosh^2(\sigma)}
\Big\{ 
- 
{\rm e}^{- 2 \sigma}
+ g^2 
\Big( 
- 4 \tau \left[ 1 + \sigma (2 + {\rm e}^{2 \sigma}) - (1 + \sinh(2\sigma)) \ln (1 + {\rm e}^{2 \sigma}) \right]
\nonumber\\
&
- 4 \sigma + 4 (1 + \sigma + \sigma \sinh (2 \sigma)) \ln (1 + {\rm e}^{2 \sigma}) - 2 (1 + \sinh (2 \sigma)) \ln^2 (1 + {\rm e}^{2 \sigma})
\Big)
+ 
O(g^4)
\Big\}
\, . 
\end{align}
This expansion will be matched below against results based on integrability to four-loop order in 't Hooft coupling. 

\subsection{Contributions from OPE}

The Grassmann degree-four $\chi_1^2 \chi_4^2$-contribution $\mathcal{W}_{6;1}$ to the hexagonal super Wilson loop with ${\bf 6}$ of SU(4) and unit charge with 
respect to U(1) in the operator product expansion channel takes the form
\begin{align}
\mathcal{W}_{6;1} = 2 \cos(\phi) {\rm e}^{- 2 \tau} \chi_1^2 \chi_4^2 \left( W_{\Psi\Psi} + W_{\rm hg} \right) + \dots
\, ,
\end{align}
where the contributions from the above two-particle states admit the following generic structure
\begin{align}
\label{GenericOPEW}
W_{{\rm pp}^\prime} = \int d \mu_{\rm p} (u) \int d \mu_{{\rm p}^\prime} (v) C_{{\rm pp}^\prime} (u, v) F_{{\rm pp}^\prime} (0|u, v) F_{{\rm p}^\prime {\rm p}} (- v, -u| 0)
\end{align}
where we included potential additional coupling form factors $C (u, v)$ not accounted by the pentagon transitions. Here in the integration measure
\begin{align}
d \mu_{\rm p} (u)
\equiv
\frac{du}{2 \pi} \mu_{\rm p} (u) {\rm e}^{- \tau \gamma_{\rm p} (u) + i \sigma p_{\rm p} (u)} 
\, ,
\end{align}
we got absorbed particle propagation ``phases'', with $ \gamma_{\rm p} (u) = E_{\rm p} (u) - 1$ being the anomalous contributions to the energy of the 
flux-tube excitation.

As we pointed out in the previous subsection, there is a more natural candidate for operator product expansion analysis, i.e., $\chi_1 \chi_3 \chi_4 \chi_6$ Grassmann 
component. A quick lowest order inspection suggests that \re{GenericOPEW} could potentially accommodate this one as well. Namely, with proper introduction of ad
hoc form factors $C_{{\rm pp}^\prime}$, the tree and one-loop NMHV amplitude can be reproduced for $C_{\Psi\Psi} = (x [u] x [v]/g^2)^2$ and
$C_{\rm hg} (u,v) = x^+[u] x^-[v]/g^2$. However, these fail to produce the correct near-collinear limit at two loops and higher, making the corresponding component
of the amplitude inappropriate for naive OPE. The question remains what determines whether a given Grassmann component is good or bad.

Before, we turn to the calculation of individual contributions, let us point out that the easiest way to infer the weak coupling expansion of the S-matrices and  
find corresponding pentagon transitions is to use the exchange relations. Since
\begin{align*}
\gamma_n \sim \widetilde\gamma_n \sim \kappa_n \sim \widetilde\kappa_n \sim O (g^n)
\, ,
\end{align*}
the perturbative solution of the flux-tube equation becomes iterative and elementary. Then everything boils down to the use of the following basic integral 
\begin{align*}
\int_0^\infty \frac{dt}{t} \frac{{\rm e}^{i w t} - 1 - i w t}{{\rm e}^t - 1} = \ln \Gamma (1 - i w) + i w \psi (1)
\end{align*}
and its derivatives with respect to $w$.

\subsection{Two-fermion states}

It appears counterintuitive from the point of view of perturbative theory for the super Wilson loop that two-fermion intermediate states in the operator
product expansion play a leading role in the weak coupling series compared to the two-particle gauge-hole flux tube excitations. Therefore, we 
discuss the fermonic case first. In addition to pentagon form factors, to properly encode the near-collinear limit of the NMHV amplitude, we have to
introduce ad hoc NMHV form factors for fermions, a power of the Zhukowski variable for each of them, i.e., $C_{\Psi\Psi} (u, v) = x [u] x[v]/g^2$.
Then
\begin{align}
\label{W2fermion}
W_{\Psi\Psi} = \int_{C} d \mu_\Psi (u) \int_{C} d \mu_\Psi (v) \frac{x[u] x[v]}{g^2} F^{{\bf 6}}_{\Psi\Psi} (0|u, v) F^{{\bf 6}}_{\Psi\Psi} (- v, -u|0) 
\, ,
\end{align}
where the integration contour $C = C_{\rm F} \cup C_{\rm f}$ runs over both fermion sheets as shown in Fig.\ \ref{FermionAnalytic}, 
see Ref.\ \cite{Basso:2014koa}. As we can see, the integration over the path on the Riemann surface can be split into three distinct contributions depending 
on the position of the integration variable on the large or small fermion sheet, such that $W_{\Psi\Psi} = W_{\rm FF} + W_{\rm fF} + W_{\rm ff}$. The intermediate 
contribution $W_{\rm fF}$ comes from the integration with respect to the rapidity $u$ over the closed infinite half-circle in the lower half-plane and receives a 
nonvanishing effect only thanks to the existence of a pole $u = v - i$ in $F^{{\bf 6}}_{\Psi\Psi}$. Due to the regularity of the resulting integrand in $v$ belonging to 
the small Riemann sheet, $W_{\rm ff} = 0$. Now, let us analyze these contributions in turn.

\subsubsection{Small-large fermion state}

According to the analysis of Ref.\ \cite{Alday:2007mf}, the small fermion acts as a supersymmetric generator such that when it acts on the large fermion
it produces a scalar accompanied by a covariant derivative, $\{ \bar{\mathcal{Q}}_{\dot\beta}^B, \psi_\alpha^A \}= i \sqrt{2} \mathcal{D}_{\alpha\dot\beta} \phi^{AB}$.
The contribution of the small-large fermion pair follows straightforwardly from Eq.\ \re{W2fermion} and reads
\begin{align}
W_{\rm fF} = \int_{\mathbb{R} + i 0_+} d \mu_{\rm fF} (v)
\, .
\end{align}
It takes the form of a single composite particle excitation
\begin{align}
d \mu_{\rm fF} (v) = \frac{du}{2 \pi} \, \mu_{\rm fF} (v) {\rm e}^{-\tau \gamma_{\rm fF} (v) + i \sigma p_{\rm fF} (v)}
\, ,
\end{align}
with the measure $\mu_{\rm fF}$ being
\begin{align}
\label{mufF}
\mu_{\rm fF} (v) = \frac{x[v]}{x[v - i]} \frac{\mu_{\rm F} (v) \mu_{\rm f} (v - i)}{P_{\rm fF} (v - i|v) P_{\rm fF} (- v + i|- v )}
\, .
\end{align}
Here we absorbed the ad hoc NMHV form factor into the measure. The anomalous energy and momentum of the resulting composite 
two-fermion excitation are
\begin{align}
\gamma_{\rm fF} (v) 
&= E_{\rm F} (v) + E_{\rm f} (v - i) - 2
= 2 g^2 \left( \psi (1 - i v) + \psi (1 + i v) - 2 \psi(1) \right) + O (g^4)
\, , \\
p_{\rm fF} (v) 
&= p_{\rm F} (v) + p_{\rm f} (v - i)
= 2v + g^2 \left( - 2 \pi \coth (\pi v) + \frac{2}{v - i} \right) + O (g^4)
\, ,
\end{align}
where we displayed the first two two terms in their perturbative expansion. The small-large fermion pentagon can be be computed to any order of 
perturbation theory from the results of the previous section and reads to $O (g^2)$
\begin{align}
P_{\rm fF} (u|v) = \frac{1}{x[u]} \left( 1 + \frac{i g^2}{u} H_{- 1 -i v} + O (g^4) \right)
\, ,
\end{align}
while the large and small fermion measures read \cite{Basso:2014koa}
\begin{align}
\label{LargeFmeasure}
\mu_{\rm F} (u)
&
=
\frac{g^2 \pi}{u \sinh(\pi u)}
\left[
1 + g^2 \left( 2 \zeta_2 + \frac{1}{u^2} + \frac{\pi}{u} \coth (\pi u) + \frac{\pi^2}{\sinh^2 (\pi u)} - H_{i u}^2 - H_{-iu}^2 \right)
\right] + O (g^6)
\, , \\
\mu_{\rm f} (v)
&
=
- 1 - \frac{g^2}{v^2} + O(g^4)
\end{align}
where $H_u = \psi (1+u) + \gamma_{\rm E}$ is the harmonic number. Combining these results together, the measure $\mu_{\rm fF} (v)$ of the 
composite excitation can be found to be
\begin{align}
\mu_{\rm fF} (v) 
&
= \frac{\pi (v - i)}{\sinh(\pi v)}
\\
&
\times\left[
g^2
+
g^4 
\left(
2 \zeta_2  - \frac{1 + i \pi \coth (\pi v)}{v (v - i)} + \frac{\pi^2}{\sinh^2 (\pi v)} - H_{i v}^2 - H_{- iv}^2
\right)
+ O(g^4)
\right]
\, . \nonumber
\end{align}
The weak coupling expansion of $W_{\rm fF} $ starts at order $g^2$, i.e., as we will see two orders earlier than the large-fermion case discussed 
in the next section,
\begin{align}
W_{\rm fF} = g^2 W_{\rm fF}^{(0)} + g^4 \left( - \tau W_{{\rm fF}; 1}^{(1)} + W_{{\rm fF}; 0}^{(1)} \right)+ O(g^6)
\, ,  
\end{align}
with
\begin{align}
W_{\rm fF}^{(0)} = \int_{\mathbb{R} + i 0_+} \frac{dv}{2 \pi} {\rm e}^{2 i v \sigma} \frac{\pi (v - i)}{\sinh (\pi v)}
=
- \frac{{\rm e}^{-2 \sigma}}{4 \cosh^2 (\sigma)}
\, ,
\end{align}
which induces, as we will see below, the twist-two ${\rm e}^{-2\tau}$ contribution to the tree-level NMHV amplitude. The two loop effects are split into the 
linear in $\tau$ term and the rest. The former arises from the anomalous energy of the composite fermion state and reads
\begin{align}
W_{{\rm fF}; 1}^{(1)} 
&
= \int_{\mathbb{R} + i 0_+} \frac{dv}{2 \pi} {\rm e}^{2 i v \sigma} \frac{2 \pi (v - i)}{\sinh (\pi v)}
\left[  \psi (1 - i v) + \psi (1 + i v) - 2 \psi(1) \right]
\\
&
=
\frac{1 + \sigma (2 + {\rm e}^{2 \sigma}) - (1 + \sinh(2 \sigma)) \ln(1 + {\rm e}^{2 \sigma}) }{\cosh^2 (\sigma)}
\, . \nonumber
\end{align}
The $\tau$-independent part comes from the corrections to the measure and momentum. The can be easily evaluated to be
\begin{align}
W_{{\rm fF}; 0}^{(1)} 
=
\frac{
- 2 \sigma + 2 (1 + \sigma + \sigma \sinh(2 \sigma)) \ln (1 + {\rm e}^{2 \sigma}) - (1 + \sinh (2\sigma)) \ln^2 (1 + {\rm e}^{2 \sigma})
}{2 \cosh^2(\sigma)}
\, .
\end{align}
As a side technical remark, let us point out that these analytical results can be easily obtained by calculating both rapidity integrals by closing them
in the lower half-plane and resuming resulting residues as was done in Ref.\ \cite{Papathanasiou:2013uoa,Hatsuda:2014oza}. Higher order contributions 
up to four loops are summarized in the attached notebook.

As we can see the small-large two-fermion state is solely responsible for inducing the tree-level and one-loop contributions to NMHV amplitude, i.e., 
both displayed term in Eq.\ \re{R6componentExp2tau}. This is counterintuitive from the point of view of perturbation theory where the Born amplitude 
comes from a scalar propagator coupled to the contour of the Wilson loop, but demonstrates that the flux-tube constituents receive a very nontrivial 
expansion in terms of elementary excitations of Yang-Mills perturbation theory. This implies that large-large and scalar-gluon effects are pushed to
two-loop order. Let us now to turn to their discussion next.

\subsubsection{Large-large fermion state}

The two large-fermion contribution is
\begin{align}
\label{W2fermionLarge}
W_{\rm FF} = 
\int_{\mathbb{R}+i 0_+} d \mu_{\rm F} (u)
\int_{\mathbb{R}+i 0_+} d \mu_{\rm F} (v)
\frac{x[u] x[v]}{g^2 [(u-v)^2+1] P_{\rm FF} (u|v) P_{\rm FF} (-u|-v)}
\, ,
\end{align}
where the large fermion measure was introduced in Eq.\ \re{LargeFmeasure}, while the leading contribution to the nonsinglet pentagon can be
read from Eq.\ \re{fPpentagon}. We immediately observe that the effect of large fermions starts at two-loop order
\begin{align}
W_{\rm FF} = g^6 W_{\rm FF}^{(2)} + O (g^8)
\end{align}
and reads
\begin{align}
W_{\rm FF}^{(2)}
&
=
\pi^3
\int_{\mathbb{R}+i 0_+} \frac{du}{2 \pi}
\int_{\mathbb{R}+i 0_+} \frac{dv}{2 \pi}
{\rm e}^{2i \sigma (u + v)}
\frac{(u - v)}{(u - v)^2 + 1} \frac{\coth (\pi v) - \coth(\pi u)}{u v \sinh(\pi u) \sinh (\pi v)}
\, .  
\end{align}

\subsubsection{Gauge-hole state}

To complete the result, we finally calculate the scalar-gluon contribution
\begin{align}
W_{\rm hg} 
= 
\int_{\mathbb{R}} d \mu_{\rm h} (u) \int_{\mathbb{R}} d \mu_{\rm g} (v)
F_{\rm hg} (0|u, v)  F_{\rm gh} (-v, -u|0)
\, .
\end{align}
Using perturbative expansion of the all-order expressions for the form factors derived in Section \ref{ScalarGluonFF} along with the
hole and gluon measures, we find that $W_{\rm hg}$ start at order $g^6$, like the large two-fermion contribution introduced in the
previous subsection,
\begin{align}
W_{\rm hg} = g^6 W_{\rm hg}^{(2)} + O (g^8)
\, .
\end{align}
We find for $W_{\rm hg}^{(2)}$,
\begin{align}
W_{\rm hg}^{(2)}
=
\pi^3 
\int_{\mathbb{R}} \frac{du}{2 \pi}
\int_{\mathbb{R}} \frac{dv}{2 \pi}
{\rm e}^{2 i (u + v) \sigma} \frac{\tanh (\pi u) - \tanh(\pi v)}{(u - v) (v^4 + \ft{1}{4}) \cosh (\pi u) \cosh (\pi v)} 
\, .
\end{align}

We computed the integrand for all contributions discussed to this point all the way to four loops in order to compare them with recent results of 
hexagon bootstrap approach \cite{Dixon:2014iba}. This consideration can be extended to any perturbative order. However, in the lack of tools to compute the
emerging integrals analytically, the resulting Fourier integrals were calculated numerically and a perfect agreement was observed\footnote{See the 
attached Mathematica notebook.} with perturbative data of Refs.\ \cite{Drummond:2008vq,Dixon:2014iba,DixVonHip15} for NMHV amplitudes to four-loop
order.

\section{Conclusions}

In this paper we analyzed twist-two contributions in the operator product expansion of the null polygonal super Wilson loop in a specific nonsinglet channel.
The latter is dual to certain Grassmann components of the NMHV amplitude. We derived nonperturbative expressions that govern the dependence
on the coupling constant and a kinematical invariant reciprocal to the momentum of the flux-tube excitation propagating on the Wilson loop world sheet.
The focus of the study was a channel with the quantum numbers of the elementary scalar that transforms in the ${\bf 6}$ of SU(4). As a consequence, there
were several two-particle states that contributed to the super Wilson loop: they are hole-gluon and two-fermion pairs. From the point of view of
conventional perturbation theory, both of them are visible at the level of the Feynman graphs. However, the  flux-tube excitations have little
to do with elementary fields of the $\mathcal{N} = 4$ SYM theory as it becomes obvious from the following observation. Namely, while the one-hole
flux-tube excitation induces the leading ${\rm e}^{- \tau}$ behavior of the NMHV component in question and it is realized in perturbation theory as
an exchange of the elementary scalar between the Wilson loop vertices, see, e.g., \cite{Basso:2013aha}. However, the latter contains all higher twist effects 
as well at leading order in 't Hooft coupling. We run then into an immediate predicament when we attempt to understand the origin of twist-two contribution to 
the tree NMHV amplitude. Where does it come from? A naive observation based on Yang-Mills perturbation theory indicates that two-particle intermediate states 
will be accompanied by at least four powers of $g$ from the coupling of excitations to the Wilson loop contour. Instead, it should be of order $O(g^2)$ since the 
tree NMHV amplitude is dual to one loop super Wilson loop. As we saw above, the resolution of this puzzle lies in the fact that the two-fermion contributions 
with one of them belonging to the small-rapidity sheet acts as a supersymmetric transformation on a large-fermion exchange rather than being a true 
propagating excitation. This decreases the perturbative order when this effect sets in from $O(g^6)$ to $O(g^2)$.

To find the main building blocks of the formalism, the nonsinglet pentagon form factors, we started from scratch by rederiving the theory of flux-tube excitations.
The initial point  of our consideration was the Baxter equation for magnons when their number tends to infinity and flux-tube excitation arise as insertions into 
this background. We hope that our approach provides a new perspective on the problem compared to the traditional technique that relies on the counting functions as a 
central object of corresponding analyses. Within our framework, we computed all two-particle S-matrices complementing and completing results available in the literature. 

Our study revealed a conundrum about the proper choice of Grassmann components of the super Wilson loop which are eagerly amenable to the operator product 
expansion analysis. As we pointed out in this paper, apparently perfect amplitudes turned out to be unsuitable for these purposes, i.e., $\chi_1 \chi_3 \chi_4 \chi_6$
versus $\chi_1^2 \chi_4^2$. The former was used in the past as a main observable in the study of twist-one hole excitations \cite{Basso:2013aha},
but we failed to find its successful description at two-particle level. On the other hand, the $\chi_1^2 \chi_4^2$ component is naturally described by the pentagon
expansion to all loop orders and we found a perfect agreement with available multiloop calculations of Refs.\ \cite{Drummond:2008vq,Dixon:2014iba}.

The refined form of the operator product expansion pursued in this paper provides a powerful nonperturbative framework to calculate all scattering 
amplitudes in maximally supersymmetric gauge theory. In order to achieve this goal, one has to compute all two-particle pentagon form factors
with other nonsinglet SU(4) quantum numbers as well as their multiparticle counterparts. The latter are known to factorize into the above single-particle 
pentagons \cite{Basso:2013vsa,Belitsky:2014rba}, however, unraveling the tensor index structure needs to be understood\footnote{See Ref.\  \cite{Basso:2014jfa}
with regards to recent progress in this direction.}. Finally, the strong coupling analysis of the operator product expansion that goes beyond the area law 
\cite{Alday:2007hr} awaits its exploration \cite{Basso:2014jfa}.

\section*{Acknowledgments}

A.B.\ would like to thank Jacob Bourjaily, Lance Dixon, Yasuyuki Hatsuda and especially Benjamin Basso for very enlightening discussions and instructive 
correspondence. This work was supported by the U.S. National Science Foundation under the grant PHY-1068286.

\appendix

\setcounter{section}{0}
\setcounter{equation}{0}
\renewcommand{\theequation}{\Alph{section}.\arabic{equation}}

\section{Integral form of flux-tube equations}
\label{OtherIntegralFTeqs}

In this appendix, we provide a compendium of integral flux-tube equations for all excitations. The ones for holes were already quoted in the main body of the paper in
Sect.\ \ref{IntegralFTeqHoles}. Their derivation is identical to the above.

\subsection{Large fermions}
\label{LargeFTappendix}

We start with large fermions. Making use of the sources \re{LFsource1}--\re{LFsource4}, we deduce from Eqs.\ \re{DefEq1}--\re{DefEqTilde2},
\begin{align}
\int_0^\infty \frac{dt}{t} \left( \cos(ut) - J_0 (2gt) \right) \gamma_{+, v}^{\rm F} (2gt)
=
&
 -
\int_0^\infty \frac{dt}{t} \frac{ \cos(ut) - J_0 (2gt) }{{\rm e}^t - 1}
\left[
\gamma_v^{\rm F} (- 2gt) + \cos (vt) - J_0 (2gt)
\right]
\nonumber\\
&
-  \frac{1}{2}
\int_0^\infty \frac{dt}{t} (\cos(ut) - J_0 (2gt)) \cos(vt)
\, , \\
\int_0^\infty \frac{dt}{t} \sin(ut)  \gamma_{-, v}^{\rm F} (2gt)
= 
&-
\int_0^\infty \frac{dt}{t} \frac{ \sin(ut) }{{\rm e}^t - 1}
\left[
\gamma_v^{\rm F} (2gt) + \cos (vt) - J_0 (2gt)
\right]
\, , 
\end{align}
for even and 
\begin{align}
&
\int_0^\infty \frac{dt}{t} \left( \cos(ut) - J_0 (2gt) \right) \widetilde\gamma_{+, v}^{\rm F} (2gt)
=
-
\int_0^\infty \frac{dt}{t} \frac{ \cos(ut) - J_0 (2gt) }{{\rm e}^t - 1}
\left[
\widetilde\gamma_v^{\rm F} (2gt) + \sin (vt)
\right]
\, , \\
&
\int_0^\infty \frac{dt}{t} \sin(ut) \widetilde\gamma_{-, v}^{\rm F} (2gt)
=
- 
\int_0^\infty \frac{dt}{t} \frac{\sin(ut)}{{\rm e}^t - 1}
\left[
- \widetilde\gamma_v^{\rm F} (- 2gt) + \sin (vt)
\right]
-
\frac{1}{2} \int_0^\infty \frac{dt}{t} \sin(ut) \sin(vt)
\, , 
\end{align}
odd $v$-parity functions, respectively. 

\subsection{Small fermions}
\label{SmallFTappendix}

Analogously for the small fermion sources \re{Skappa1} and \re{Skappa2}, we get
\begin{align}
\int_0^\infty \frac{dt}{t} ( \cos(ut)  - J_0 (2gt) ) \gamma^{\rm f}_{+, v} (2gt) 
&= 
-
\int_0^\infty \frac{dt}{t} \frac{\gamma^{\rm f}_v (- 2gt) }{{\rm e}^t - 1} (\cos(ut) - J_0 (2gt) )
\\
&
+
\frac{1}{2} \int_0^\infty \frac{dt}{t} (\cos (ut) - J_0 (2gt)) \cos(vt)
\, , \nonumber\\
\int_0^\infty \frac{dt}{t} \sin(ut) \gamma^{\rm f}_{-, v} (2gt) 
&= -
\int_0^\infty \frac{dt}{t} \frac{\gamma^{\rm f}_v (2gt) }{{\rm e}^t - 1} \sin(ut) 
\, , \\
\int_0^\infty \frac{dt}{t} ( \cos(ut) -J_0(2gt) )\widetilde\gamma^{\rm f}_{+, v} (2gt) 
&= -
\int_0^\infty \frac{dt}{t} \frac{\widetilde\gamma^{\rm f}_v (2gt) }{{\rm e}^t - 1} \left( \cos(ut) - J_0 (2gt) \right)
\, , \\
\int_0^\infty \frac{dt}{t} \sin(ut) \widetilde\gamma^{\rm f}_{-, v} (2gt) 
&= 
\int_0^\infty \frac{dt}{t} \frac{\widetilde\gamma^{\rm f}_v (- 2gt) }{{\rm e}^t - 1} \sin(ut)
+
\frac{1}{2} \int_0^\infty \frac{dt}{t} \sin (ut) \sin(vt)
\, .
\end{align}

\subsection{Gauge fields}
\label{GaugeFTappendix}

Finally using Eqs.\ \re{GluonKappa}, we find
\begin{align}
&
\int_0^\infty \frac{dt}{t} \left( \cos(ut) 
- J_0 (2gt) \right) \gamma^{\rm g}_{+, v} (2gt)
\\
&\qquad\qquad\qquad\quad
= -
\int_0^\infty \frac{dt}{t ({\rm e}^t - 1)} \left( \cos (ut) - J_0 (2gt) \right) 
\left[
\gamma^{\rm g}_v (- 2gt)
+
{\rm e}^{t/2} \cos (vt) - J_0 (2gt)
\right]
\, , \nonumber\\
&
\label{GFMinus}
\int_0^\infty \frac{dt}{t} \sin(ut) \gamma^{\rm g}_{-, v} (2gt)
\\
&\qquad\qquad\qquad\quad
= -
\int_0^\infty \frac{dt}{t ({\rm e}^t - 1)} \sin (ut)
\left[
\gamma^{\rm g}_v (2gt)
+
{\rm e}^{- t/2} \cos (vt) - J_0 (2gt)
\right]
\, . \nonumber
\end{align}
For odd parity functions we can obtain analogously from \re{GluonKappaTilde}
\begin{align}
\label{GFtildePlus}
&
\int_0^\infty \frac{dt}{t} \left( \cos(ut) 
- J_0 (2gt) \right) \widetilde\gamma^{\rm g}_{+, v} (2gt)
\\
&\qquad\qquad\qquad\quad
= -
\int_0^\infty \frac{dt}{t ({\rm e}^t - 1)} \left( \cos (ut) - J_0 (2gt) \right) 
\left[
\widetilde\gamma^{\rm g}_v (2gt)
+
{\rm e}^{- t/2} \sin (vt) 
\right]
\, , \nonumber\\
&
\label{GFtildeMinus}
\int_0^\infty \frac{dt}{t} \sin(ut) \widetilde\gamma^{\rm g}_{-, v} (2gt)
\\
&\qquad\qquad\qquad\quad
= -
\int_0^\infty \frac{dt}{t ({\rm e}^t - 1)} \sin (ut)
\left[
- \widetilde\gamma^{\rm g}_v (- 2gt)
+
{\rm e}^{t/2} \sin (vt)
\right]
\, . \nonumber
\end{align}
This concludes the list of  integral flux-tube equations used in this paper.

\section{From large to small fermion}
\label{LargeSmallAppendix}

\begin{figure}[t]
\begin{center}
\mbox{
\begin{picture}(0,250)(190,0)
\put(0,-170){\insertfig{22}{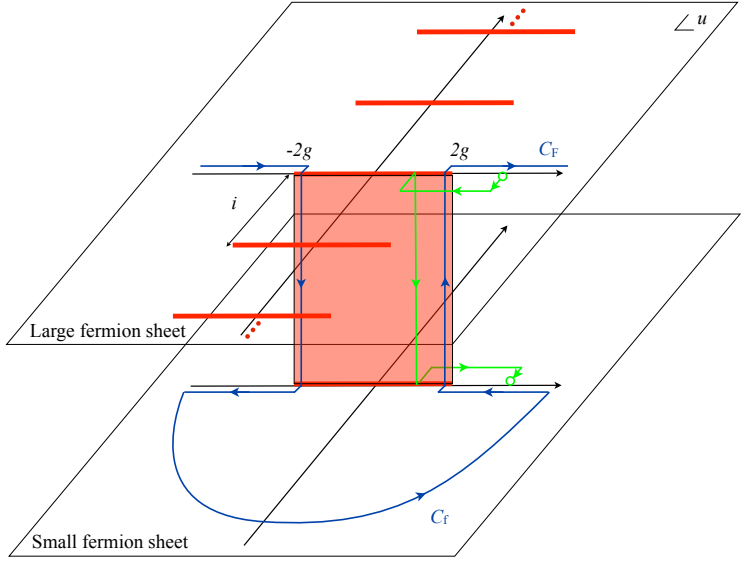}}
\end{picture}
}
\end{center}
\caption{ \label{FermionAnalytic} The path (in green) of analytic continuation from large to small fermion. In blue, we display the integration 
contour $C = C_{\rm F} \cup C_{\rm f}$ for integrals in Eq.\ \re{W2fermion}.}
\end{figure}

Since we use it in the main text, let us recall the path of analytic continuation from the large to the small fermion domain. We demonstrate it making
use of the dispersion relation, while in the main text we apply it to scattering matrices. Namely, the large-fermion energy and momentum read
as a function of the rapidity $u$
\begin{align}
\label{LargeFermEnergy}
E_{\rm F} (u) 
&= 1 + \int_0^\infty \frac{dt}{t} \frac{\gamma^{\o}_+ (2gt) - \gamma^{\o}_- (2gt)}{{\rm e}^t - 1} \left( \cos(ut) - J_0 (2gt) \right)
+
\frac{1}{2} \int_0^\infty \frac{dt}{t} \cos(ut) \gamma^{\o}_+ (2gt)
\, , \\
\label{LargeFermMomentum}
p_{\rm F} (u) 
&= 2u - \int_0^\infty \frac{dt}{t} \frac{\gamma^{\o}_+ (2gt) + \gamma^{\o}_- (2gt)}{{\rm e}^t - 1} \sin(ut) 
-
\frac{1}{2} \int_0^\infty \frac{dt}{t} \sin(ut) \gamma^{\o}_- (2gt)
\, ,
\end{align}
both valid for $|u|>2g$ as will be clarified below. Here the $\gamma^{\o}$ functions entering these expressions are the coefficient of the logarithmic 
dependence of the vacuum solution
\begin{align}
\gamma^{\O} = \gamma^{\o} \ln \bar\eta + \delta \gamma^{\o}
\, ,
\end{align}
in agreement with notations adopted in Ref.\ \cite{Basso:2010in} and the source \re{kappaO} decomposed accordingly, $\kappa^{\O} = \kappa^{\o} \ln \bar\eta 
+ \delta \kappa^{\o}$. 

To start with, let us first recall the analytic properties of the energy and momentum. It is sufficient to analyze just one terms to draw conclusions about the rest. Namely, 
using the generic series representation of, say, the first term in $E_{\rm lf} (u)$, we find
\begin{align}
 \int_0^\infty \frac{dt}{t} 
\frac{\gamma^{\o}_+ (2gt)}{{\rm e}^t - 1} {\rm e}^{i u t }
=
\sum_{n = 1}^\infty \lambda_+ \left(u + i n \right)
\, ,
\end{align}
where $\lambda_+ (u)$ has the following square-root cut  structure
\begin{align}
\lambda_+ (u) \equiv \int_0^\infty \frac{dt}{t} {\rm e}^{i u t} \gamma^{\o}_+ (2gt)
=
\alpha_+ (u) \sqrt{u^2 - (2g)^2} + \beta_+ (u)
\, ,
\end{align}
with analytic functions $\alpha_+ (u)$ and $\beta_+ (u)$ of $u$, according to the analysis in Ref.\ \cite{Kostov:2008ax}. Thus we observe that this particular term 
possess an infinite number of square root cuts $[-2g, 2g]$ starting from $\Im{\rm m}[u] = i$ and going all the way to infinity with $i$-interval along imaginary axis.
While the last term in \re{LargeFermEnergy} possesses a cut on the real axis. Thus, extending the same consideration to other terms, we conclude that both
energy and momentum have an infinite number of equidistant cuts in the rapidity plane as shown in Fig.\ \ref{FermionAnalytic}.

As was demonstrated in Ref.\ \cite{Basso:2010in}, to pass to the small fermion kinematics, one has to cross the cut $[-2g,2g]$ and pass to another sheet of the
Riemann surface, see Fig.\ \ref{FermionAnalytic}. As it is obvious from the integral representations \re{LargeFermEnergy} and \re{LargeFermMomentum}, we 
have to continue the last term only. We show it for energy. Namely, on the real axis for $|u|>2g$, we have
\begin{align}
\label{lambdaPlus}
\int_0^\infty \frac{dt}{t} \cos(ut) \gamma^{\o}_+ (2gt) = \int_0^\infty \frac{dt}{t} {\rm e}^{\pm i u t} \gamma^{\o}_+ (2gt)
\, , 
\end{align}
where we used the fact that $\lambda_+^\ast (u) = \lambda_+ (u^\ast)$ and choose the negative sign in the exponent that will define the function in the lower 
half plane of $u$. Next we move inside the strip $|\Im{\rm m} [u]| \leq 2g$ in order to cross the cut, i.e.,  $u \to u+i 0_+$. However, one cannot perform this step since the integral 
is not convergent due to growing exponent. To overcome this problem, we use Eq.\ \re{AnContInt} at first and then the vacuum flux-tube equations. The latter 
are found from generic Eq.\ \re{GKPequation1} with the logarithmically enhanced contribution $\kappa^{\o}$ to the source \re{kappaO}. With the help of the 
Jacobi-Anger transformation these can be cast into the form
\begin{align}
&
\label{VacuumGamma}
\int_0^\infty \frac{dt}{t} \left( \cos (ut) - J_0 (2gt) \right) \left[  \frac{\gamma^{\o}_+ (2gt)}{1 - {\rm e}^{-t}} - \frac{\gamma^{\o}_- (2gt)}{{\rm e}^{t} - 1} \right] = 0
\, , \\
&
\label{VacuumTildeGamma}
\int_0^\infty \frac{dt}{t}  \sin (ut) \left[  \frac{\gamma^{\o}_- (2gt)}{1 - {\rm e}^{-t}} + \frac{\gamma^{\o}_+ (2gt)}{{\rm e}^{t} - 1} \right] = 2 u
\, ,
\end{align}
valid for $|u|<2g$. Applying the above relations to the even/odd part of $\gamma^{\o}$, we find
\begin{align}
\label{VacuumGammaPlus}
\int_0^\infty \frac{dt}{t} \gamma^{\o}_+ (2gt) {\rm e}^{-iut}
&
=
- \int_0^\infty \frac{dt}{t} \gamma^{\o}_+ (2gt) {\rm e}^{iut}
-
2 \int_0^\infty \frac{dt}{t} \frac{\cos (ut) - J_0 (2gt)}{{\rm e}^t - 1} \gamma^{\o} (-2gt) 
\, ,\\
\label{VacuumGammaMinus} 
\int_0^\infty \frac{dt}{t} \gamma^{\o}_- (2gt) {\rm e}^{-iut}
&
=
\int_0^\infty \frac{dt}{t} \gamma^{\o}_- (2gt) {\rm e}^{iut}
-
4i u
+
2i \int_0^\infty \frac{dt}{t} \frac{\sin (ut)}{{\rm e}^t - 1} \gamma^{\o} (2gt) 
\, ,
\end{align}
where the first relation is relevant for energy, while the second one to the momentum. Using Eq. \re{VacuumGammaPlus} we find
\begin{align}
E_{\rm f} (u + i 0_+) = 1 - \frac{1}{2} \int_0^\infty \frac{dt}{t} {\rm e}^{i u t} \gamma^{\o}_+ (2gt)
\, .
\end{align}
Going outside the strip $|\Im{\rm m}[u]|\leq 2g$, we can then move back to the real axis by means of \re{lambdaPlus} and get the final expression 
for small fermion energy. Identical considerations using Eq.\ \re{VacuumGammaMinus} yield the small fermion momentum. These read
\begin{align}
E_{\rm f} (u) 
&= 1 - \frac{1}{2} \int_0^\infty \frac{dt}{t} \cos(ut) \gamma^{\o}_+ (2gt)
\, , \\
p_{\rm f} (u) 
&
= \frac{1}{2} \int_0^\infty \frac{dt}{t} \sin(ut) \gamma^{\o}_- (2gt)
\, .
\end{align}

\section{Mirror kinematics: holes}
\label{ScalarMirrorApp}

Here we will recapitulate the path of the analytic continuation to the mirror kinematics advocated in Ref.\  \cite{Basso:2011rc}.
Though a detailed account on this transformation is available in the literature \cite{Basso:2013pxa}, we nevertheless provide 
it here to make the current presentation self-contained. 

\begin{figure}[t]
\begin{center}
\mbox{
\begin{picture}(0,250)(190,0)
\put(0,-170){\insertfig{22}{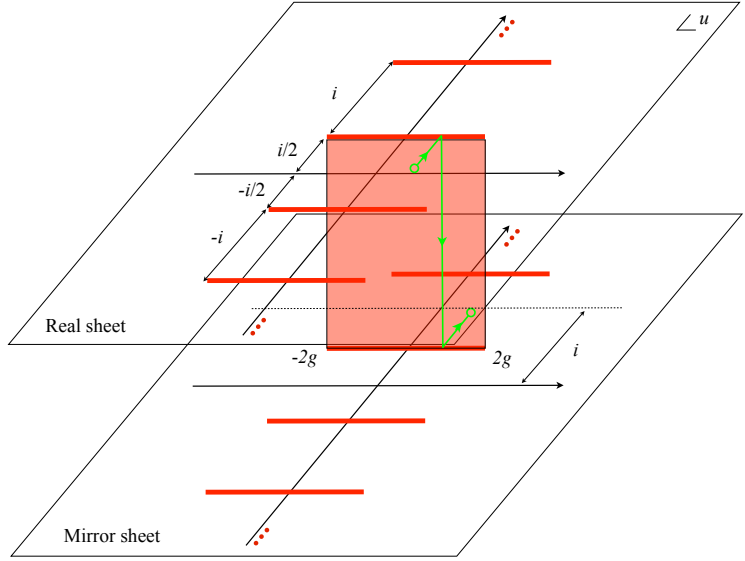}}
\end{picture}
}
\end{center}
\caption{ \label{ScalarMirror} The path of analytic continuation to the mirror kinematics for scalars.}
\end{figure}

The energy and momentum for the hole read
\begin{align}
E_{\rm h} (u)
&=
1 + \int_0^\infty \frac{dt}{t}  \frac{\gamma^{\o}_+ (2gt) - \gamma^{\o}_- (2gt)}{{\rm e}^t - 1} \left( {\rm e}^{t/2} \cos (ut) - J_0 (2gt) \right)
\, , \\
p_{\rm h} (u)
&=
2 u - \int_0^\infty \frac{dt}{t} \frac{\gamma^{\o}_+ (2gt) + \gamma^{\o}_- (2gt)}{{\rm e}^t - 1} {\rm e}^{t/2} \sin (ut) 
\, .
\end{align}
Studying the analytical structure of the dispersion relation for the hole in the complex $u$-plane along the same lines as in the previous Appendix, we conclude that the 
real sheet contains an infinite number of equidistant cuts as demonstrated in Fig.\ \ref{ScalarMirror}.

To reach the mirror kinematics by moving to another sheet of the Riemann surface of the energy/momentum as a function of complex rapidity $u$
as was shown in Ref.\ \cite{Basso:2011rc}, one first moves to $u \to u + i/2 - i 0_+$
\begin{align}
\label{ShiftedPhole}
p_{\rm h} (u^+ - i0_+)
= 2u + i 
- \int_0^\infty \frac{dt}{t} \frac{\gamma^{\o} (2gt)}{{\rm e}^t - 1} \sin (ut) 
- \frac{i}{2} \int_0^\infty \frac{dt}{t} {\rm e}^{-iut} \gamma^{\o} (2gt)
\, .
\end{align}
Then substituting the relations \re{VacuumGammaPlus} and \re{VacuumGammaMinus} into Eq.\ \re{ShiftedPhole}, we cross the cut to another Riemann sheet,
\begin{align}
p_{\rm h} (u^+ + i0_+)
= 
i 
+ i \int_0^\infty \frac{dt}{t} \frac{\gamma^{\o} (- 2gt)}{{\rm e}^t - 1} \left( \cos (ut) - J_0 (2gt) \right)
+ \frac{i}{2}\int_0^\infty \frac{dt}{t} {\rm e}^{iut} \gamma^{\o} (- 2gt)
\, .
\end{align}
Finally, we move $u \to u + i/2$, to get
\begin{align}
p_{\rm h} (u^+ + i0_+ + i/2) = i E_{\rm h} (u)
\, .
\end{align}
Identical considerations hold for the energy. Thus, under the above mirror transformations, the two interchange and acquire an $i$.

\section{Mirror kinematics: gauge field and bound states}
\label{GaugeMirrorApp}

\begin{figure}[t]
\begin{center}
\mbox{
\begin{picture}(0,250)(190,0)
\put(0,-170){\insertfig{22}{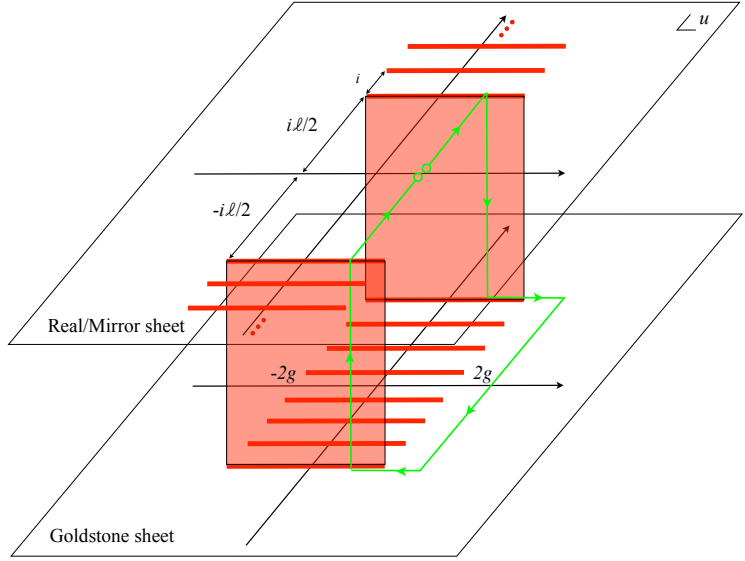}}
\end{picture}
}
\end{center}
\caption{ \label{GluonMirror} The path of analytic continuation to the mirror kinematics for gluons. For simplicity, the real and mirror sheets are shown to
coincide here. They do not.}
\end{figure}

In this appendix we review the analytic continuation to the mirror kinematics for the gauge field bound states \cite{Basso:2011rc}.
We will presently demonstrate it using dispersion relations as an example. We will use the same path to continue the
scattering matrix in gluon rapidity in the main text. The energy and momentum of gluon bound states is given by
\begin{align}
\label{gaugeBSenergy}
E_{\rm g} (u) 
&
= \ell + \int_0^\infty \frac{dt}{t} 
\left[
\frac{\gamma^{\o}_+ (2gt)}{1 - {\rm e}^{- t}} - \frac{\gamma^{\o}_- (2gt)}{{\rm e}^{t} - 1}
\right]
\left(
{\rm e}^{- \ell t/2} \cos (ut) - 1
\right)
\, , \\
\label{gaugeBSmomentum}
p_{\rm g} (u)
&
=
2 u
- 
\int_0^\infty \frac{dt}{t} 
\left[
\frac{\gamma^{\o}_- (2gt)}{1 - {\rm e}^{- t}} + \frac{\gamma^{\o}_+ (2gt)}{{\rm e}^{t} - 1}
\right]
{\rm e}^{- \ell t/2} \sin (ut)
\, .
\end{align}
Performing the consideration identical to the one for the scalars, we conclude that $E_{\rm g} (u)$ and $p_{\rm g} (u)$ in real kinematics possess an infinite 
number of square root cuts $[-2g, 2g]$ starting from $|u| = i \ell/2$ and going all the way to infinity with an $i$-interval along imaginary axis, as shown in 
Fig.\ \ref{GluonMirror}.

The path to the mirror kinematics was found in Ref.\ \cite{Basso:2011rc}. First, one has to cross the cut for $|u| \leq 2g$ in the upper half-plane
down to the Goldstone sheet. This is done in two steps. Namely, one shifts $u \to u + i (\ell/2 - 0_+)$ just below the cut,
\begin{align}
p_{\rm g} \left( u^{[\ell]} - i0_+ \right)
=
2 u + i \ell 
+ 
\frac{i}{2} \int_0^\infty \frac{dt}{t} \left[ \gamma^{\o}_- (2gt) + \frac{\gamma^{\o} (2gt)}{{\rm e}^t - 1} \right]
\left(
{\rm e}^{i u t - \ell t} - {\rm e}^{-iut}
\right)
\, .
\end{align}
We can clearly see that the current expression cannot be use to cross the cut, i.e.,  $u \to u + i 0_+$ since the cross-term $\gamma^{\o} {\rm e}^{-iut}$
becomes divergent. However, this can be overcome by using the relation \re{VacuumGammaMinus}, such that one gets for $u + i (\ell/2 + 0_+)$,
\begin{align}
p_{\rm Goldstone} (u)
\equiv
p_{\rm g} (u^{[\ell]} + i 0_+)
=
i \ell + \frac{i}{2} \int_0^\infty \frac{dt}{t} {\rm e}^{i u t}  \left[ \frac{\gamma^{\o}_+ (2gt)}{{\rm e}^t - 1} +  \frac{\gamma^{\o}_- (2gt)}{1 - {\rm e}^{-t}} \right]
\left( {\rm e}^{- \ell t} - 1 \right)
\, .
\end{align}
This way we pass to another sheet of the Riemann surface, as shown in Fig.\ \ref{GluonMirror}, which is dubbed as the Goldstone sheet.
Notice that compared to the real sheet, $p_{\rm Goldstone} (u)$ has only a finite number of cuts due to partial cancellation between denominator
and denominator, for instance,
\begin{align}
I_+ (u)
\equiv
\int_0^\infty \frac{dt}{t}  {\rm e}^{i u t} \gamma^{\o}_+ (2gt) \frac{ {\rm e}^{- \ell t} - 1}{{\rm e}^t - 1} = - \sum_{n = 1}^\ell \lambda_+ (u + i n)
\, .
\end{align}
And analogously for the rest. This exhibits a finite number of cuts equidistant from each other along the imaginary axis as shown in Fig.\ \ref{GluonMirror}.
Since there are no other singularities, $p_{\rm Goldstone} (u)$ is analytic outside the strip $[-2g, 2g]$.

Next, we move outside the strip $[-2g, 2g]$, i.e., $|u| \geq 2g$ and go just below the bottom cut in the lower half-plane, $u \to u - i \ell$, on the Goldstone sheet. 
This is accomplished with the formula
\begin{align}
I_+ (u - i \ell) 
= 
- \sum_{n = 1}^\ell \lambda_+ (u - i (\ell - n)) = -  \sum_{n = 0}^{\ell - 1} \lambda_+ (u - i n) 
=
\int_0^\infty \frac{dt}{t} {\rm e}^{-iut} \gamma^{\o}_+ (2gt) \frac{{\rm e}^{- \ell t} - 1}{1 - {\rm e}^{- t}}
\, .
\end{align}
Here we used the fact that $\lambda_+ (u)$ is real for $|u|>2g$, see Eq.\ \re{lambdaPlus}. Analogously
\begin{align}
I_- (u)
\equiv
- i
\int_0^\infty \frac{dt}{t}  {\rm e}^{i u t} \gamma^{\o}_- (2gt) \frac{ {\rm e}^{- \ell t} - 1}{1 - {\rm e}^{-t}} 
\quad
\to
\quad
I_- (u - i \ell)
=
i
\int_0^\infty \frac{dt}{t}  {\rm e}^{- i u t} \gamma^{\o}_- (2gt) \frac{ {\rm e}^{- \ell t} - 1}{{\rm e}^{t} - 1} 
\, ,
\end{align}
since
\begin{align}
\lambda_- (u) 
= -i \int_0^\infty \frac{dt}{t} {\rm e}^{iut} \gamma^{\o}_- (2gt) =  \int_0^\infty \frac{dt}{t} \sin(ut) \gamma^{\o}_- (2gt) =  i \int_0^\infty \frac{dt}{t} {\rm e}^{- iut} \gamma^{\o}_- (2gt)
\, .
\end{align}
So that we get
\begin{align}
\label{pGoldBelow}
p_{\rm Goldstone} (u - i \ell)
=
i \ell + \frac{i}{2} \int_0^\infty \frac{dt}{t} 
\left[ 
\frac{\gamma^{\o}_+ (2gt)}{1 - {\rm e}^{-t}}
-
\frac{\gamma^{\o}_- (2gt)}{{\rm e}^{t} - 1}
\right] 
{\rm e}^{-iut} \left( {\rm e}^{- \ell t} - 1 \right)
\, ,
\end{align}
and move under the cut $[-2g - i \ell /2, +2g - i \ell/2]$ by assuming that $|u| < 2g$.

To cross the cut and move back into the real sheet, we have to make the shift $u \to u+i 0_+$. To this end we recognize that only the term involving $\gamma^{\o}_+$
and the last exponent in the brackets in Eq.\ \re{pGoldBelow} needs to be addressed. This is handled again with the help of \re{AnContInt} and the flux-tube equations 
\re{VacuumGamma}, yielding Eq.\ \re{VacuumGammaPlus}, and thus
\begin{align}
p_{\rm g} (u - i \ell/2) 
&
\equiv p_{\rm Goldstone} (u +i 0_+ - i \ell)
\nonumber\\
&
=
i \ell + \frac{i}{2} \int_0^\infty \frac{dt}{t} 
\left[ 
\frac{\gamma^{\o}_+ (2gt)}{1 - {\rm e}^{-t}}
-
\frac{\gamma^{\o}_- (2gt)}{{\rm e}^{t} - 1}
\right] 
\left( {\rm e}^{- \ell t/2} \cos (u^{[ -\ell]} t) - J_0 (2gt) \right)
\, .
\end{align}
The final step is trivial, $u \to u + i \ell/2$ and we find
\begin{align}
p_{\rm g} (u^\gamma) = i E_{\rm g} (u)
\, ,
\end{align}
by comparing the final expression with Eq.\ \re{gaugeBSenergy}.


\end{document}